%% file: ver_resubmission.tex
\documentclass[a4paper,dvipdfmx]{pasj01}
\draft

\usepackage{lscape}

\usepackage{longtable}

\usepackage{comment}

\usepackage{natbib, aas_macros}
\usepackage{color}
\begin{document} 
\Received{}
\Accepted{}

\title{A systematic study of ULIRGs using near-infrared absorption bands reveals a strong UV environment in their star-forming regions}

\author{Ryosuke \textsc{Doi},\altaffilmark{1,2} Takao \textsc{Nakagawa},\altaffilmark{1} Naoki \textsc{Isobe},\altaffilmark{1} Shunsuke \textsc{Baba},\altaffilmark{1} Kenichi \textsc{Yano},\altaffilmark{1,2} Mitsuyoshi \textsc{Yamagishi},\altaffilmark{1}}
\altaffiltext{1}{Institute of Space and Astronautical Science (ISAS), Japan Aerospace Exploration Agency (JAXA), 3-1-1 Yoshinodai, Chuo-ku, Sagamihara, Kanagawa 252-5210, Japan}
\altaffiltext{2}{Department of Physics, The University of Tokyo, 7-3-1 Hongo, Bunkyo-ku, Tokyo 133-0033, Japan}
\email{r-doi@ir.isas.jaxa.jp}


\maketitle

\begin{abstract}
\input{./input/abst_12}
\end{abstract}

\section{Introduction}
\input{./input/intro_10}

\section{Observation and spectral analysis}
\input{./input/obsana_15_draft}

\section{Results}
\input{./input/result_21_draft}

\section{Discussion}
\input{./input/discussion_2_15_draft}

\section{Summary and conclusion}
\input{./input/conclusion_10}


\input{./input/references_pasj}
\end{document}

%% file: input/abst_12.tex
 We present a systematic study of the 3.0 $\mathrm{\,\mu m}$ H$_{2}$O ice and the 3.4 $\mathrm{\,\mu m}$ aliphatic carbon absorption features toward 48 local ultraluminous infrared galaxies (ULIRGs) using spectra obtained by the AKARI Infrared Camera to investigate the UV environment in their star-forming regions.
 All the ULIRGs in our sample exhibit a ratio of optical depth of H$_{2}$O ice to silicate dust ($\tau_{3.0}/\tau_{9.7}$) that is lower than that in the Taurus dark cloud.
  This implies that ULIRGs cannot be described as an ensemble of low-mass star-forming regions and that a significant amount of high-mass star-forming regions contribute to star-forming clouds in local ULIRGs.
  The results also show that the ratios of optical depth of aliphatic carbon to silicate dust, $\tau_{3.4}/\tau_{9.7}$,  exhibit diverse values.
  We investigate two effects that can affect this ratio: the geometric temperature gradient (which increases the ratio) and the intense UV environment (which decreases it).
  The geometric temperature gradient is typically considered as a sign of active galactic nuclei (AGN). ULIRGs with AGN signs (optical classification, near-infrared color, and a PAH emission strength of 3.3 $\mathrm{\,\mu m}$) indeed tend to exhibit a large $\tau_{3.4}/\tau_{9.7}$ ratio. However, we find that the presence of buried AGN is not the only cause of the geometric temperature gradient, because the enhancement of the ratio is also evident in pure starburst-like ULIRGs without these AGN signs.
  Regarding the intense UV environment in star-forming regions, the correlation between the aliphatic carbon ratio and the ratio of the [C \emissiontype{II}] 158 $\mathrm{\,\mu m}$ line luminosity to the far-infrared luminosity ($L_{[\mathrm{C} \emissiontype{II}]}/L_{\mathrm{FIR}}$), which represents the UV environment in photodissociation regions, implies that the intense UV environment causes the decrease of the aliphatic carbon ratio.
   We find that an intense UV environment ($G/n_{\mathrm{H}}>3$) in star-forming regions is needed for the aliphatic carbon ratio to be suppressed.

%% file: input/intro_10.tex
 Ultraluminous infrared galaxies (ULIRGs) are dust-obscured galaxies luminous in the infrared region ($L_{\mathrm{IR}}>10^{12}\,L_{\odot}$). 
 ULIRGs are characterized in spectroscopy by very deep silicate dust absorption features, which indicate the substantial presence of dust inside ULIRGs \citep{2007ApJ...655L..77H}.
 From the perspective of galactic evolution, local ULIRGs are regarded as galaxies transitioning from a merger phase to a quasar phase \citep{1988ApJ...325...74S}. It is also believed that in their evolutionary path, the star formation rate (SFR) peaks in the ULIRG stage (SFR $\gtrsim 100 M_{\odot}/\mathrm{yr}$; \citealt{2008ApJS..175..356H}) .
  It is thus important to understand physical conditions in star-forming clouds in ULIRGs.
  
 The ultraviolet (UV) environment is one such condition, because it not only heats dust but also can change the chemistry that affects the temperature in clouds. 
 In spite of its importance, the UV environment in star-forming clouds in ULIRGs is not well understood because of heavy extinction (in terms of $A_{V}$, $A_{V} \gtrsim 15 \mathrm{\,mag}$;  \citealt{1998ApJ...498..579G}).
 Although studies of the forbidden line from ionized carbon atoms at 157.7 $\mathrm{\,\mu m}$ have suggested that ULIRGs have a strong UV environment (e.g., \citealt{1998ApJ...504L..11L}; \citealt{2003ApJ...594..758L}; \citealt{2013ApJ...774...68D}; \citealt{2001ApJ...561..766M}; \citealt{2013ApJ...776...38F}), assumptions such as that regarding the origin of the line make interpretations somewhat complex and indirect.
 
 Another way to observe the UV environment to which dust in star-forming regions in ULIRGs is exposed is observing features related to dust itself.
 This study focuses on two near-infrared dust absorption features: the H$_{2}$O ice absorption (peaking at 3.0$\mathrm{\,\mu m}$) and the aliphatic carbon absorption (3.4$ \mathrm{\,\mu m}$).
 In the Milky Way, H$_{2}$O ice is observationally detected only toward regions with sufficient extinction. The threshold extinction of H$_{2}$O ice has been investigated for various clouds in the Milky Way. For example, the Taurus dark cloud exhibits a threshold of $A_{V} >$ several mag (e.g., \citealt{2001ApJ...547..872W}; \citealt{2013ApJ...777...73B}). A larger threshold was claimed for a high-mass star-forming region, reflecting a stronger UV environment \citep{1990ApJ...352..724T}.
 The existence of thresholds means that the H$_{2}$O ice absorption traces a region where the UV environment is well attenuated.
 
 On the other hand, aliphatic carbon has been detected toward the Galactic center (GC) or Cygnus OB2 (\citealt{1990MNRAS.243..400A}; \citealt{1994ApJ...437..683P}), but it is suppressed in molecular clouds such that H$_{2}$O ice is detected \citep{1996ApJ...472..665C}.
 This means that the aliphatic carbon absorption of 3.4 $\mathrm{\,\mu m}$ traces a region with $A_{V} \lesssim$ several mag, a UV environment different from that traced by the H$_{2}$O ice absorption.

 Although the H$_{2}$O ice and aliphatic carbon absorptions are useful for investigating the UV environment, the evaluation of these absorption features in ULIRGs has faced some difficulties.
 First, the shorter wavelength side of the H$_{2}$O ice absorption spectra ($\lambda \lesssim 2.8\mathrm{\,\mu m}$) is difficult to observe from the ground due to atmospheric absorption. 
 Therefore, previous ground-based observations toward ULIRGs (e.g., \citealt{2006ApJ...637..114I}; \citealt{2006MNRAS.365..303R}) that detected these features have often needed to use \textit{extrapolated} continuum levels from the longer wavelength side of the spectra to estimate the optical depth of H$_{2}$O ice. However, an \textit{interpolated} continuum level is preferable for a robust estimate of optical depth, which is difficult from ground-based observations.
 Second, ULIRGs usually exhibit an emission feature at 3.3$\mathrm{\,\mu m}$, which is attributed to C--H bonds of polycyclic aromatic hydrocarbons (PAHs; \citealt{1981MNRAS.196..269D}).
 This feature partly overlaps with the 3.0$\mathrm{\,\mu m}$ H$_{2}$O ice and the $3.4\mathrm{\,\mu m}$ aliphatic carbon absorption features. 
 In determining the optical depths, it is thus important to consider the PAH emission simultaneously with the H$_{2}$O ice and aliphatic carbon absorptions.
Observing spectra ranging from $2.5\mathrm{\,\mu m}$ to $4.0\mathrm{\,\mu m}$ is needed to overcome these difficulties.

 AKARI, the Japanese infrared telescope launched in 2006 \citep{2007PASJ...59S.369M}, offers the best opportunity to observe the H$_{2}$O ice and the aliphatic carbon absorption features. The Infrared Camera (IRC) onboard AKARI has a near-infrared grism spectroscopic mode that provides continuous 2.5--5.0$\mathrm{\,\mu m}$ spectral coverage free from telluric absorption, with good sensitivity for observing ULIRGs and enough spectral resolution to detect the features; the spectral resolution of IRC is $\lambda/\Delta\lambda \simeq 120$ at $3.6\mathrm{\,\mu m}$ \citep{2007PASJ...59S.411O}, whereas these features need $\lambda/\Delta\lambda\gtrsim 20$ to be resolved.
 
 We present the 2.5--4.0$\mathrm{\,\mu m}$ (rest frame) spectra of 48 local ULIRGs obtained by the AKARI IRC.
 A summary of observations and a detailed description of our spectral fitting analysis are presented in section 2.  The results of spectral fitting are discussed in section 3. A comparison with Galactic sources and interpretations are presented in section 4. Section 5 provides a summary and conclusion.

%% file: input/obsana_15_draft.tex

\subsection{Observation and data reduction}
\label{sec:obs_red}

Because of our spectral range of interest (2.5--4.0$\mathrm{\,\mu m}$), we used observations made by the AKARI IRC's near-infrared channel, in which grism spectroscopic observations cover 2.5--5.0 $\mathrm{\,\mu m}$ (\citealt{2007PASJ...59S.401O}; \citealt{2007PASJ...59S.411O}).
 The spectra of local ULIRGs we analyzed were taken from the AKARI mission program ``AGNUL'' (P.I. T. Nakagawa) which was focused on study of nearby ULIRGs and AGN. The objects observed were mostly selected from the IRAS 1 Jy sample \citep{1998ApJS..119...41K}. In general, AKARI near-infrared spectra have better quality in phases 1 and 2, during which the entire instrument was cooled with a combination of liquid helium and mechanical coolers.
 In this paper, we analyzed the spectra observed in this period. The targets are listed in table \ref{tab:target}. 
 IRC data reduction was carried out with the IRC Spectroscopy Toolkit version 20150331 provided by the AKARI team.
 Reduction was carried out using the procedure they recommended. Eight raw frames were stacked for constructing a 2D image after dark subtraction and bad pixels correction. The 2D image was then dispersed in the 1$' \times$ 1$'$ aperture and a spectrum was extracted within $\sim$ 5 pixels ($\sim$ 5$\times$1$''$.46; \citealt{2007PASJ...59S.401O}) along a direction perpendicular to the dispersion direction.
 An example of the spectra we obtained is shown in figure \ref{fig:spec_ex}.
 \citet{2008PASJ...60S.489I} also presented a part of the AGNUL spectra presented in this paper. However, the reduction toolkit has since been greatly improved recently in such as the spectral response function. The updated toolkit thereby reduces spectral fringes and provides spectra with a better signal-to-noise ratio. We therefore adopted the spectra reduction we ourselves performed.

\begin{longtable}{cccccc}
\caption{Target data.}
 \\ \hline
\label{tab:target}
		Object Name	& Observation ID	& Redshift$^{*}$ 	&	R.A. (J2000)	&	Dec.	(J2000)	&	Optical Classification$^{\dagger}$ \\ \hline	  
		\endfirsthead
		
		\multicolumn{6}{p{140mm}}{(Continued)} \\
	\hline
	Object Name	& Observation ID	& Redshift$^{*}$ 	&	R.A. (J2000)	&	Dec.	(J2000)	&	Optical Classification$^{\dagger}$ \\ 
	\hline
	\endhead
	\multicolumn{6}{p{140mm}}{$^{*}$ The values are taken from the NASA/IPAC Extragalactic Database (NED) or SIMBAD.
		$^{\dagger}$ Optical classification by \citet{1999ApJ...522..113V} unless otherwise noted. 
		$^{\ddagger}$ Classification by \citet{1997AAS..124..533D}. }
		\endlastfoot
    IRAS 00456$-$2904    &   1100221.1    &   0.110    &   00 48 06.7    &   $-$28 48 18.0    &   H\emissiontype{II}       \\
    IRAS 00482$-$2721    &   1100036.1    &   0.129    &   00 50 40.4    &   $-$27 04 37.9    &   LINER        \\
    IRAS 01199$-$2307    &   1100209.1    &   0.156    &   01 22 20.8    &   $-$22 52 05.2    &   H\emissiontype{II}        \\
    IRAS 01298$-$0744    &   1100226.1    &   0.136    &   01 32 21.4    &   $-$07 29 08.2    &   H\emissiontype{II}        \\
    IRAS 01355$-$1814    &   1100018.1    &   0.192    &   01 37 57.4    &   $-$17 59 20.0    &   H\emissiontype{II}        \\
    IRAS 01494$-$1845    &   1100215.1    &   0.158    &   01 51 51.4    &   $-$18 30 47.2    &   Unclassified        \\
    IRAS 01569$-$2939    &   1100225.1    &   0.140    &   01 59 13.7    &   $-$29 24 33.8    &   H\emissiontype{II}        \\
    IRAS 02480$-$3745    &   1100030.1    &   0.165    &   02 50 01.7    &   $-$37 32 44.9    &   Unclassified        \\
    IRAS 03209$-$0806    &   1100210.1    &   0.166    &   03 23 22.8    &   $-$07 56 15.0    &   H\emissiontype{II}        \\
    IRAS 03521$+$0028   &   1100200.1    &   0.152    &   03 54 42.2    &   $+$00 37 01.9    &   LINER        \\
    IRAS 04074$-$2801    &   1100201.1    &   0.154    &   04 09 30.5    &   $-$27 53 44.2    &   LINER        \\
    IRAS 04313$-$1649    &   1100031.1    &   0.268    &   04 33 37.1    &   $-$16 43 32.2    &   Unclassified        \\
    IRAS 05020$-$2941    &   1100003.1    &   0.154    &   05 04 00.7    &   $-$29 36 54.0    &   LINER        \\
    IRAS 05189$-$2524    &   1100129.1    &   0.043    &   05 21 01.5    &   $-$25 21 45.0    &   Seyfert 2        \\
    IRAS 06035$-$7102    &   1100130.1    &   0.079    &   06 02 54.0    &   $-$71 03 10.1    &   H\emissiontype{II}$^{\ddagger}$        \\
    IRAS 08572$+$3915   &   1100049.1    &   0.058    &   09 00 25.4    &   $+$39 03 55.1    &   LINER        \\
    IRAS 08591$+$5248   &   1100121.1    &   0.157    &   09 02 48.9    &   $+$52 36 24.1    &   Unclassified        \\
    IRAS 09463$+$8141   &   1100004.1    &   0.156    &   09 53 00.1    &   $+$81 27 28.1    &   LINER        \\
    IRAS 09539$+$0857   &   1100267.1    &   0.129    &   09 56 34.3    &   $+$08 43 05.2    &   LINER        \\
    IRAS 10035$+$2740   &   1100216.1    &   0.166    &   10 06 26.4    &   $+$27 25 45.8    &   Unclassified        \\
    IRAS 10091$+$4704   &   1100122.1    &   0.246    &   10 12 16.7    &   $+$46 49 43.0    &   LINER        \\
    IRAS 10494$+$4424   &   1100266.1    &   0.092    &   10 52 23.5    &   $+$44 08 46.0    &   LINER        \\
    IRAS 10594$+$3818   &   1100021.1    &   0.158    &   11 02 14.0    &   $+$38 02 34.1    &   H\emissiontype{II}        \\
    IRAS 11028$+$3130   &   1100006.1    &   0.199    &   11 05 37.5    &   $+$31 14 30.8    &   LINER        \\
    IRAS 11180$+$1623   &   1100202.1    &   0.166    &   11 20 41.7    &   $+$16 06 56.2    &   LINER        \\
    IRAS 11387$+$4116   &   1100269.1    &   0.149    &   11 41 22.0    &   $+$40 59 49.9    &   H\emissiontype{II}        \\
    IRAS 12447$+$3721   &   1100022.1    &   0.158    &   12 47 07.8    &   $+$37 05 37.0    &   H\emissiontype{II}        \\
    Mrk 231    		       &   1100271.1    &   0.042    &   12 56 14.3    &   $+$56 52 25.0    &   Seyfert 1     \\
    Mrk 273    		       &   1100273.1    &   0.038    &   13 44 42.1    &   $+$55 53 12.8    &   Seyfert 2        \\
    IRAS 13469$+$5833   &   1100023.1    &   0.158    &   13 48 40.1    &   $+$58 18 51.8    &   H\emissiontype{II}        \\
    IRAS 13539$+$2920   &   1100235.1    &   0.108    &   13 56 09.9    &   $+$29 05 35.2    &   H\emissiontype{II}        \\
    IRAS 14121$-$0126    &   1100011.1    &   0.150    &   14 14 45.5    &   $-$01 40 55.9    &   LINER        \\
    IRAS 14202$+$2615   &   1100212.1    &   0.159    &   14 22 31.3    &   $+$26 02 04.9    &   H\emissiontype{II}        \\
    IRAS 14394$+$5332   &   1100283.1    &   0.105    &   14 41 04.3    &   $+$53 20 08.2    &   Seyfert 2        \\
    IRAS 15043$+$5754   &   1100213.1    &   0.151    &   15 05 39.6    &   $+$57 43 09.1    &   H\emissiontype{II}        \\
    IRAS 16333$+$4630   &   1100013.1    &   0.191    &   16 34 52.4    &   $+$46 24 52.9    &   LINER        \\
    IRAS 16468$+$5200   &   1100249.1    &   0.150    &   16 48 01.7    &   $+$51 55 44.0    &   LINER        \\
    IRAS 16487$+$5447   &   1100247.1    &   0.104    &   16 49 46.8    &   $+$54 42 34.9    &   LINER        \\
    IRAS 17028$+$5817   &   1100248.1    &   0.106    &   17 03 41.9    &   $+$58 13 44.0    &   LINER        \\
    IRAS 17044$+$6720   &   1100297.1    &   0.135    &   17 04 28.4    &   $+$67 16 28.9    &   LINER        \\
    IRAS 17068$+$4027   &   1100026.1    &   0.179    &   17 08 32.0    &   $+$40 23 28.0    &   H\emissiontype{II}        \\
    IRAS 17179$+$5444   &   1100253.1    &   0.147    &   17 18 54.4    &   $+$54 41 48.1    &   Seyfert 2        \\
    IRAS 19254$-$7245   &   1100132.1    &   0.062    &   19 31 21.6    &   $-$72 39 20.9    &   Seyfert 2        \\
    IRAS 21477$+$0502   &   1100207.1    &   0.171    &   21 50 16.4    &   $+$05 16 03.0    &   LINER        \\
    IRAS 22088$-$1831   &   1100214.1    &   0.170    &   22 11 33.8    &   $-$18 17 06.0    &   H\emissiontype{II}       \\
    IRAS 23129$+$2548   &   1100015.1    &   0.179    &   23 15 21.4    &   $+$26 04 32.9    &   LINER       \\
    IRAS 23128$-$5919   &   1100294.1    &   0.045    &   23 15 46.8    &   $-$59 03 15.8    &   H\emissiontype{II}$^{\ddagger}$        \\
    IRAS 23498$+$2423   &   1100287.1    &   0.212    &   23 52 26.1    &   $+$24 40 16.0    &   Seyfert 2        \\
		\hline
\end{longtable}

\subsection{Spectral analysis}
\label{sec:spec_ana}
The observed spectra ($\lambda_{\mathrm{rest}} = $ 2.5--4.0$\mathrm{\,\mu m}$) contain various interstellar medium (ISM) features, which can also be found in the Milky Way (e.g., \citealt{2014ApJ...784...53M}), in addition to the H$_{2}$O ice absorption and the aliphatic carbon absorption features. Some features overlap and need to be analyzed consistently. 

 We thus carried out spectral fitting with various features together. The features considered in fitting are summarized in table \ref{tab:feature}. The PAH emission that peaks at 3.3 $\mathrm{\,\mu m}$ is the most prominent feature among them.  It is known that this feature is often accompanied by weaker features (hereafter PAH subfeatures) that peak at 3.41$\mathrm{\,\mu m}$ and 3.48$\mathrm{\,\mu m}$ \citep{2014ApJ...784...53M}.  The exact origin of the PAH subfeatures is unclear, but -- CH$_{3}$ and/or -- CH$_{2}$ bonds on PAH molecules are thought to be involved \citep[][and references therein]{2012ApJ...760L..35L}.

 The Br$\beta$ line ($\lambda_{\mathrm{rest}}=$2.63$\mathrm{\,\mu m}$) can also be seen in this spectral range. Other weaker hydrogen recombination lines (Pfund series) could be seen, but not as clearly as Br$\beta$. For example,  Pf$\gamma$, peaking at $3.74\mathrm{\,\mu m}$, is the second strongest feature after Br$\beta$ in the wavelength range, but in the case B, $T=10^{4}\,\mathrm{K}$ condition, the Pf$\gamma$/Br$\beta$ intensity ratio is 0.16 \citep{1987MNRAS.224..801H}.

\begin{figure}
 \begin{center}
 \hspace{-50mm}
 \raisebox{-100mm}{\includegraphics[width=0.9\linewidth]{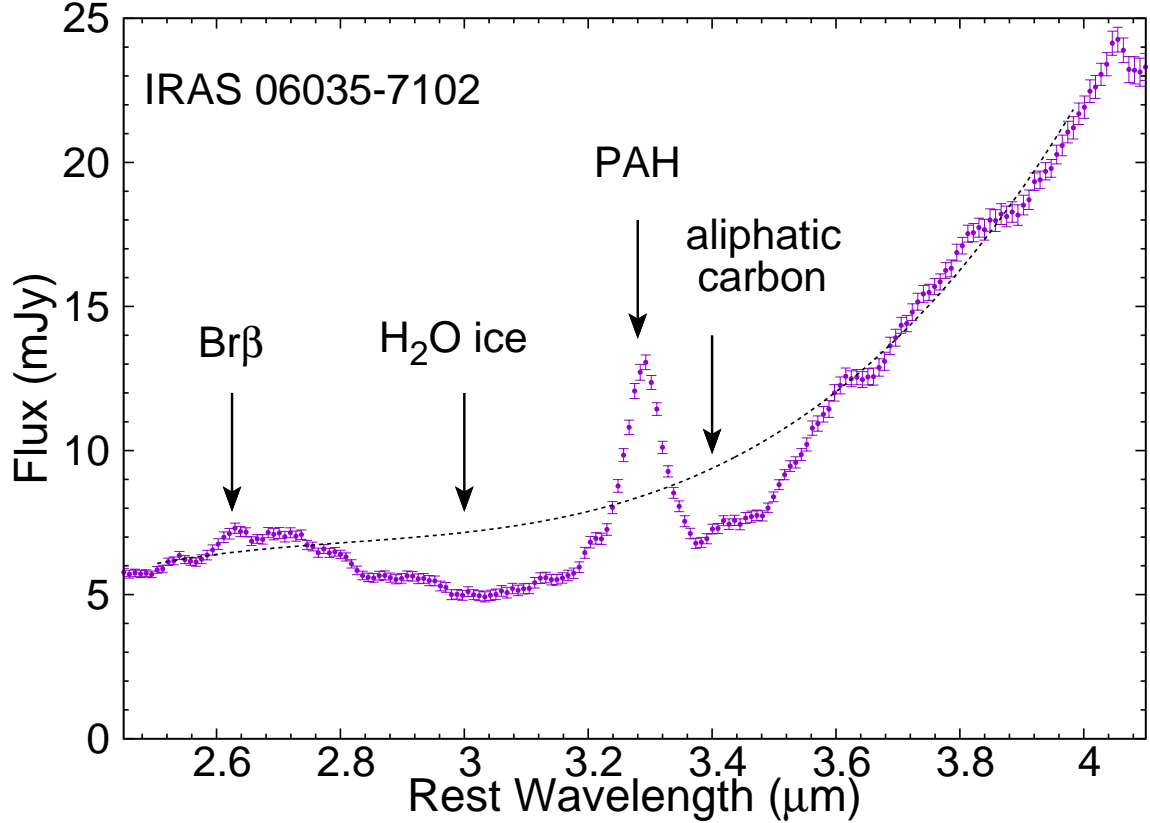}}
 \end{center}
 \caption{An example of the spectra obtained by the AKARI IRC grism spectroscopic observations (IRAS 06035$-$7102). The spectra contain various spectral features such as hydrogen recombination lines and PAH emissions, in addition to the H$_{2}$O ice absorption and the aliphatic carbon absorption. The dotted black line denotes the continuum component of the best-fit spectrum (see section \ref{sec:spec_ana} for details).}
 \label{fig:spec_ex}
 \end{figure}

The function with which we performed spectral fitting is expressed by

\begin{eqnarray}
	F_{\nu}(\lambda)&=& \displaystyle \left[ \sum_{i=0}^{3}a_{i}\lambda ^{i} + \sum_{j} b_{j}f_{j}(\lambda)\right] \nonumber \\
							&\times&\exp\left[-c_{\mathrm{ice}}\tau_{\mathrm{ice}}(\lambda) \right]\exp\left[-c_{\mathrm{ali}}\tau_{\mathrm{ali}} (\lambda)\right],
	\label{eq:function}
\end{eqnarray}
\noindent
where $a_{i}$ represent the coefficients of the polynomial, $f_{j}$ the shape of the emission features, $b_{j}$ the scaling factors for the emission features,  $\tau_{\mathrm{ice}}(\lambda)$ the optical depth profile of the H$_{2}$O ice absorption, $c_{\mathrm{ice}}$ the scaling factor for the H$_{2}$O ice absorption, $\tau_{\mathrm{ali}} (\lambda)$ the optical depth profile of the aliphatic absorption, and $c_{\mathrm{ali}}$ the scaling factor for the aliphatic carbon absorption. The fitting parameters are $a_{i}$, $b_{j}$, $c_{\mathrm{ice}}$, and $c_{\mathrm{ali}}$.

\begin{figure}
 \begin{center}
 \hspace{-50mm}
 \raisebox{-100mm}{\includegraphics[width=0.9\linewidth]{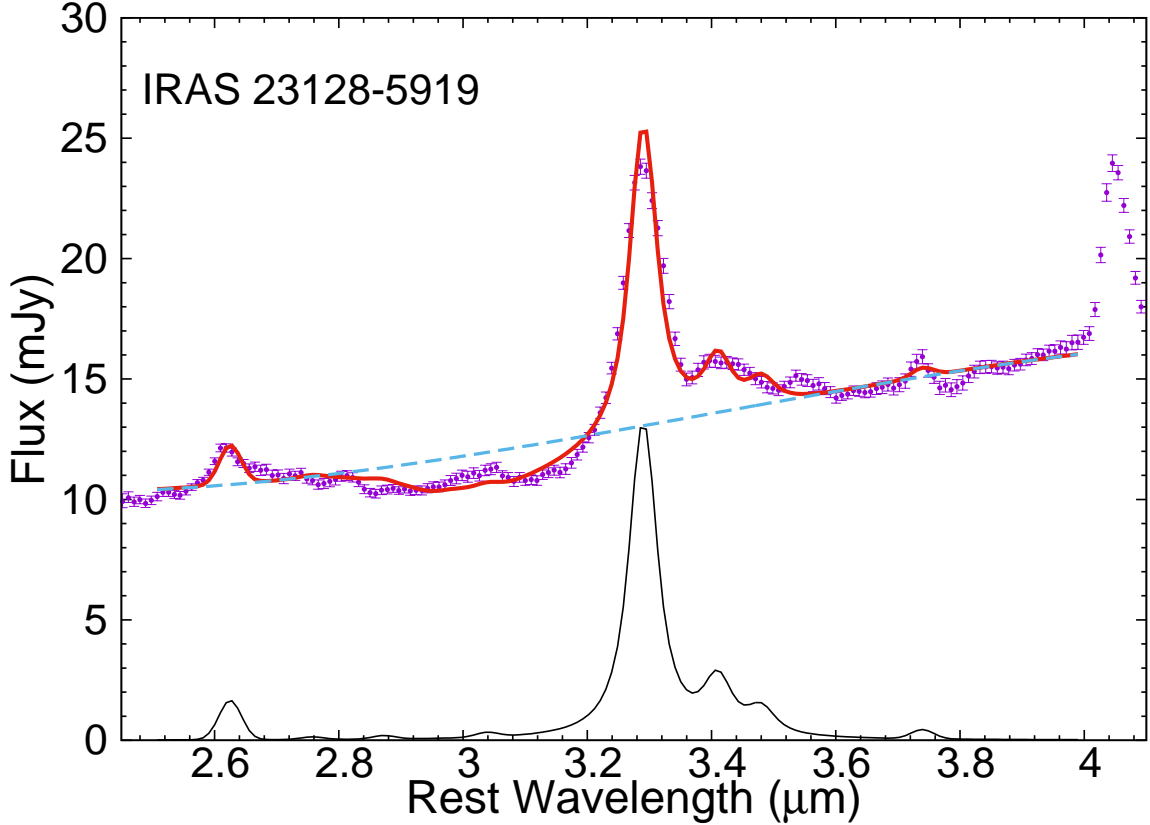}}
 \end{center}
 \caption{The spectrum used for the estimate of the proportional coefficients of the PAH subfeatures. Upper panel: The points with error bars represent the observed flux. The solid red line, the dashed blue line, and the black line respectively represent the best-fit spectrum, its continuum component, and its emission components (PAH emissions and hydrogen recombination lines). See section \ref{sec:spec_ana} for details of the fitting.}
 \label{fig:23128}
 \end{figure}
 
 We adopted the Lorentzian profile as $f_{j}$ for emissions associated with PAH. The spectral resolution of AKARI IRC NIR grism spectroscopy ($\lambda / \Delta\lambda\simeq 120$) is not sufficient to distinguish the PAH subfeatures from the aliphatic carbon absorption feature; therefore, these features cannot be determined separately. We thus assumed that the intensities of the PAH subfeatures were proportional to that of the 3.3$\mathrm{\,\mu m}$ PAH feature, which is clearly evident in the spectra (figure \ref{fig:spec_ex}). 
 The $I_{3.4}/I_{3.3}$ ratio, where the numerator is the intensity of the PAH subfeature at 3.4$\mathrm{\,\mu m}$ and the denominator is the intensity of the PAH feature at 3.3$\mathrm{\,\mu m}$, has not been established yet in ULIRGs.
 We therefore estimated the proportional coefficients from the spectrum of IRAS 23128$-$5919 (figure \ref{fig:23128}), which exhibits the most evident PAH subfeatures among our sample. The estimated intensity ratio is $I_{3.4}/I_{3.3}=0.17 \pm 0.01$.
 We assume $I_{3.4}/I_{3.3}=0.17$ for other ULIRGs in our sample.
 In order to check the validity of the assumption, we then estimated the $I_{3.4}/I_{3.3}$ ratio for another ULIRG, IRAS 00456$-$2904, which also exhibits the PAH subfeatures relatively clearly (figure \ref{fig:hii_spec}), and found that $I_{3.4}/I_{3.3}=0.14 \pm 0.01$.
 These values ($I_{3.4}/I_{3.3}=0.17$ in IRAS 23128$-$5919, $0.14$ in IRAS 00456$-$2904) are close to those found in typical Galactic sources ($I_{3.4}/I_{3.3}\sim0.2$ on average; \citealt{1996ApJ...458..610J}).
 We hence adopt $I_{3.4}/I_{3.3}=0.17$ found in IRAS 23128$-$5919, because the PAH subfeatures are most evident in this ULIRG. The effect of a possible variation of the $I_{3.4}/I_{3.3}$ ratio on the fitting will be discussed further in section \ref{sec:aof}.
 In order to make the fitting converge properly, all the full width at half maximums (FWHMs) in the fitting were also fixed. The FWHM for PAH was the averaged value determined by fitting, where we let it be a free parameter in addition to the other free parameters, described below. We assumed that the profiles of the PAH subfeatures were also the Lorentzian, and that their FWHMs were same as the 3.3 $\mathrm{\,\mu m}$ PAH main feature.
 
 We adopted the Gaussian profiles as $f_{j}$ for hydrogen recombination lines. We included Br$\beta$ and other weaker recombination lines in the fitting.
 The ratios of the intensity of the weak hydrogen recombination lines to Br$\beta$ were fixed on the basis of the case B condition in $T=10^{4}\,\mathrm{K}$.
 The assumption regarding temperature does not seriously affect the results because of the weak temperature dependence on the intensity of these features \citep[$T=5000$ -- $20000$ K; ][]{1987MNRAS.224..801H}. 
 Regarding the hydrogen recombination lines, the widths were determined by the spectral resolution. Note that the spectral resolution of AKARI IRC NIR grism spectroscopy corresponds to $\Delta v \simeq 2,500\,\mathrm{km/s}$, which is much larger than the typical rotational speed of a galaxy. It is thus expected that the hydrogen recombination lines will not have a larger FWHM than the value we adopted.

 We used laboratory and astronomical data to represent the H$_{2}$O ice and the aliphatic carbon absorption features, respectively.
 For the H$_{2}$O ice absorption feature, we adopted a laboratory extinction spectrum of NH$_{3}$:H$_{2}$O=100:20, $T=10\mathrm{\,K}$ \citep{1996AA...312..289G} since in Galactic molecular clouds, observed 3.0 $\mathrm{\,\mu m}$ H$_{2}$O ice spectra need NH$_{3}$ contribution to express the ice profile (\citealt{1982ApJ...260..141K}). NH$_{3}$ forms ammonium hydrate groups, which make the absorption somewhat deeper in the 3.2--3.7 $\mathrm{\,\mu m}$ region.
 
\begin{figure}
 \begin{center}
 \hspace{-50mm}
 \raisebox{-100mm}{\includegraphics[width=0.9\linewidth]{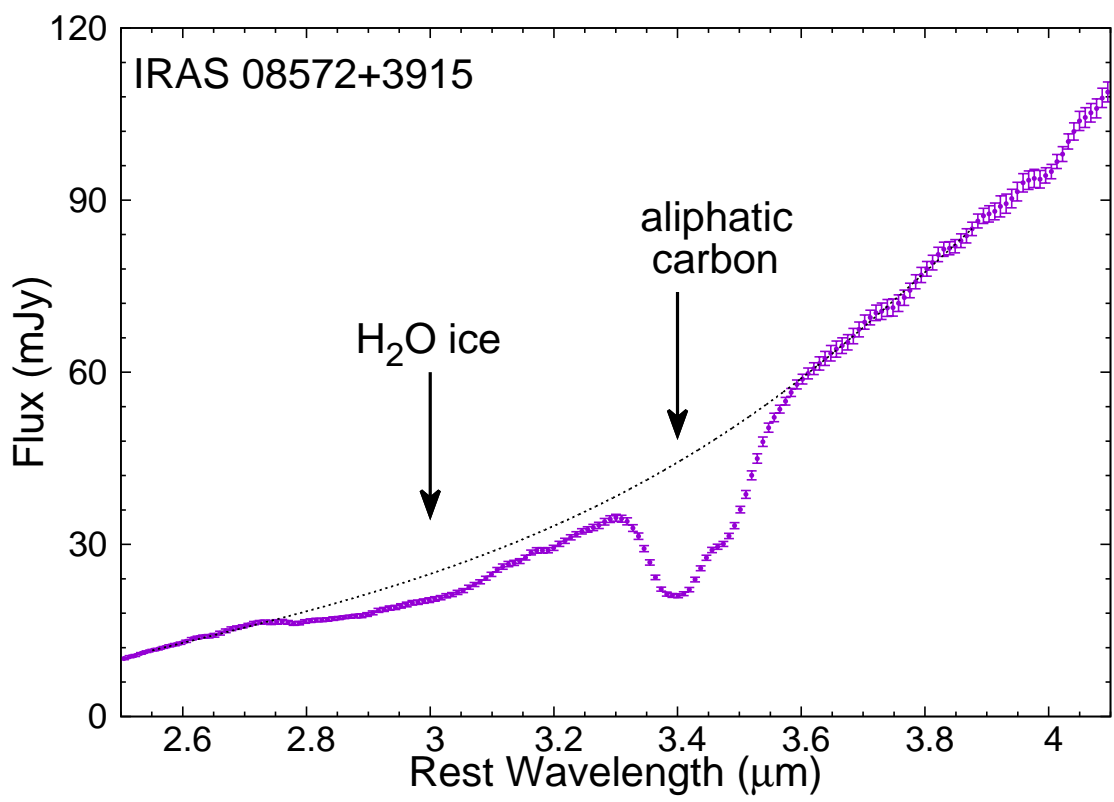}}
 \end{center}
 \caption{The spectrum of IRAS 08572$+$3915 that is used to extract a profile of the aliphatic carbon absorption. The dotted black line denotes the continuum component.}
 \label{fig:hac_obj}
 \end{figure}

\begin{figure}
 \begin{center}
 \hspace{-50mm}
 \raisebox{-100mm}{\includegraphics[width=0.9\linewidth]{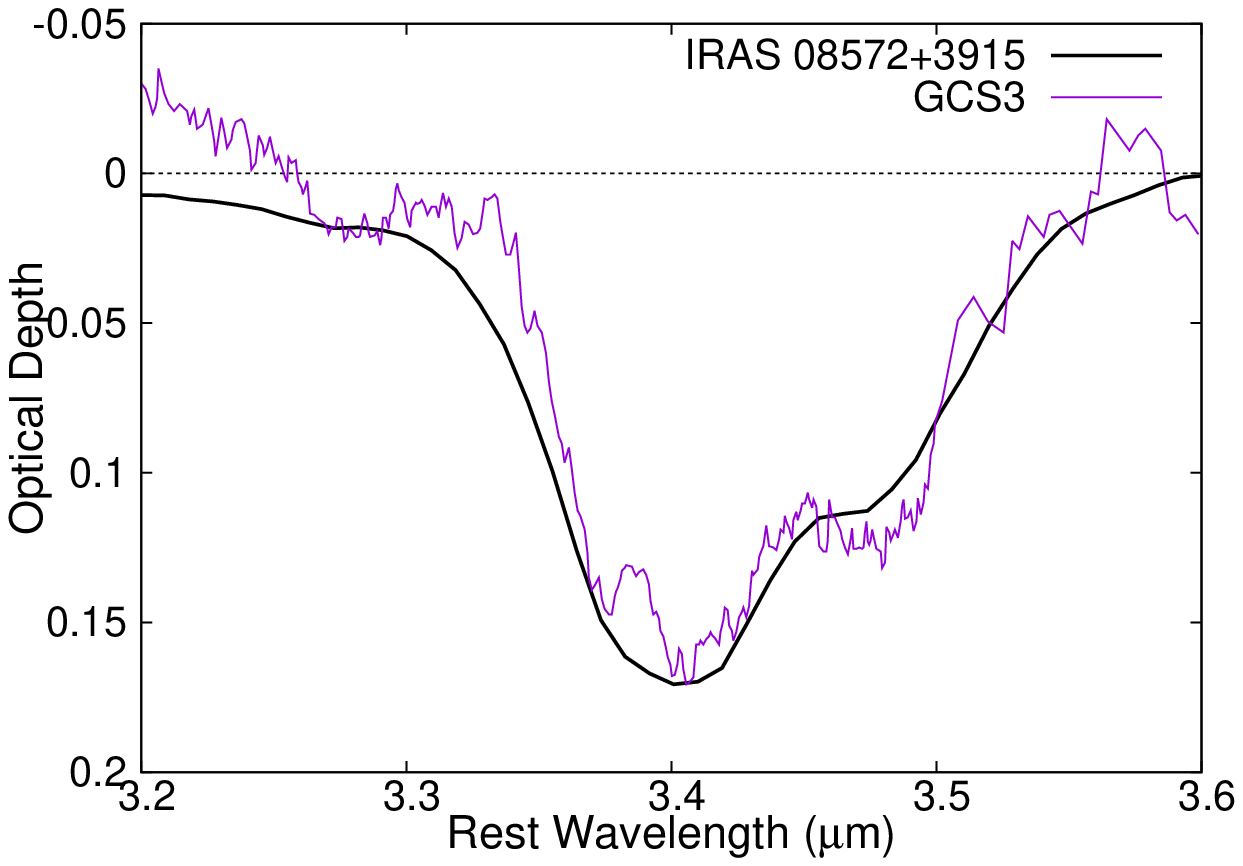}}
 \end{center}
 \caption{The profile of the aliphatic carbon absorption. Thick black line: the profile extracted from IRAS 08572$+$3915. Thin purple line: the profile observed in GCS 3 \citep{2000ApJ...537..749C}. The profile of IRAS 08572$+$3915 is x0.17 scaled to that of GCS 3.}
 \label{fig:hac_spec}
 \end{figure}

 The astronomical spectrum of IRAS 08572$+$3915 obtained by the IRC was used to represent the shape of $\tau_{\mathrm{ali}} (\lambda)$, because it exhibited no obvious PAH emissions that would contaminate the aliphatic carbon absorption. The continuum component was determined using a cubic spline. Four pivots ($\lambda_{\mathrm{rest}}=2.55$, $2.73$, $3.62$, $3.88\mathrm{\,\mu m}$), which were chosen avoiding the potential Br$\alpha$ and Br$\beta$ recombination lines, were used to determine the spline curve. The profile was extracted after removal of absorption by H$_{2}$O ice. The spectrum of IRAS 08572$+$3915 and obtained $\tau_{\mathrm{ali}} (\lambda)$ are shown in figure \ref{fig:hac_obj} and figure \ref{fig:hac_spec}, respectively. 
 Figure \ref{fig:hac_spec} also shows for reference the profile obtained toward a Galactic source GCS 3  \citep{2000ApJ...537..749C}, which is located near the GC and known as a member of the Quintuplet cluster (\citealt{1990ApJ...351...83N}; \citealt{1990ApJ...351...89O}). Those profiles have similar peak positions and shapes. We therefore use the profile seen in IRAS 08572$+$3915 as the template of $\tau_{\mathrm{ali}} (\lambda)$ in this study.

All the free parameters in the fitting were $a_{i}$ ($i=0$--3), $b_{\mathrm{3.3\,\mu m\,PAH}},\,b_{\mathrm{Br\beta}}$, $c_{\mathrm{ice}},\,c_{\mathrm{ali}}$, and a least squares fit was applied to the spectra. The hydrogen recombination lines were included in the fitting function only if their inclusion significantly improved $\chi^{2}$ with a 99\% confidence level ($\chi^{2}_{\mathrm{w/o\,H}}-\chi^{2}_{\mathrm{w/\,H}}\gtrsim 7$ for the typical degrees of freedom $\simeq$ 170; \citealt{2003drea.book.....B}). The other features were included in the fitting regardless of their detection levels.

\begin{table*}
\caption{The features considered in the fitting.}
\label{tab:feature}
   \begin{center}
	\begin{tabular}{cccc} \hline
		Feature 					&  Wavelength ($\mathrm{\mu m}$)  	&	Free 	parameter						&	Fixed parameter \\ \hline
		H {\footnotesize I} Br$\beta$	&  2.625 						&		strength ($b_{\mathrm{3.3\,\mu m\,PAH}}$)					& FWHM$^{*}$\\ 
		H {\footnotesize I} Pf$\eta$	&  2.758						&		\rule[0.8mm]{0.5cm}{0.15mm} 	& relative strength to H {\footnotesize I} Br$\beta\,^{\dagger}$\\ 
		H {\footnotesize I} Pf$\zeta$	&  2.872 						&		 \rule[0.8mm]{0.5cm}{0.15mm} 	& relative strength to H {\footnotesize I} Br$\beta\,^{\dagger}$\\ 
		H$_{2}$O ice				&  3.0 						&		peak optical depth ($c_{\mathrm{ice}}$) 			& shape\\ 
		H {\footnotesize I} Pf$\epsilon$ 	&  3.038						&		\rule[0.8mm]{0.5cm}{0.15mm} 	& relative strength to H {\footnotesize I} Br$\beta\,^{\dagger}$\\ 
		PAH						& 3.29						&		strength ($b_{\mathrm{Br\beta}}$)					& FWHM$^{*}$\\ 
		Aliphatic Carbon			& 3.4							&		peak optical depth ($c_{\mathrm{ali}}$)			& shape\\ 
		H-added PAH				& 3.41						&		 \rule[0.8mm]{0.5cm}{0.15mm} & relative strength to PAH$\,^{\ddagger}$\\ 
		H-added PAH				& 3.48						&		 \rule[0.8mm]{0.5cm}{0.15mm} & relative strength to PAH$\,^{\ddagger}$\\ 
		H {\footnotesize I} Pf$\gamma$ 	&  3.740					&		\rule[0.8mm]{0.5cm}{0.15mm} & relative strength to H {\footnotesize I} Br$\beta\,^{\ddagger}$\\ \hline
		\multicolumn{4}{p{146mm}}{Parentheses indicate the variables in equation \ref{eq:function}. $^{*}\,0.0382\mathrm{\,\mu m}$ for the hydrogen recombination lines and $0.0482\mathrm{\,\mu m}$ for the PAH emission features.  $^{\dagger}\,$ Under the assumption of the case B, $T=10^{4}\mathrm{\,K}$ condition. $^{\ddagger}\,$The proportional coefficient was estimated from the spectrum of $\mathrm{IRAS\,23128-5919}$. See text for details.}
	\end{tabular}
   \end{center}
\end{table*}

%% file: input/result_21_draft.tex

\subsection{Spectral characteristics}
\label{sec:sc}
 The observed spectra and the results of the fitting are shown in figure \ref{fig:hii_spec}.
 The derived optical depths, namely, $\tau_{3.0}$ (= $\tau_{\mathrm{ice}}(3.0)$; equation \ref{eq:function}) and $\tau_{3.4}$ (= $\tau_{\mathrm{ali}}(3.4)$) are summarized in table \ref{tab:result_sil}.
 In this section, we also present a brief overview of the spectra.
 
 The observed spectra exhibit wide diversity in their shapes.  As shown in figure \ref{fig:hii_spec}, the spectra generally exhibit a prominent PAH emission feature at $3.3\mathrm{\,\mu m}$, indicating vigorous star-forming activity (e.g., \citealt{1998ApJ...498..579G}; \citealt{2004ApJ...613..986P}). Hydrogen recombination lines are also evident in the PAH-prominent ULIRGs, although they are less apparent than the PAH features.
 The aliphatic carbon absorption feature is not immediately apparent in the PAH-prominent ULIRGs. However, some of them have flattened spectra around 3.4$\mathrm{\,\mu m}$, which indicate that the PAH subfeatures (emission) are canceled out by the aliphatic carbon absorption. We discuss this further below (section \ref{sec:aof}).
 On the other hand, some objects exhibit red continuums and weak PAH features, suggesting a significant AGN contribution to the spectra (e.g., IRAS 08572$+$3915; \citealt{2008MNRAS.385L.130N}).
 The aliphatic carbon absorption sometimes appears clearly in spectra when PAH emissions are relatively weak.

\subsection{Assumptions regarding features}
\label{sec:aof}
The H$_{2}$O ice absorption, the aliphatic carbon absorption, and the emissions from PAH molecules all overlap. Therefore, assumptions regarding these features will affect optical depth estimates, especially for $\tau_{3.4}$, because at 3.4$\mathrm{\,\mu m}$ there exist three major features: (1) the PAH subfeatures in emission, (2) the aliphatic carbon absorption, and (3) the 3.2--3.7 $\mathrm{\,\mu m}$ wing of the H$_{2}$O ice absorption, which is observed in some Galactic sources (e.g.,  \citealt{1989ApJ...344..413S}; \citealt{1990ApJ...349..107P}; \citealt{2000ApJ...536..347G}; \citealt{2011ApJ...729...92B}). 

 We examined the effect of assumptions regarding the PAH subfeatures and the red wing of the H$_{2}$O ice absorption on the estimate of $\tau_{3.4}$.
 The assumption regarding the aliphatic carbon absorption profile was not altered, because the profile we used to carry out the fitting was similar to the one observed in the Galactic sources (e.g., GCS 3; figure \ref{fig:hac_spec}. The aliphatic carbon optical depth $\tau_{3.4}$ changes typically by no more than 5$\%$ even if the GCS 3 profile is used instead).
 The intensity ratio of the PAH subfeature at 3.4$\mathrm{\,\mu m}$ to the PAH main feature at 3.3$\mathrm{\,\mu m}$, $I_{3.4}/I_{3.3}$, ranges from $\sim 0.06$ to $\sim 0.34$ in typical Galactic sources \citep{1996ApJ...458..610J}. We therefore repeated the fitting procedure, allowing $I_{3.4}/I_{3.3}$ to take a value in the range of 0.06--0.34.
 To investigate how the red wing of the H$_{2}$O ice absorption affects the results, we employed the profile observed in the Galactic field star 2MASS J21240614 \citep{2011ApJ...729...92B} as a sample profile of the red wing. \citet{2011ApJ...729...92B} reported that profiles showing the red wing are similar and independent of extinction. Therefore, we expect that we can safely use the ice profile of 2MASS J21240614 as a sample of the red wing.
 
 To summarize, we repeated spectral fitting, allowing the intensity ratio of the PAH main feature to the subfeature, $I_{3.4}/I_{3.3}$,  to take a different value ($I_{3.4}/I_{3.3}=$ 0.06--0.34) and the ice profile to take a different profile (the laboratory profile or the profile observed in the Galactic field star 2MASS J21240614).
 We found that the changes of assumptions resulted in systematic uncertainty in the optical depth of aliphatic carbon of  $\sim 25\%$. The average statistical error in the fitting is $\sim 12\%$. The following discussions are based on the results with our original assumptions ($I_{3.4}/I_{3.3}=0.17$, laboratory ice profile) for simplicity. These changes of assumptions do not vitiate the following discussion.

 \begin{figure}
 \begin{center}
 \hspace{-50mm}
 \raisebox{-100mm}{\includegraphics[width=0.9\linewidth]{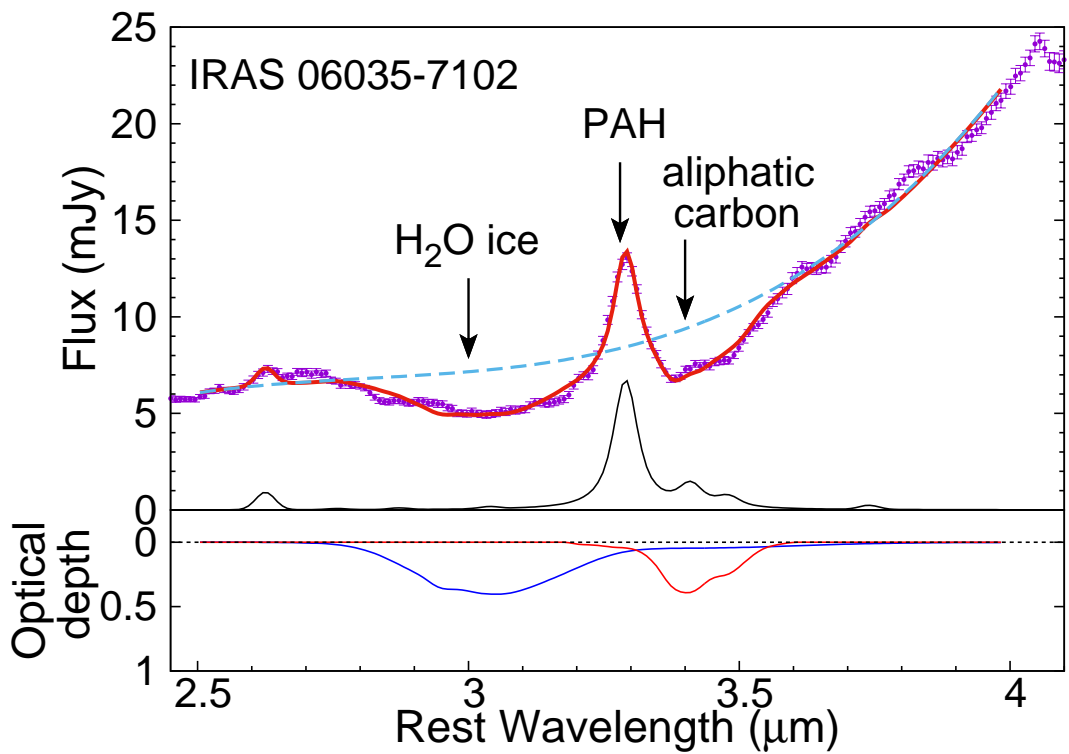}}
 \end{center}
 \caption{The result of the model fitting for the spectrum shown in figure \ref{fig:spec_ex} (IRAS 06035$-$7102). Successive features between $\lambda_{\mathrm{rest}}\simeq2.6$ and $\lambda_{\mathrm{rest}}\simeq3.6\mathrm{\,\mu m}$ were nicely fitted by equation \ref{eq:function} ($\chi_{\nu}=2.28$). Upper panel: The points with error bars represent the observed flux. The solid red line, the dashed blue line, and the black line represent the best-fit spectrum, its continuum component, and its emission components (PAH emissions and hydrogen recombination lines), respectively. Lower panel: The blue and the red lines represent the absorption profiles of H$_{2}$O ice and aliphatic carbon, respectively.}
 \label{fig:fit_ex}
 \end{figure}
\begin{figure*}
	\begin{center}
		\begin{minipage}{0.28\textwidth}
			\begin{center}
				\includegraphics[width=1.08\textwidth]{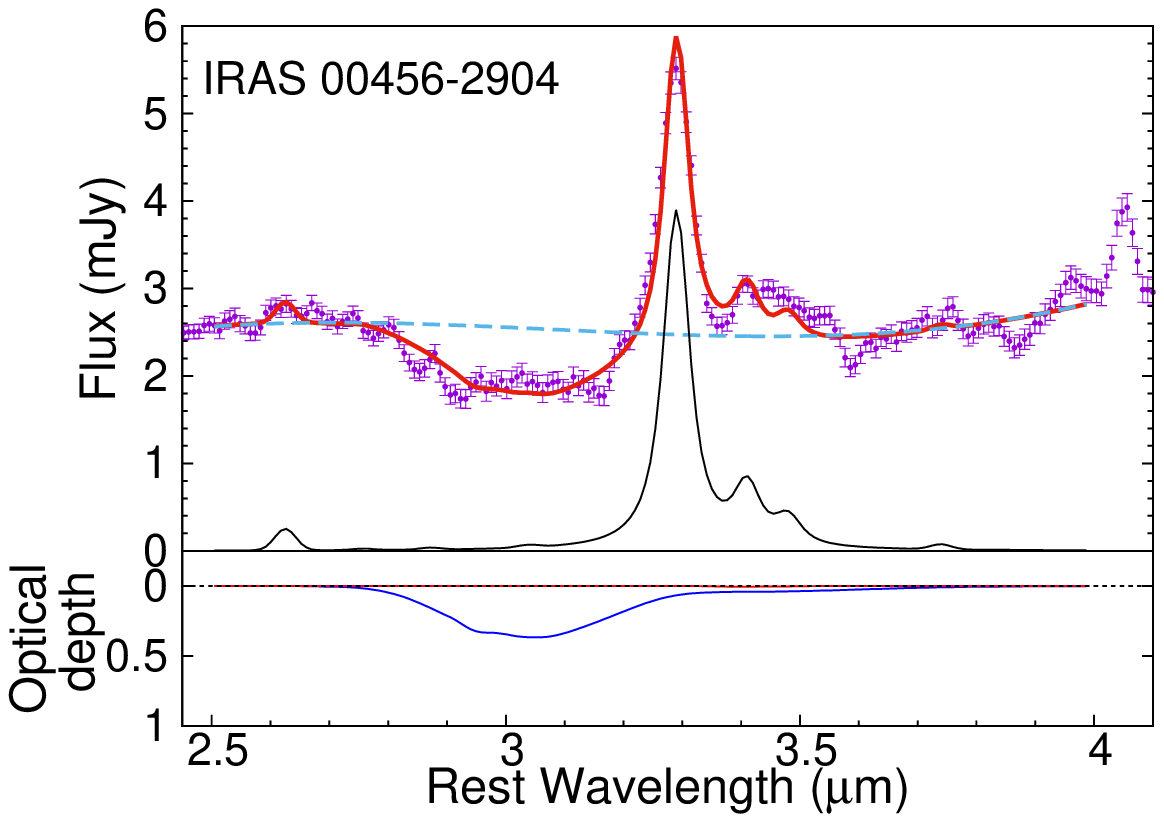}
			\end{center}
		\end{minipage}
		\begin{minipage}{0.28\textwidth}
			\begin{center}
				\includegraphics[width=1.08\textwidth]{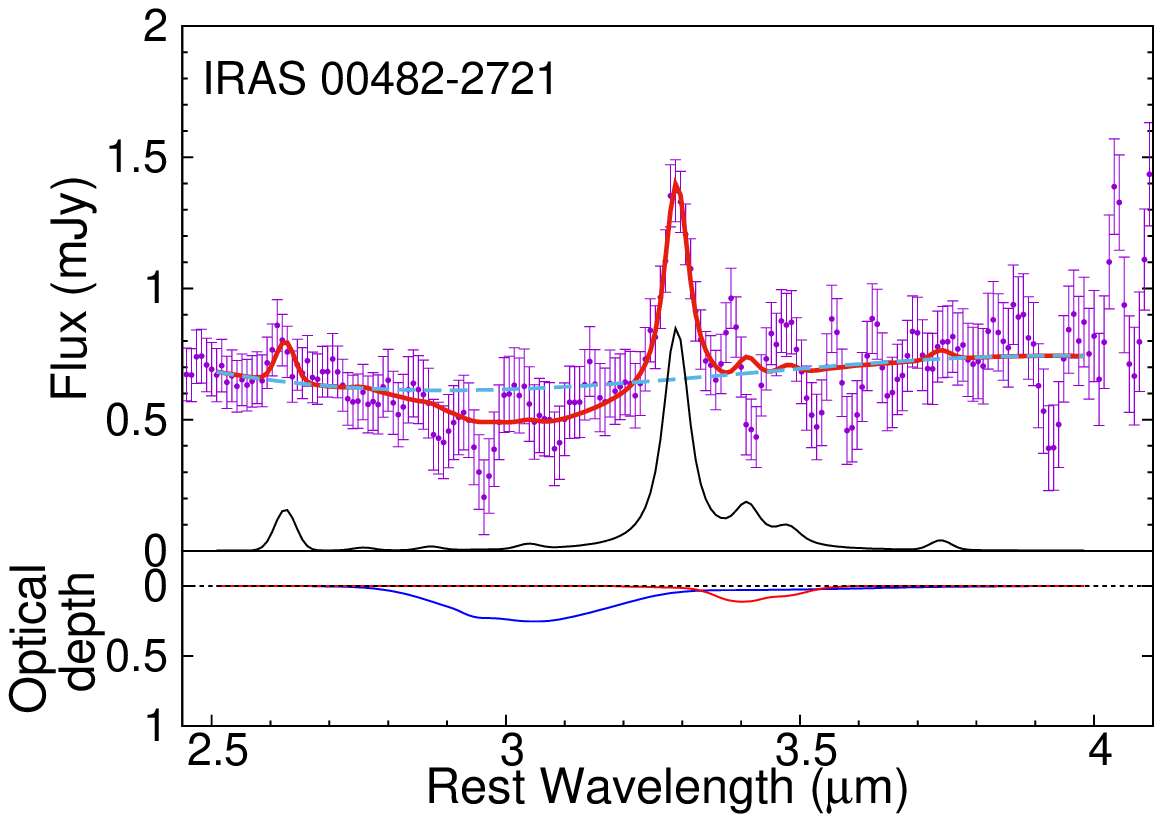}
			\end{center}
		\end{minipage}
		\begin{minipage}{0.28\textwidth}
			\begin{center}
				\includegraphics[width=1.08\textwidth]{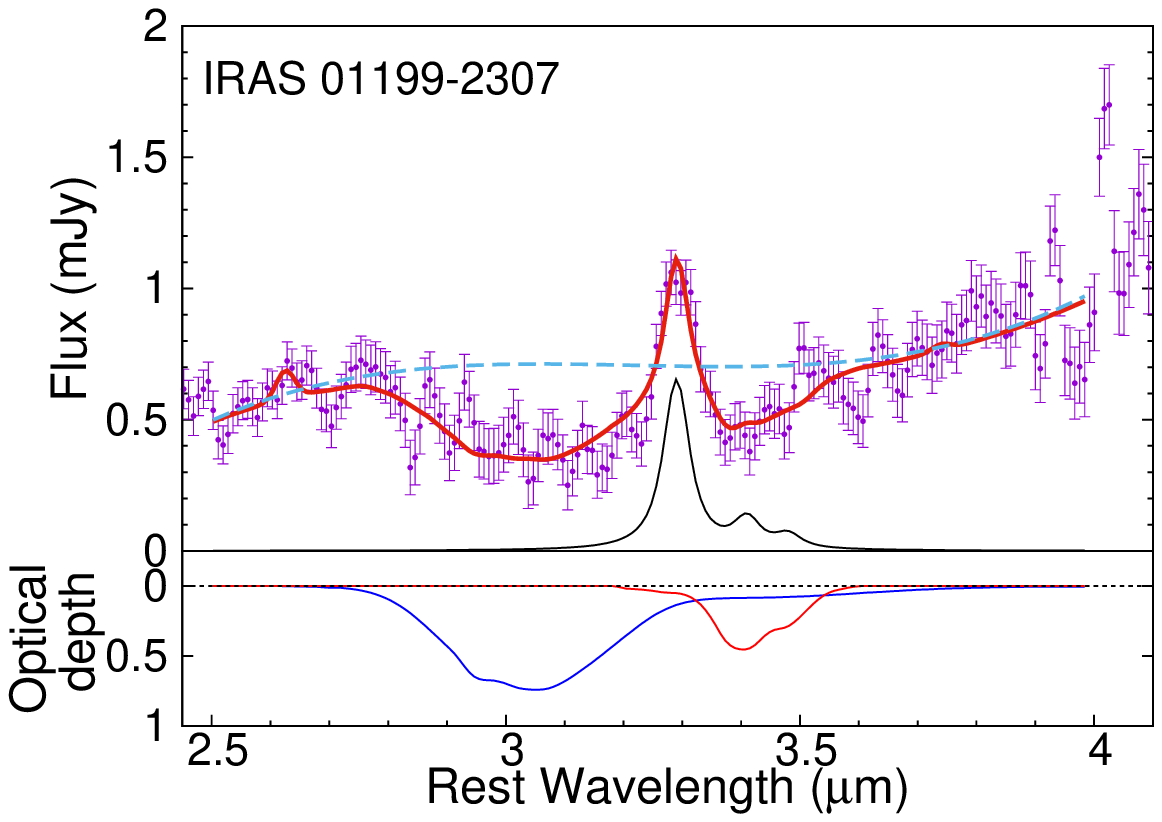}
			\end{center}
		\end{minipage}
	\end{center}
	
	\begin{center}
		\begin{minipage}{0.28\textwidth}
			\begin{center}
				\includegraphics[width=1.08\textwidth]{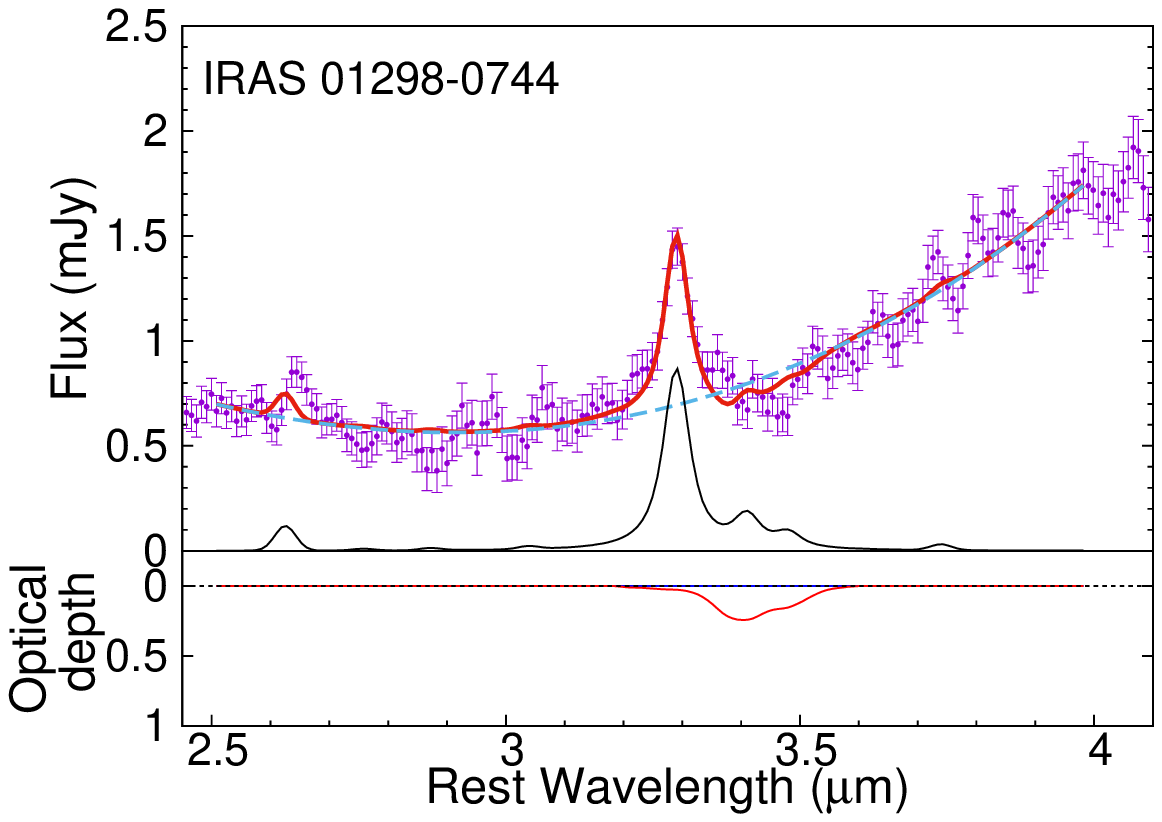}
			\end{center}
		\end{minipage}
		\begin{minipage}{0.28\textwidth}
			\begin{center}
				\includegraphics[width=1.08\textwidth]{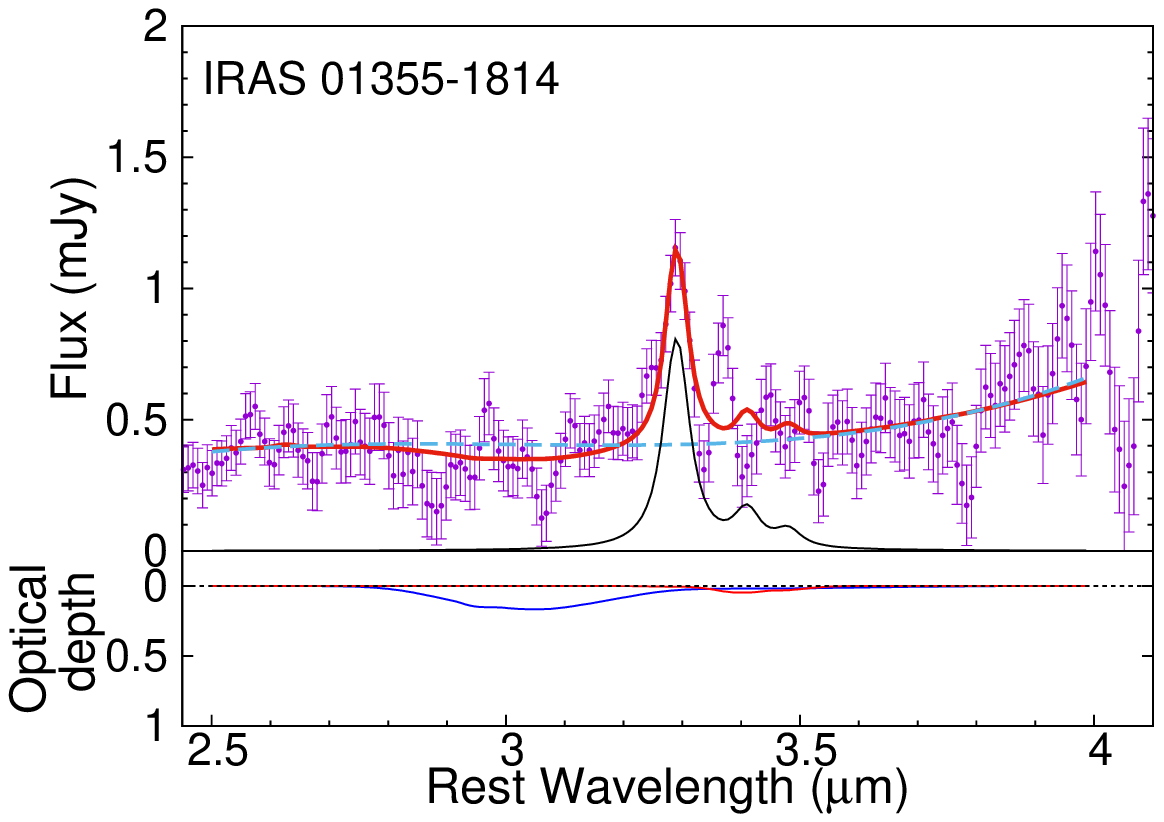}
			\end{center}
		\end{minipage}
		\begin{minipage}{0.28\textwidth}
			\begin{center}
				\includegraphics[width=1.08\textwidth]{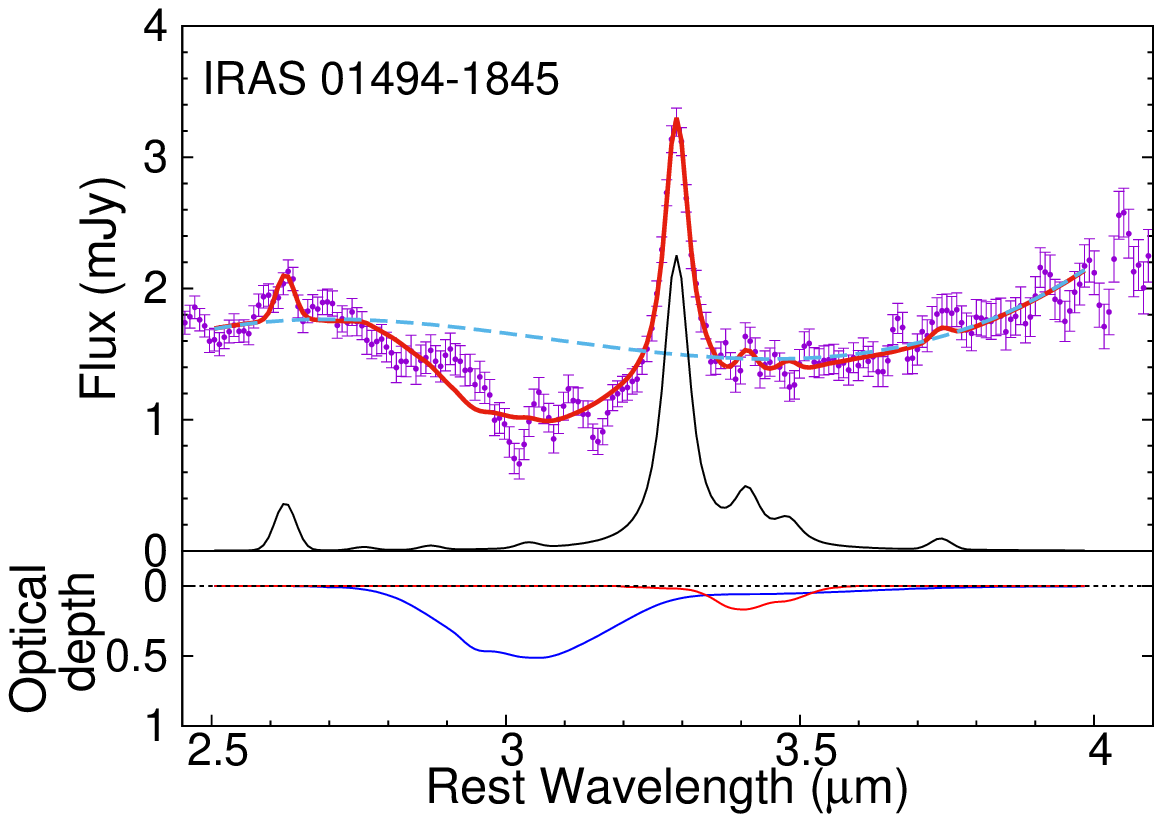}
			\end{center}
		\end{minipage}
	\end{center}
	
	\begin{center}
		\begin{minipage}{0.28\textwidth}
			\begin{center}
				\includegraphics[width=1.08\textwidth]{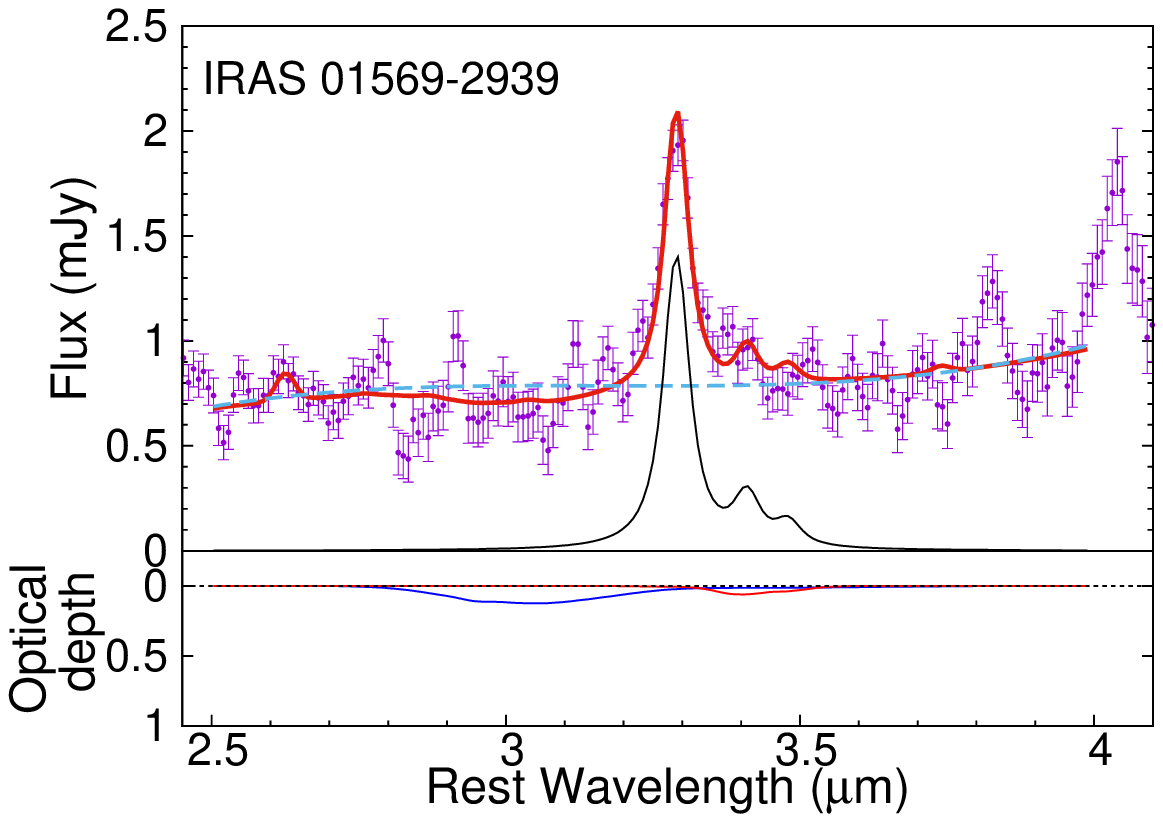}
			\end{center}
		\end{minipage}
		\begin{minipage}{0.28\textwidth}
			\begin{center}
				\includegraphics[width=1.08\textwidth]{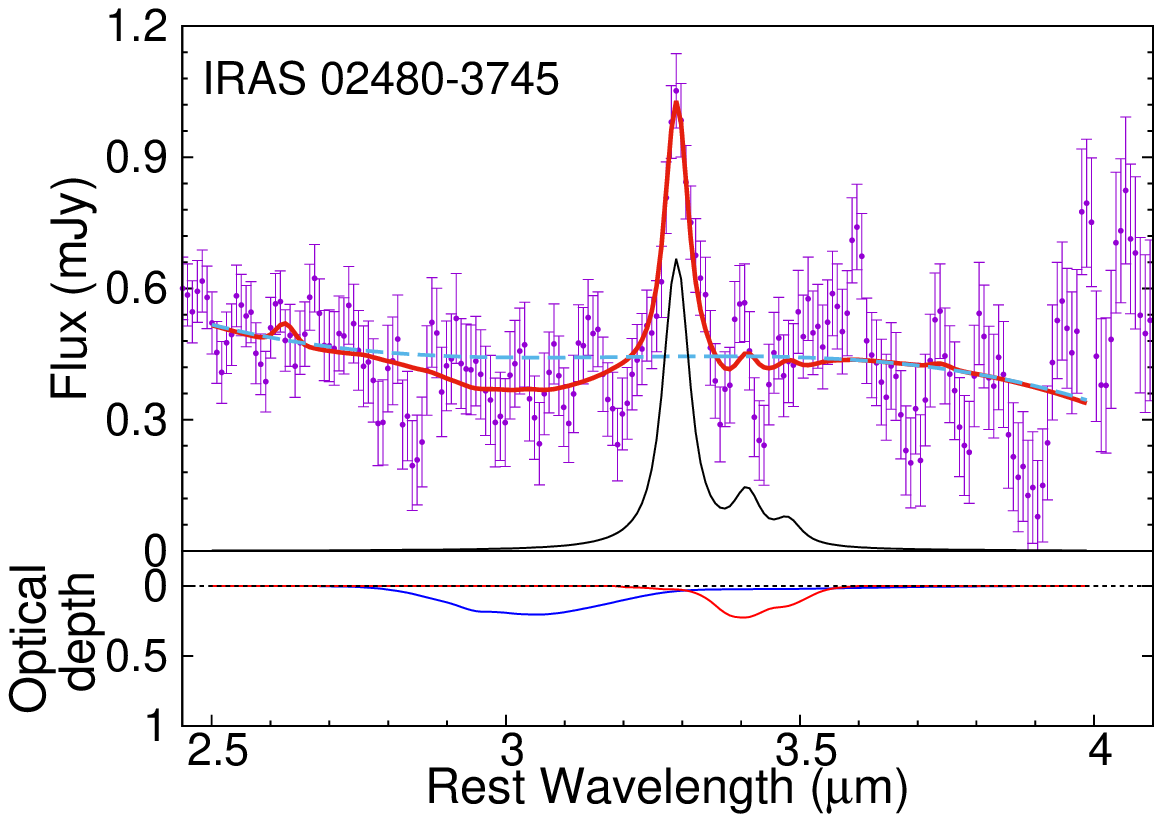}
			\end{center}
		\end{minipage}
		\begin{minipage}{0.28\textwidth}
			\begin{center}
				\includegraphics[width=1.08\textwidth]{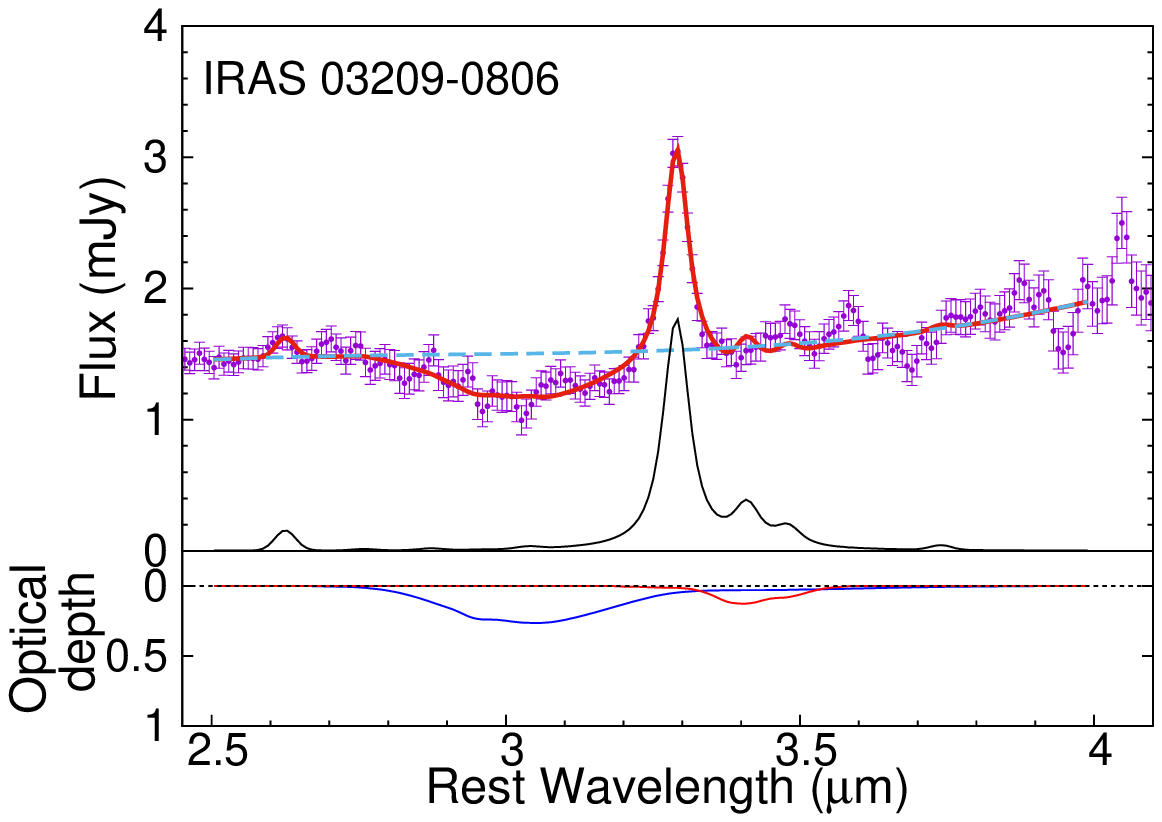}
			\end{center}
		\end{minipage}
	\end{center}
	
	\begin{center}
		\begin{minipage}{0.28\textwidth}
			\begin{center}
				\includegraphics[width=1.08\textwidth]{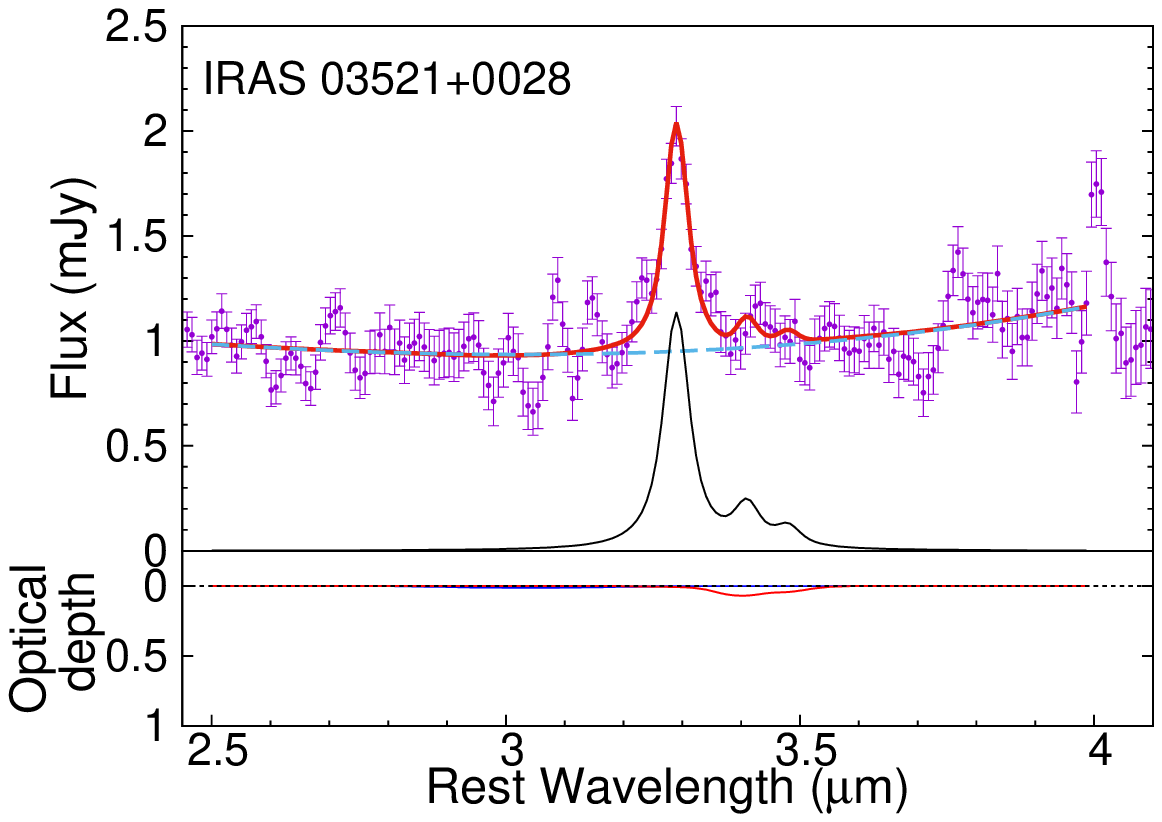}
			\end{center}
		\end{minipage}
		\begin{minipage}{0.28\textwidth}
			\begin{center}
				\includegraphics[width=1.08\textwidth]{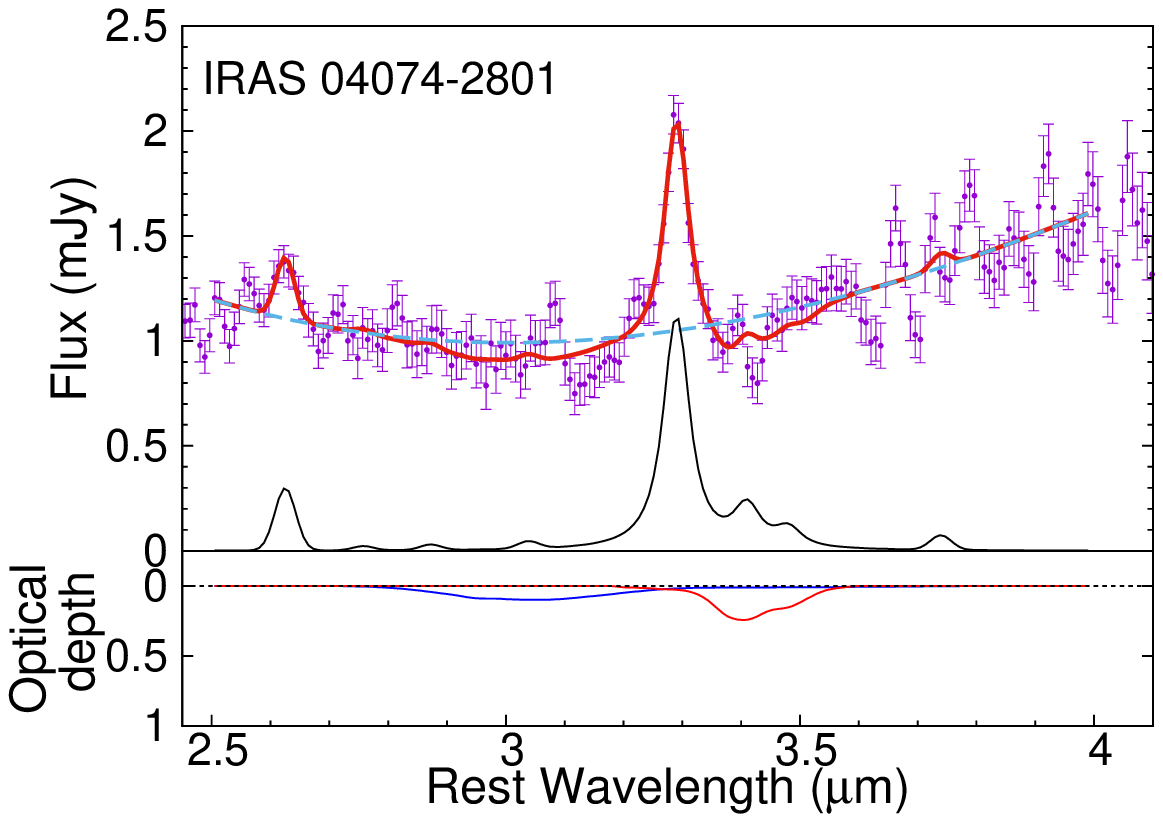}
			\end{center}
		\end{minipage}
		\begin{minipage}{0.28\textwidth}
			\begin{center}
				\includegraphics[width=1.08\textwidth]{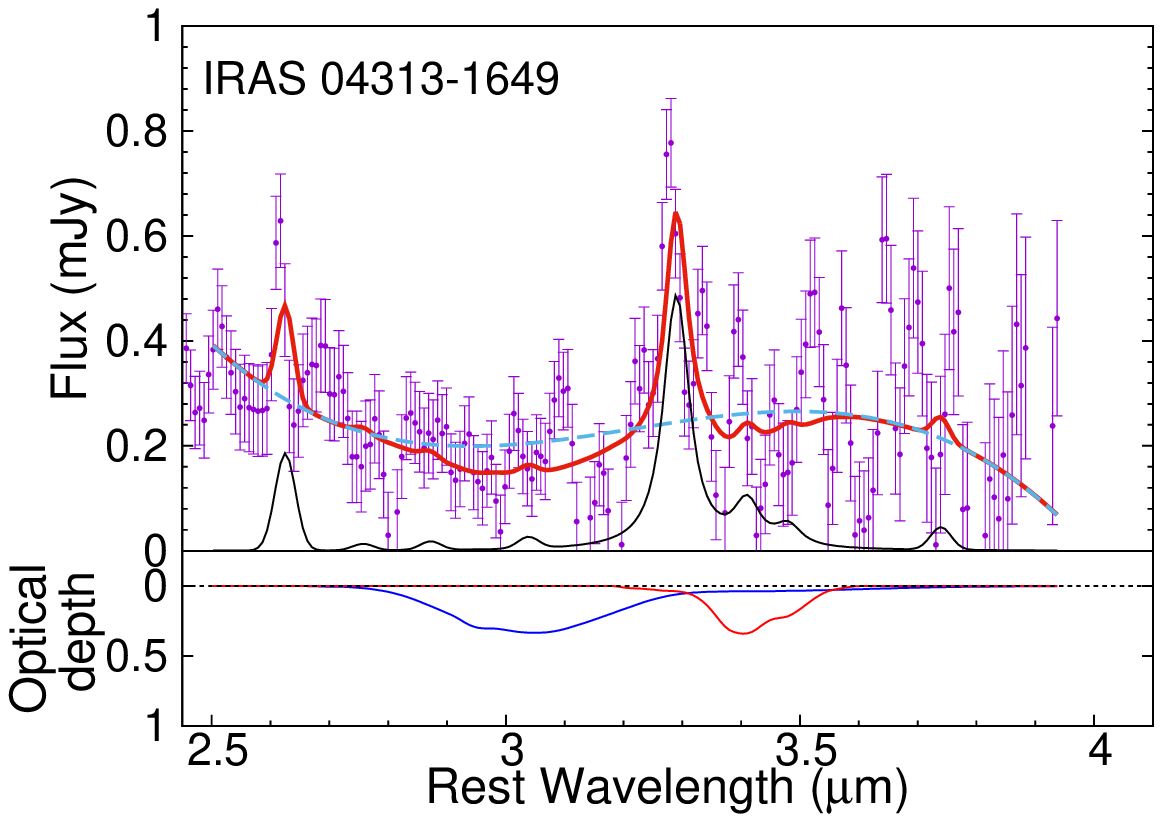}
			\end{center}
		\end{minipage}
	\end{center}
	
	\begin{center}
		\begin{minipage}{0.28\textwidth}
			\begin{center}
				\includegraphics[width=1.08\textwidth]{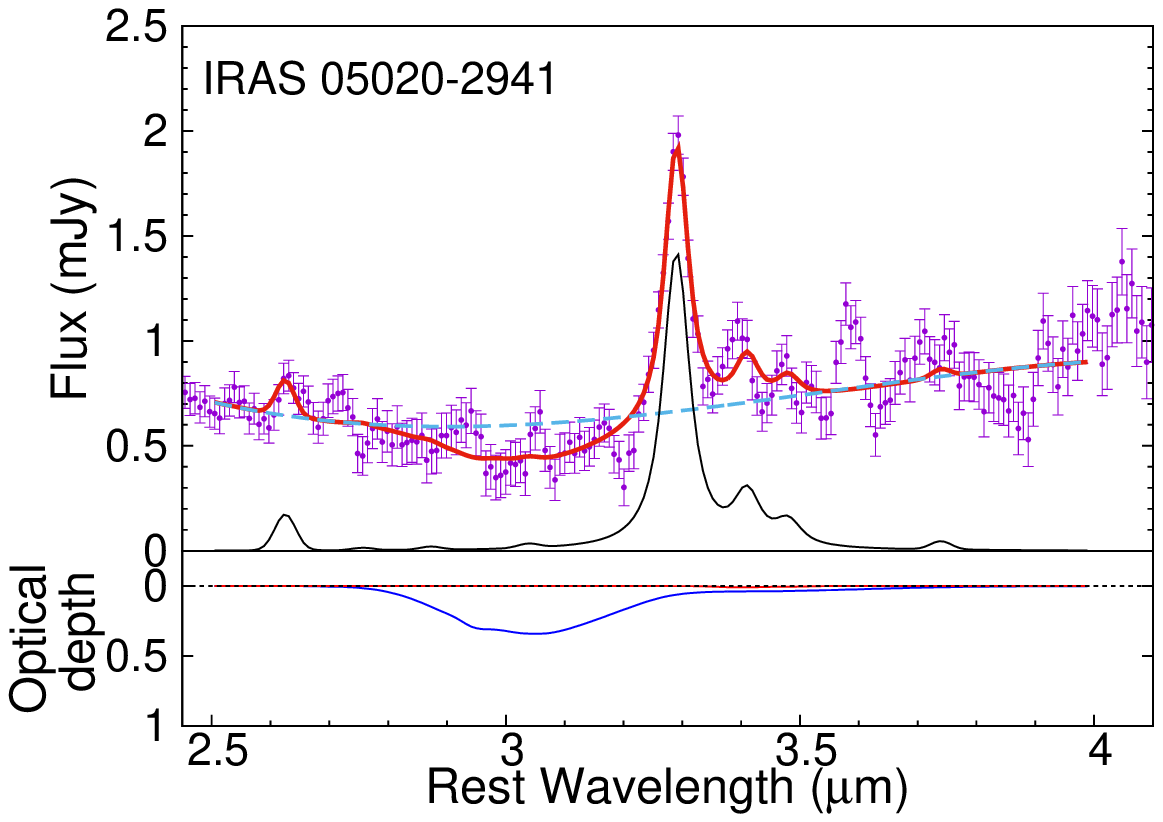}
			\end{center}
		\end{minipage}
		\begin{minipage}{0.28\textwidth}
			\begin{center}
				\includegraphics[width=1.08\textwidth]{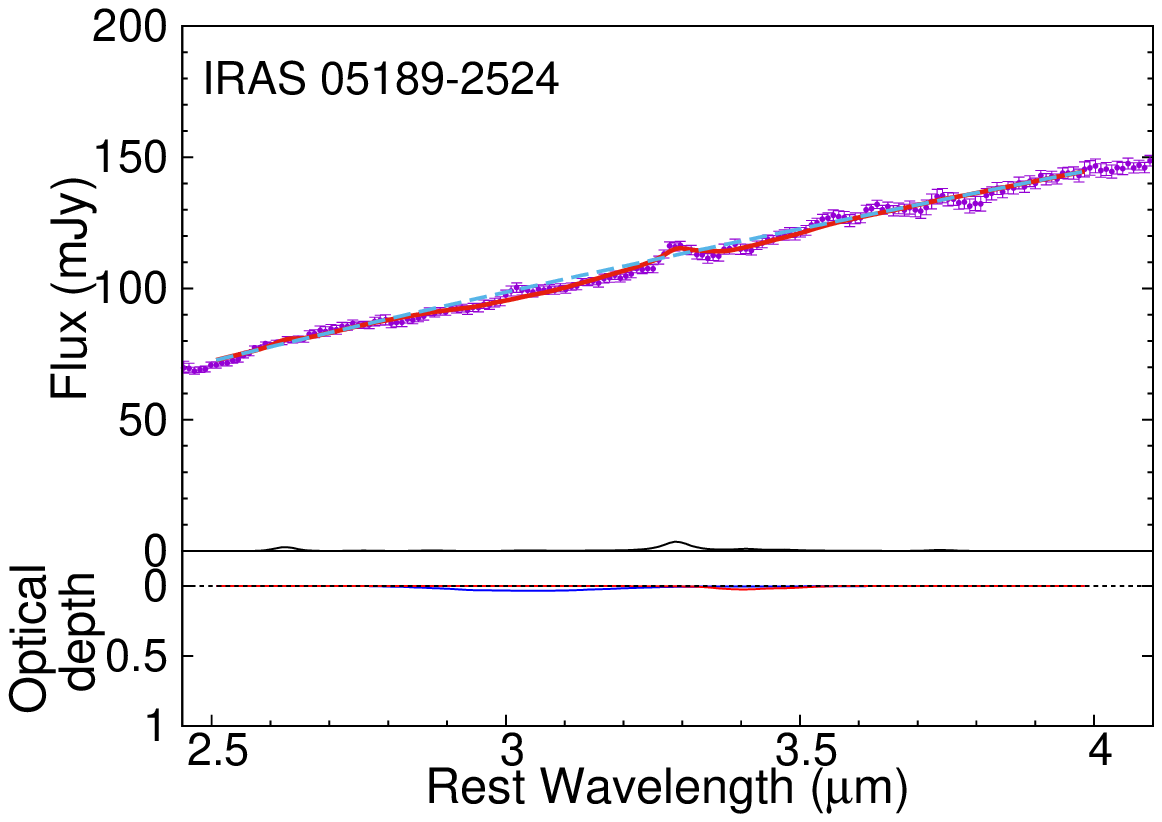}
			\end{center}
		\end{minipage}
		\begin{minipage}{0.28\textwidth}
			\begin{center}
				\includegraphics[width=1.08\textwidth]{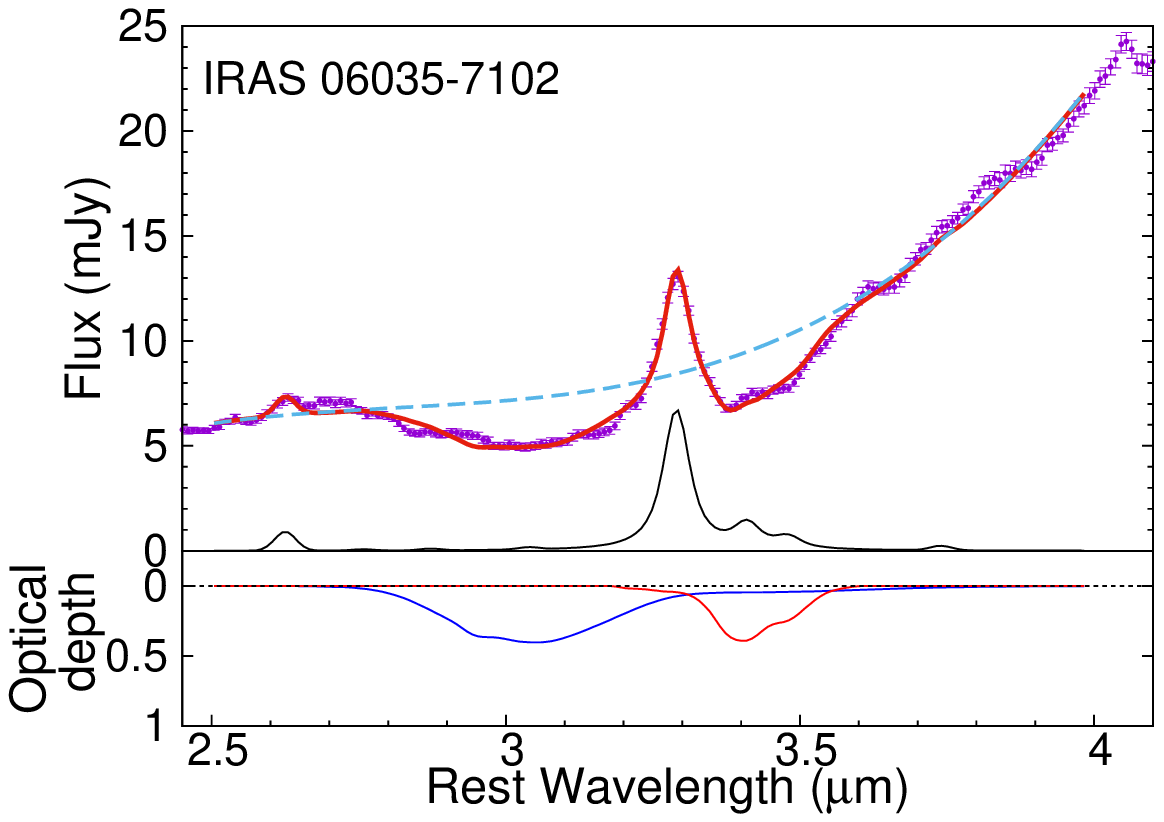}
			\end{center}
		\end{minipage}
		\vspace{2.0mm}
		\caption{The results of the model fitting for the spectra. The points and the lines represent the same as figure \ref{fig:fit_ex}.}
		\label{fig:hii_spec}
	\end{center}

\end{figure*}

\setcounter{figure}{5}
\begin{figure*}

	\begin{center}
		\begin{minipage}{0.28\textwidth}
			\begin{center}
				\includegraphics[width=1.08\textwidth]{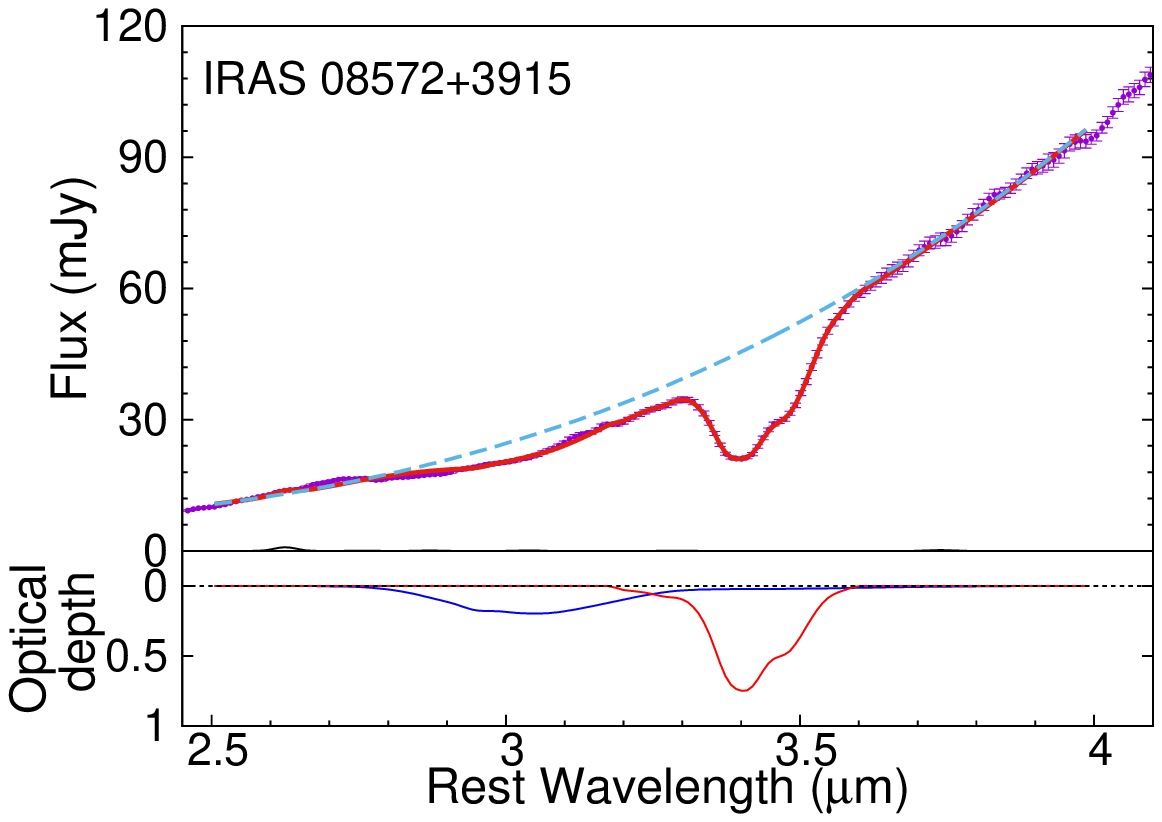}
			\end{center}
		\end{minipage}
		\begin{minipage}{0.28\textwidth}
			\begin{center}
				\includegraphics[width=1.08\textwidth]{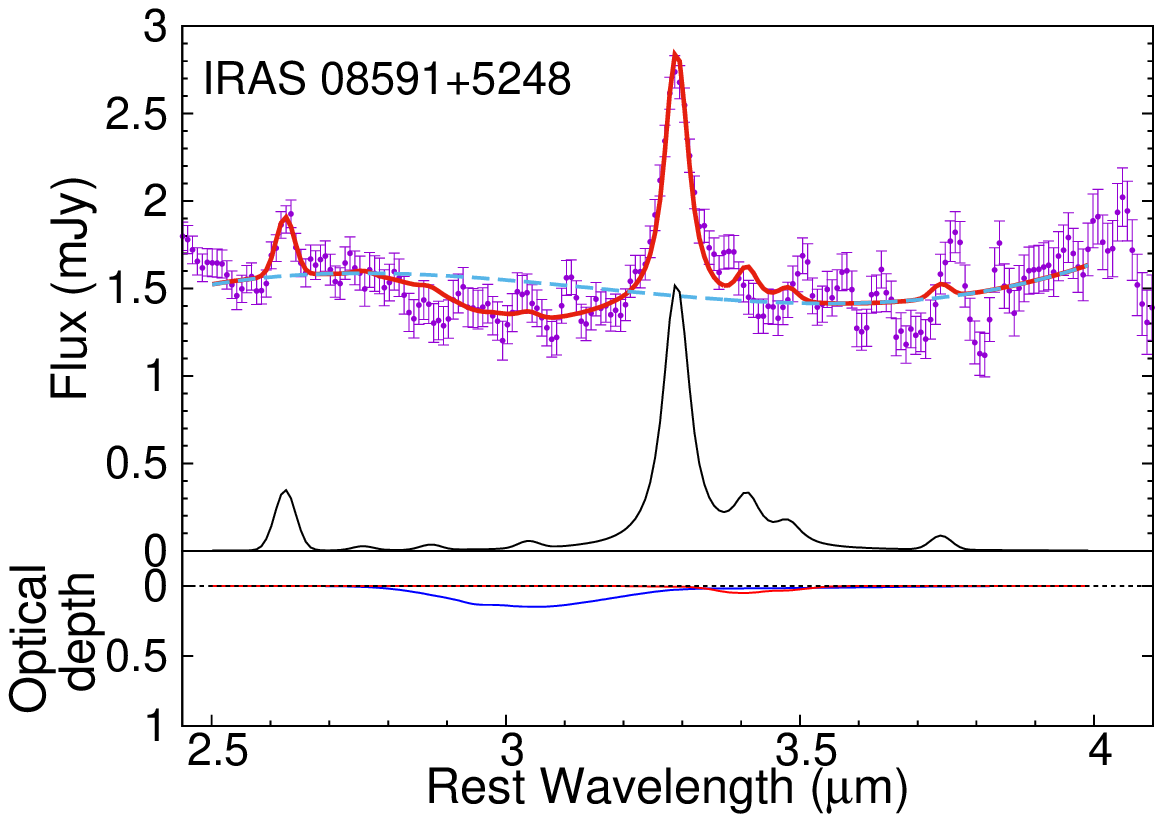}
			\end{center}
		\end{minipage}
		\begin{minipage}{0.28\textwidth}
			\begin{center}
				\includegraphics[width=1.08\textwidth]{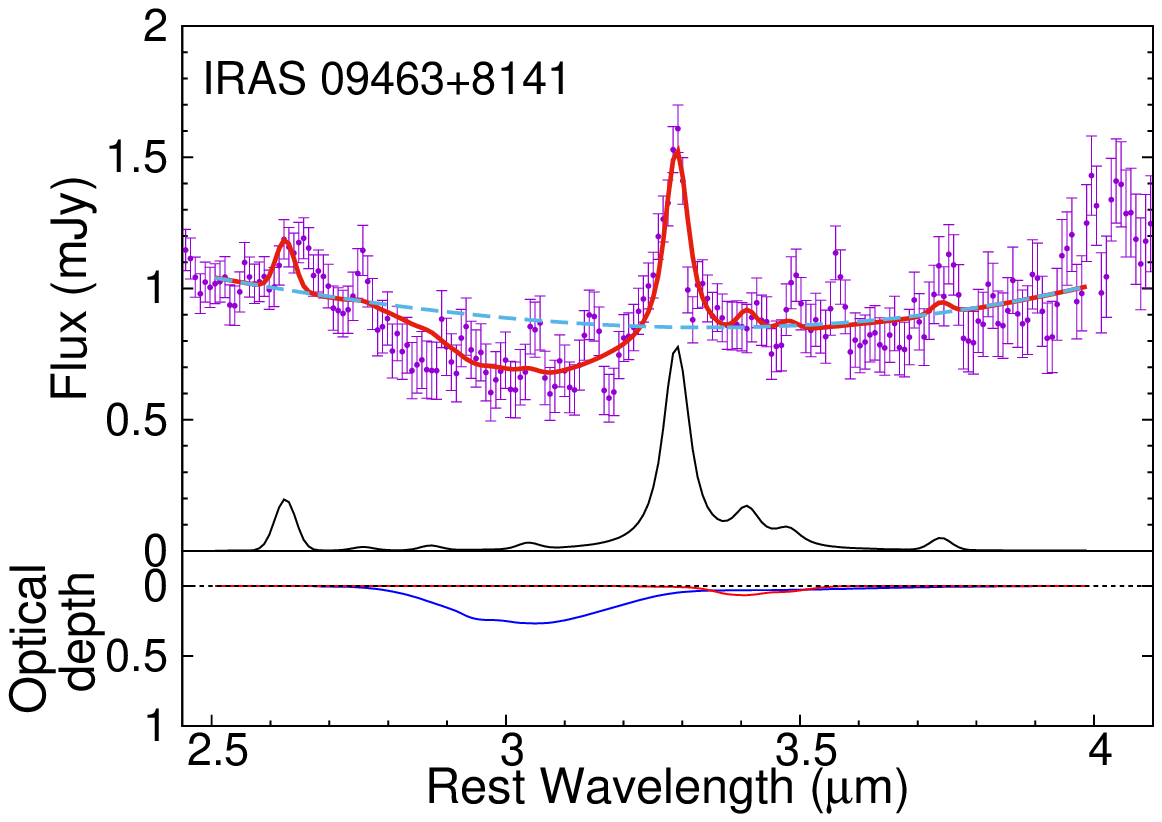}
			\end{center}
		\end{minipage}
	\end{center}
	
	\begin{center}
		\begin{minipage}{0.28\textwidth}
			\begin{center}
				\includegraphics[width=1.08\textwidth]{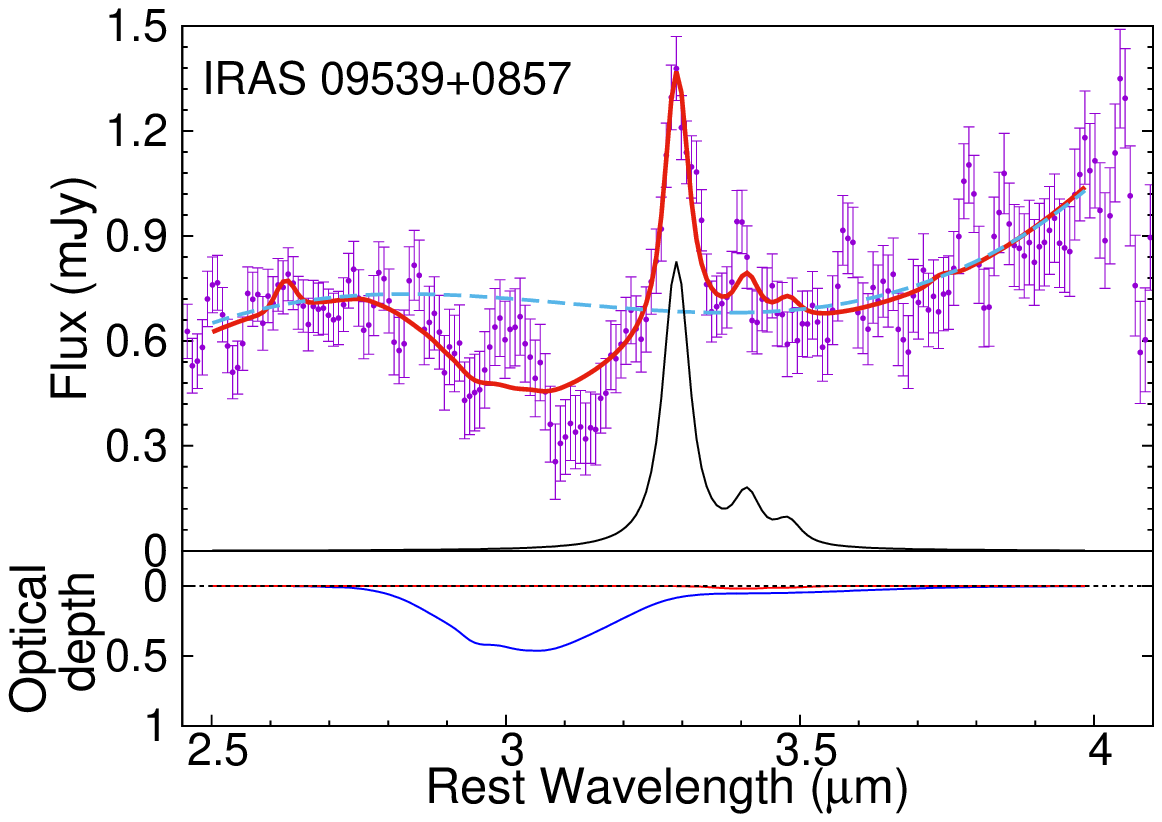}
			\end{center}
		\end{minipage}
		\begin{minipage}{0.28\textwidth}
			\begin{center}
				\includegraphics[width=1.08\textwidth]{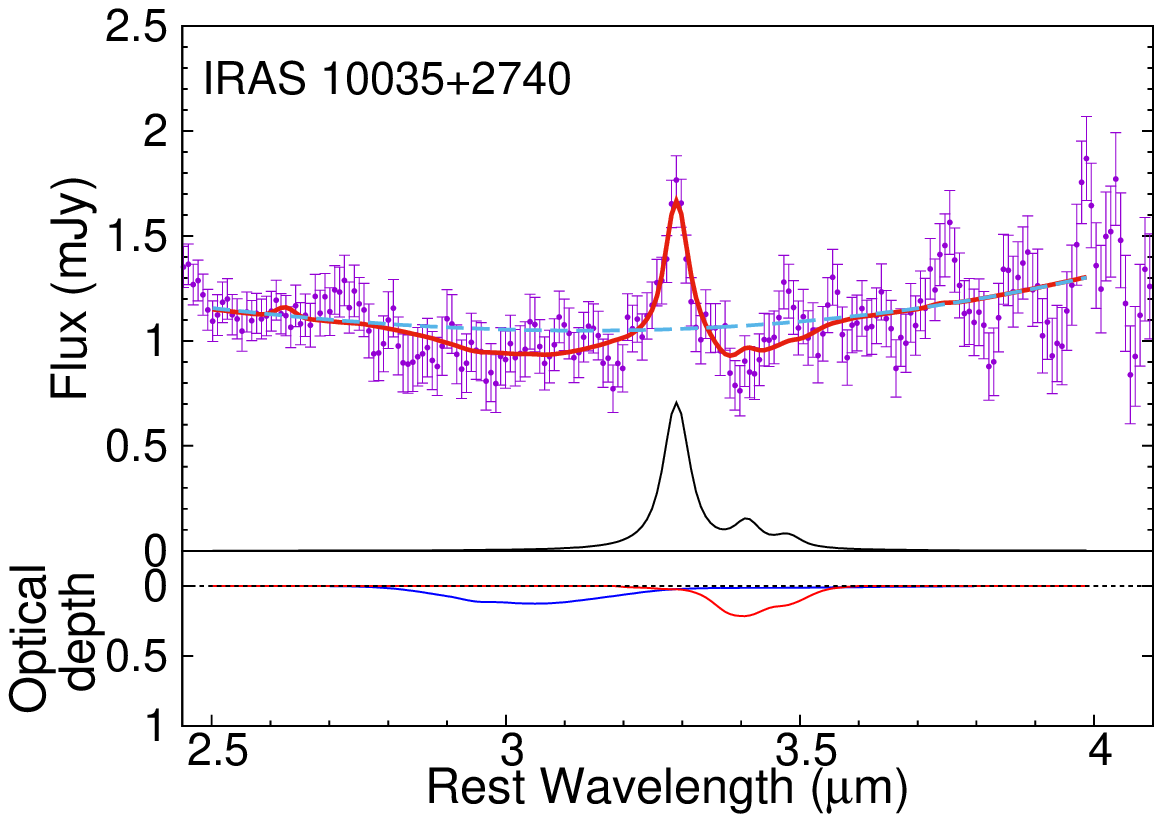}
			\end{center}
		\end{minipage}
		\begin{minipage}{0.28\textwidth}
			\begin{center}
				\includegraphics[width=1.08\textwidth]{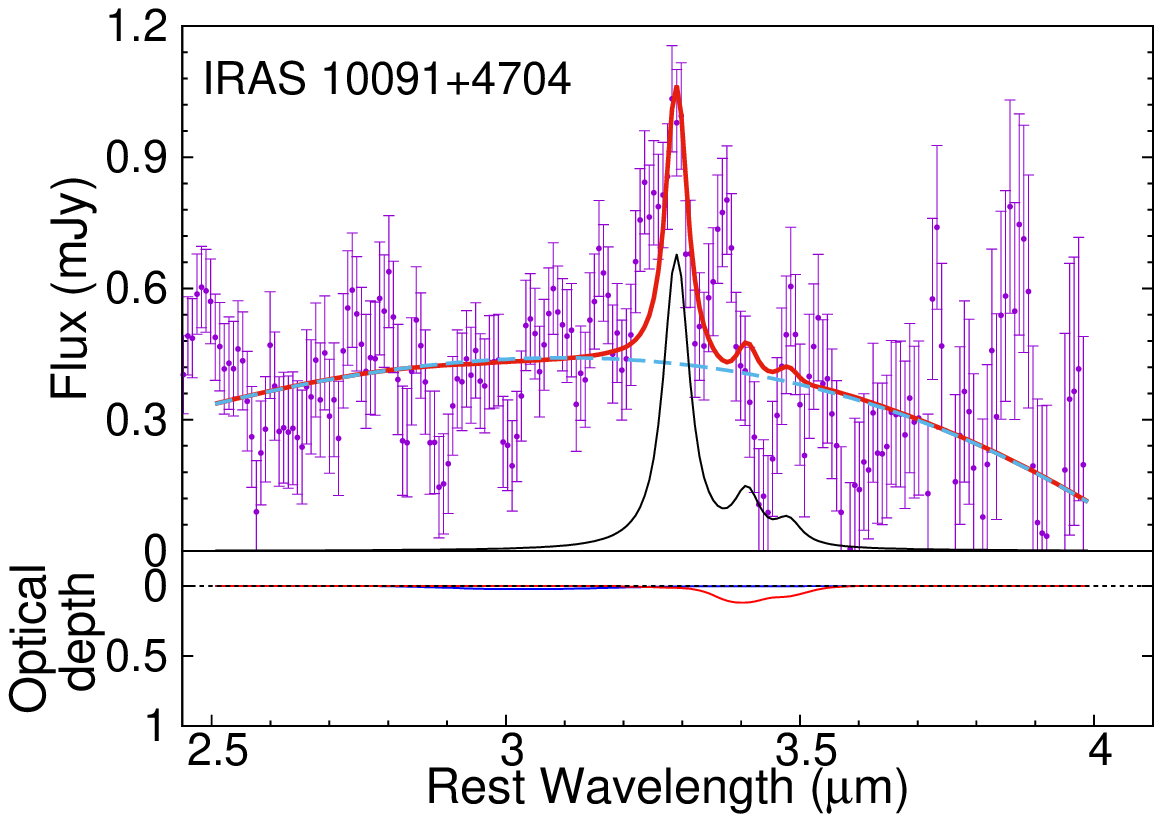}
			\end{center}
		\end{minipage}
	\end{center}
	
	\begin{center}
		\begin{minipage}{0.28\textwidth}
			\begin{center}
				\includegraphics[width=1.08\textwidth]{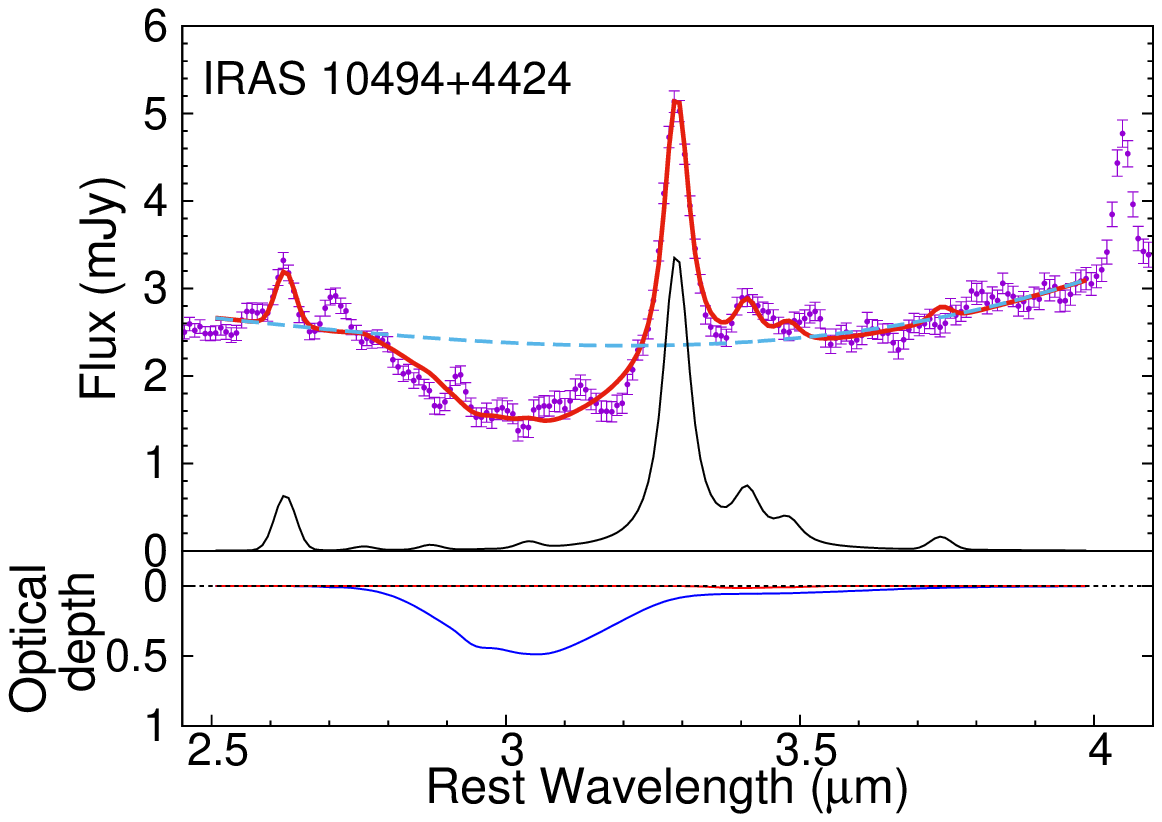}
			\end{center}
		\end{minipage}
		\begin{minipage}{0.28\textwidth}
			\begin{center}
				\includegraphics[width=1.08\textwidth]{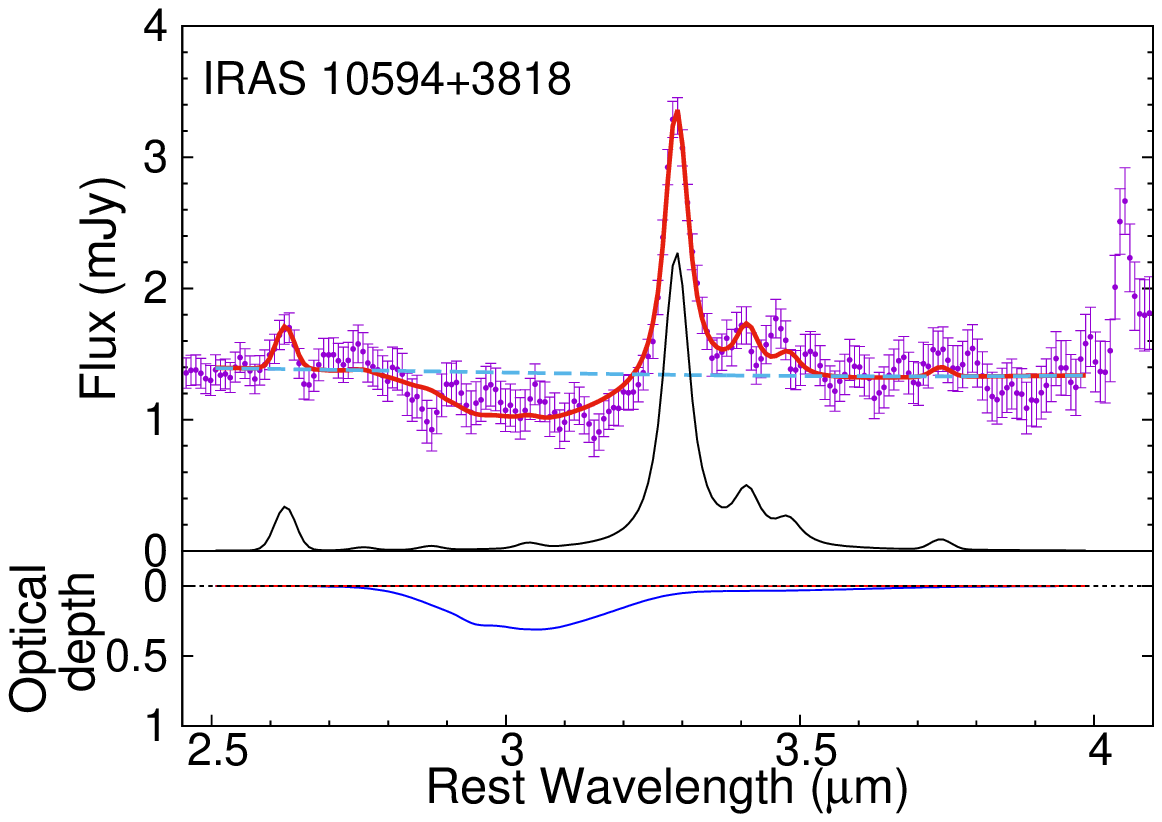}
			\end{center}
		\end{minipage}
		\begin{minipage}{0.28\textwidth}
			\begin{center}
				\includegraphics[width=1.08\textwidth]{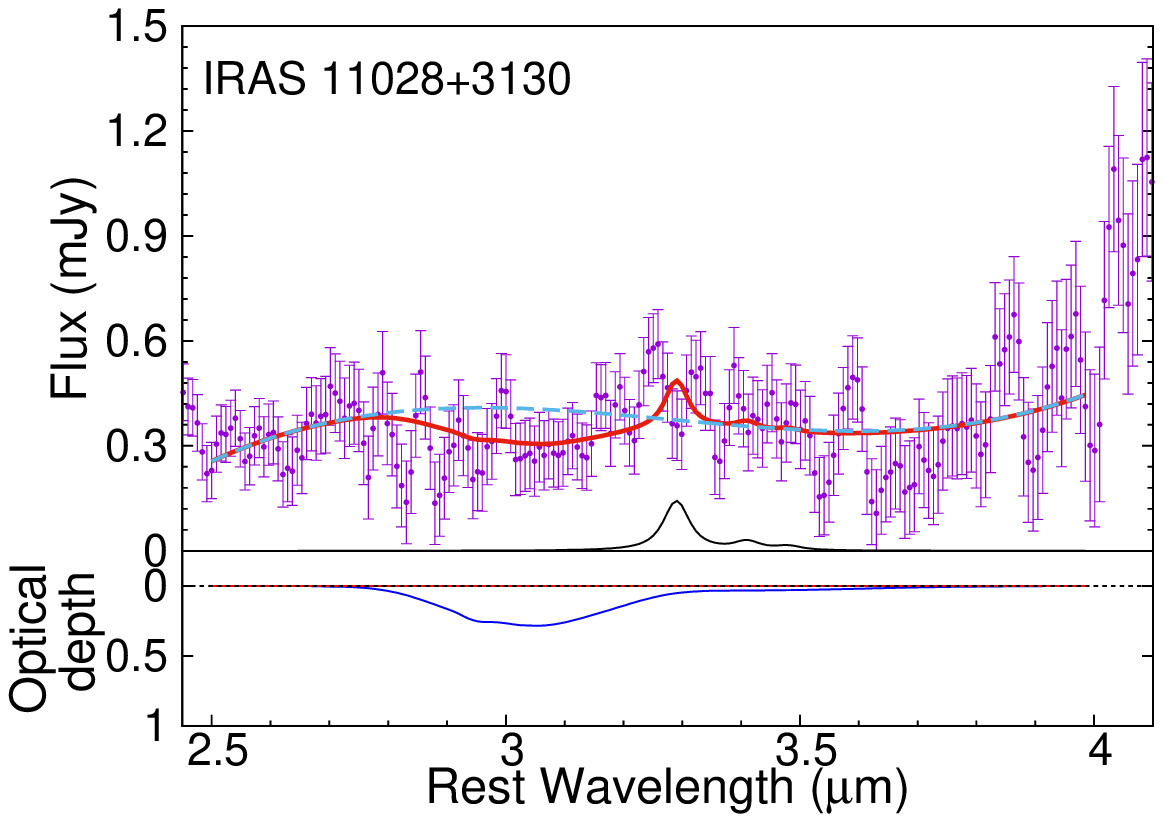}
			\end{center}
		\end{minipage}
	\end{center}
	
	\begin{center}
		\begin{minipage}{0.28\textwidth}
			\begin{center}
				\includegraphics[width=1.08\textwidth]{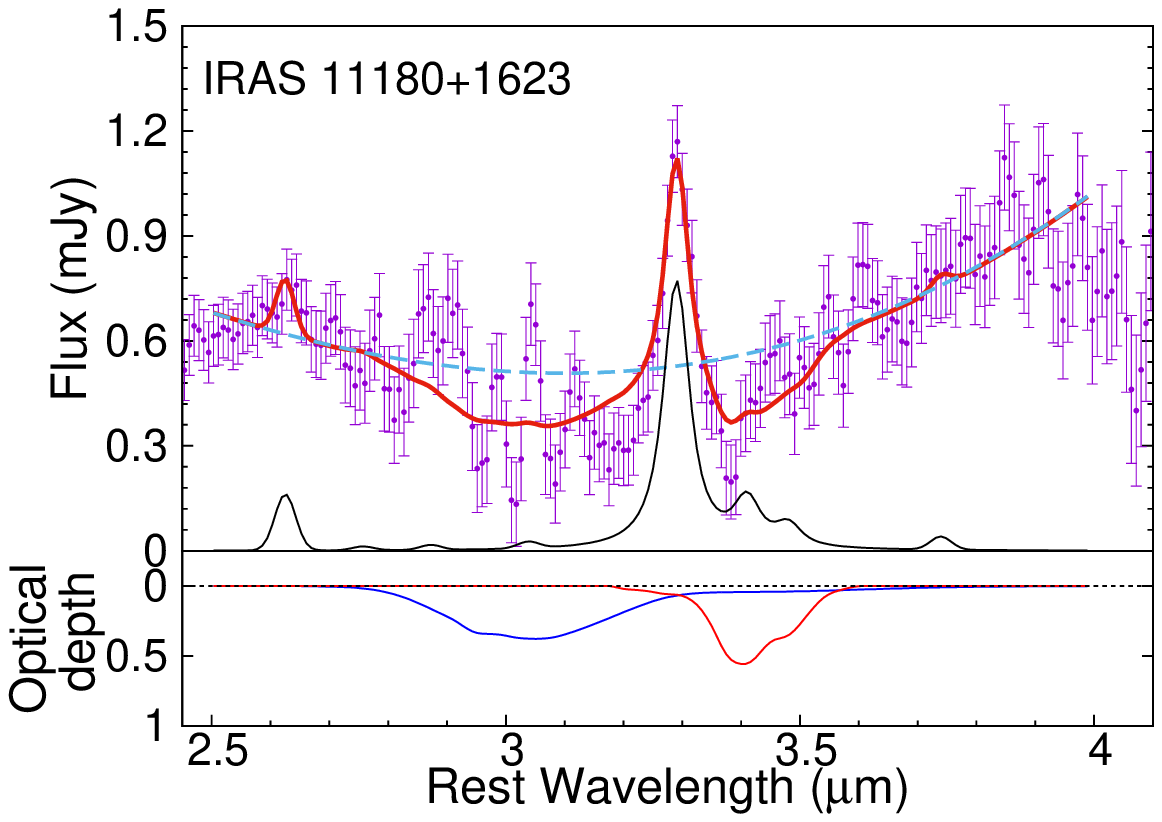}
			\end{center}
		\end{minipage}
		\begin{minipage}{0.28\textwidth}
			\begin{center}
				\includegraphics[width=1.08\textwidth]{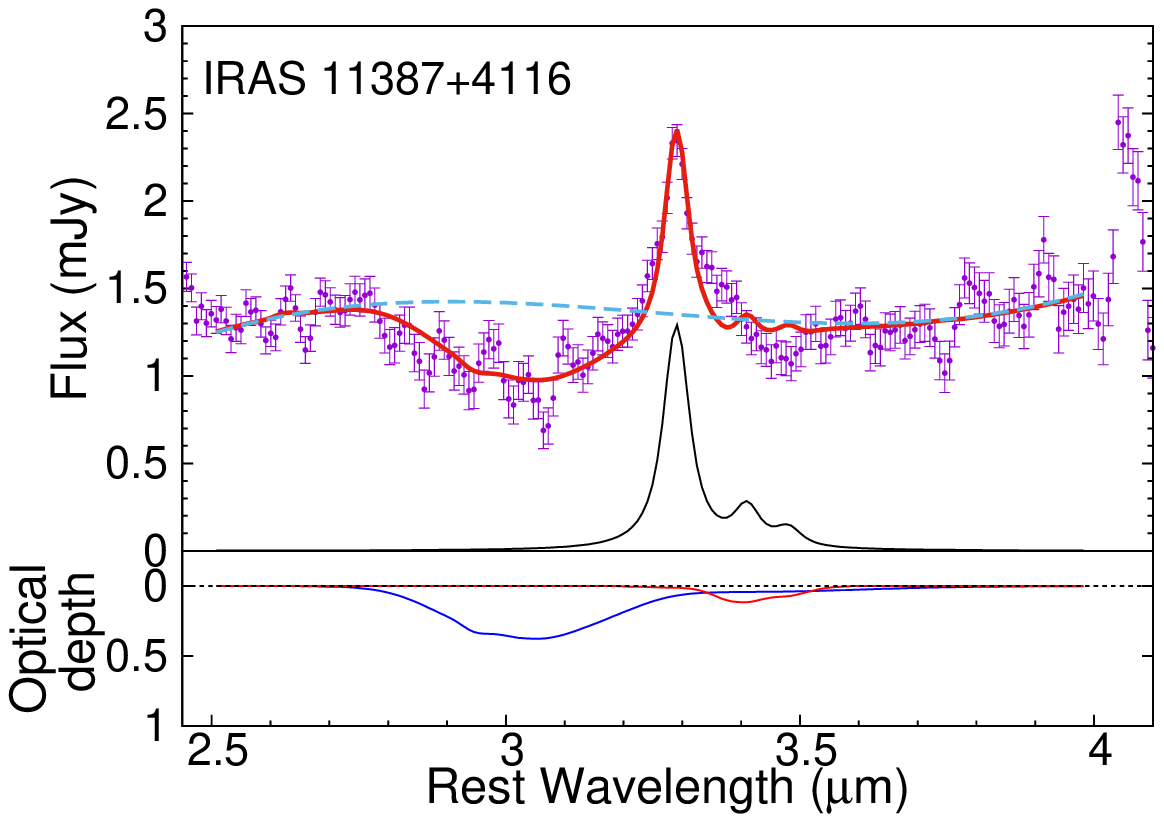}
			\end{center}
		\end{minipage}
		\begin{minipage}{0.28\textwidth}
			\begin{center}
				\includegraphics[width=1.08\textwidth]{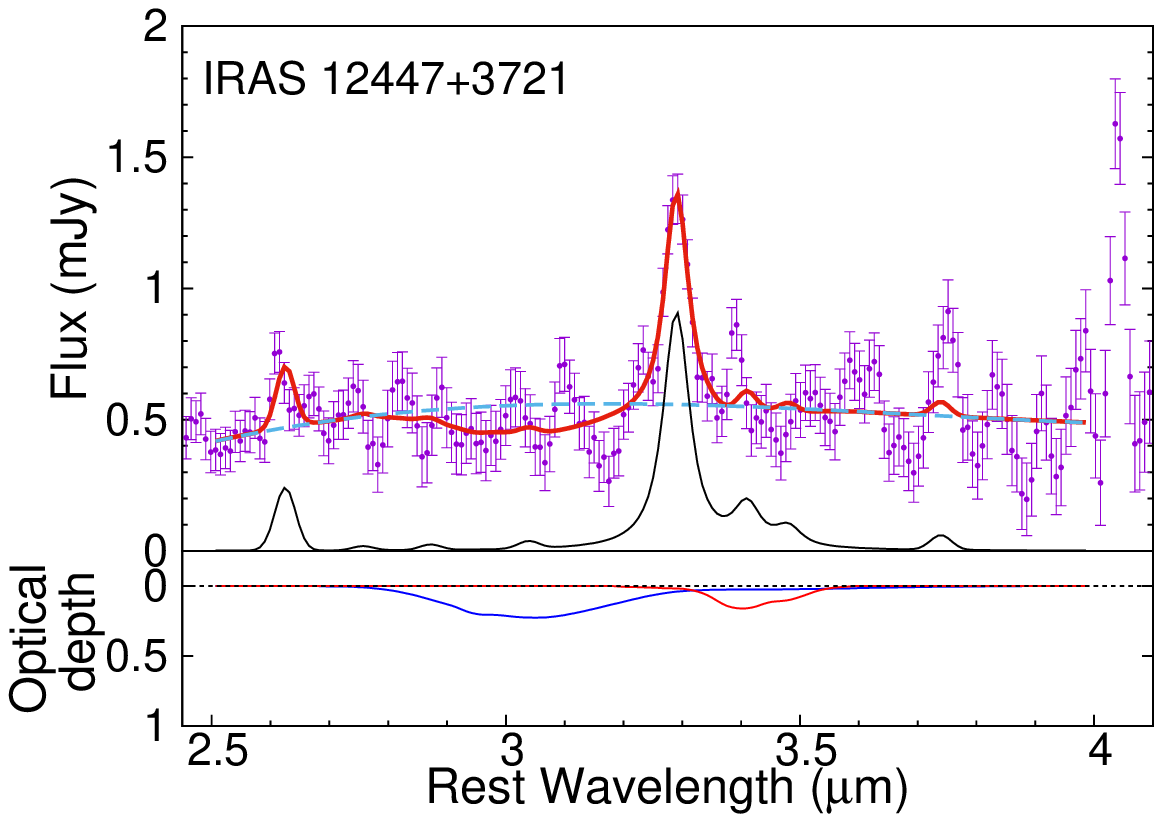}
			\end{center}
		\end{minipage}
	\end{center}
	
	\begin{center}
		\begin{minipage}{0.28\textwidth}
			\begin{center}
				\includegraphics[width=1.08\textwidth]{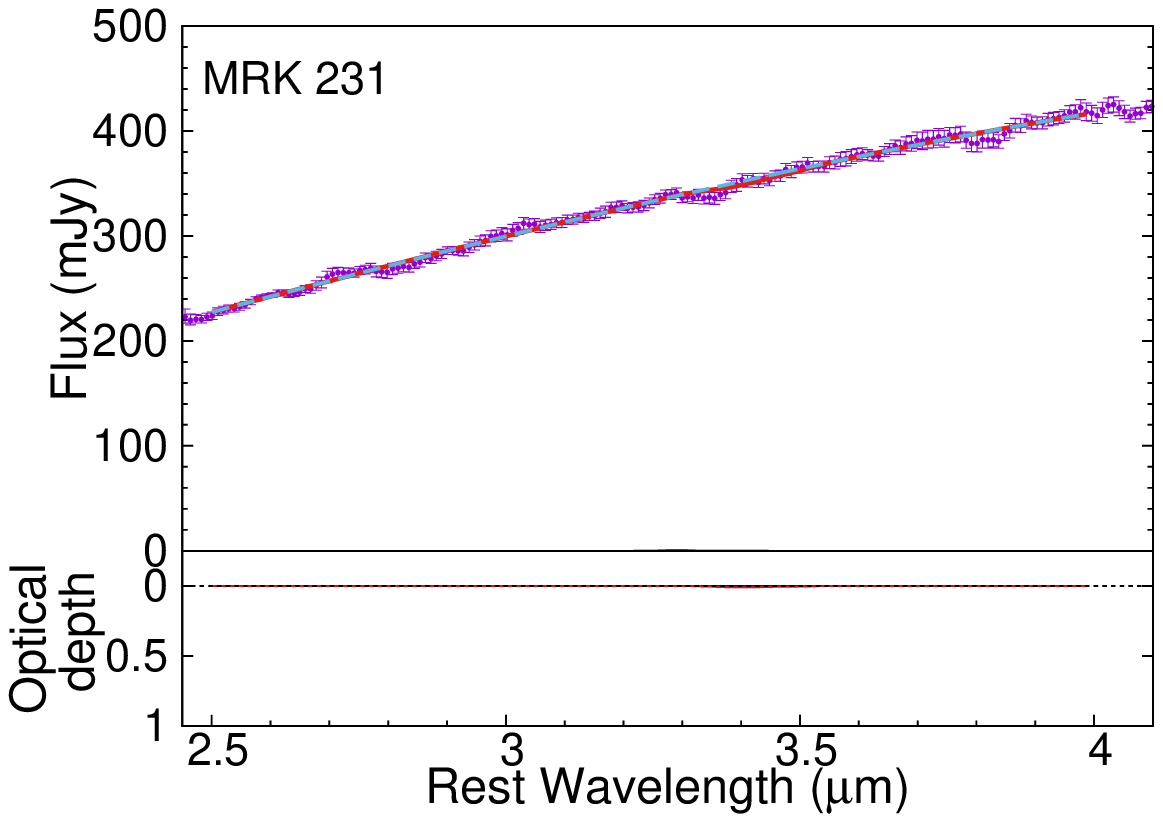}
			\end{center}
		\end{minipage}
		\begin{minipage}{0.28\textwidth}
			\begin{center}
				\includegraphics[width=1.08\textwidth]{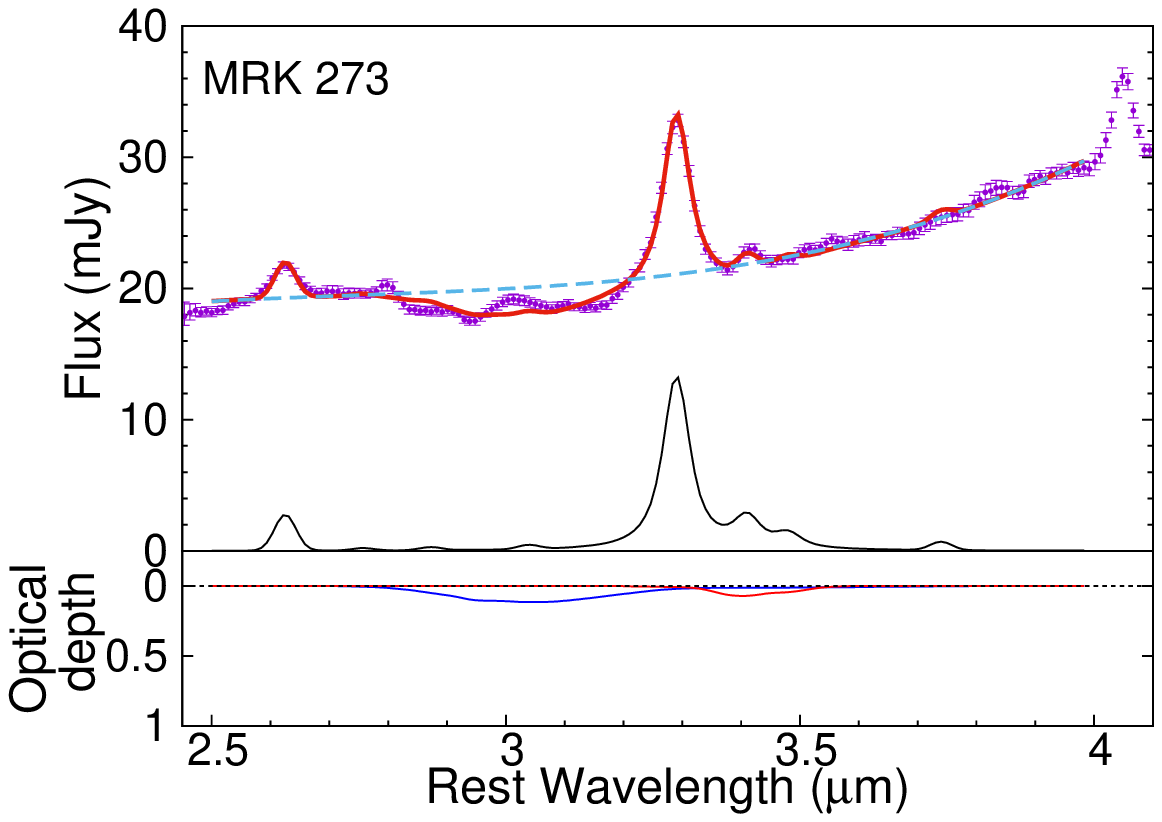}
			\end{center}
		\end{minipage}
		\begin{minipage}{0.28\textwidth}
			\begin{center}
				\includegraphics[width=1.08\textwidth]{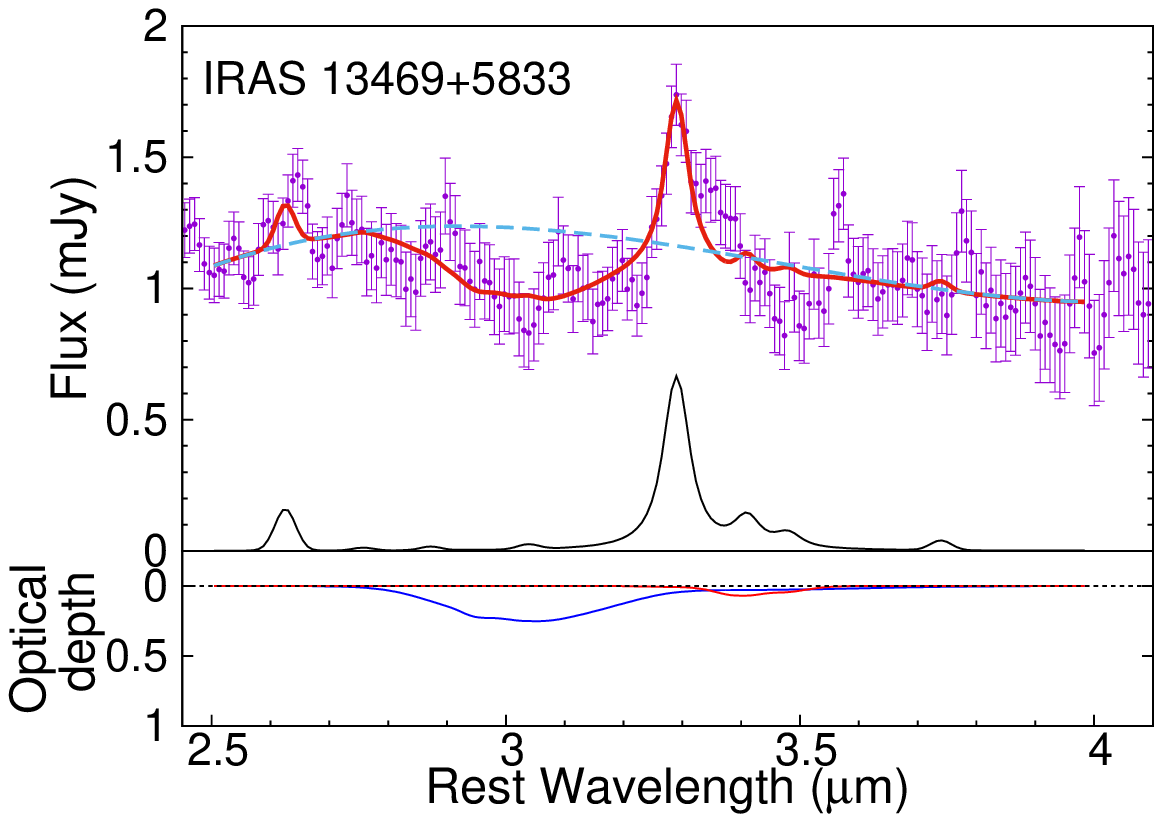}
			\end{center}
		\end{minipage}
		\vspace{2.0mm}
		\caption{(Continued)}
	\end{center}
\end{figure*}

\setcounter{figure}{5}
\begin{figure*}

	\begin{center}
		\begin{minipage}{0.28\textwidth}
			\begin{center}
				\includegraphics[width=1.08\textwidth]{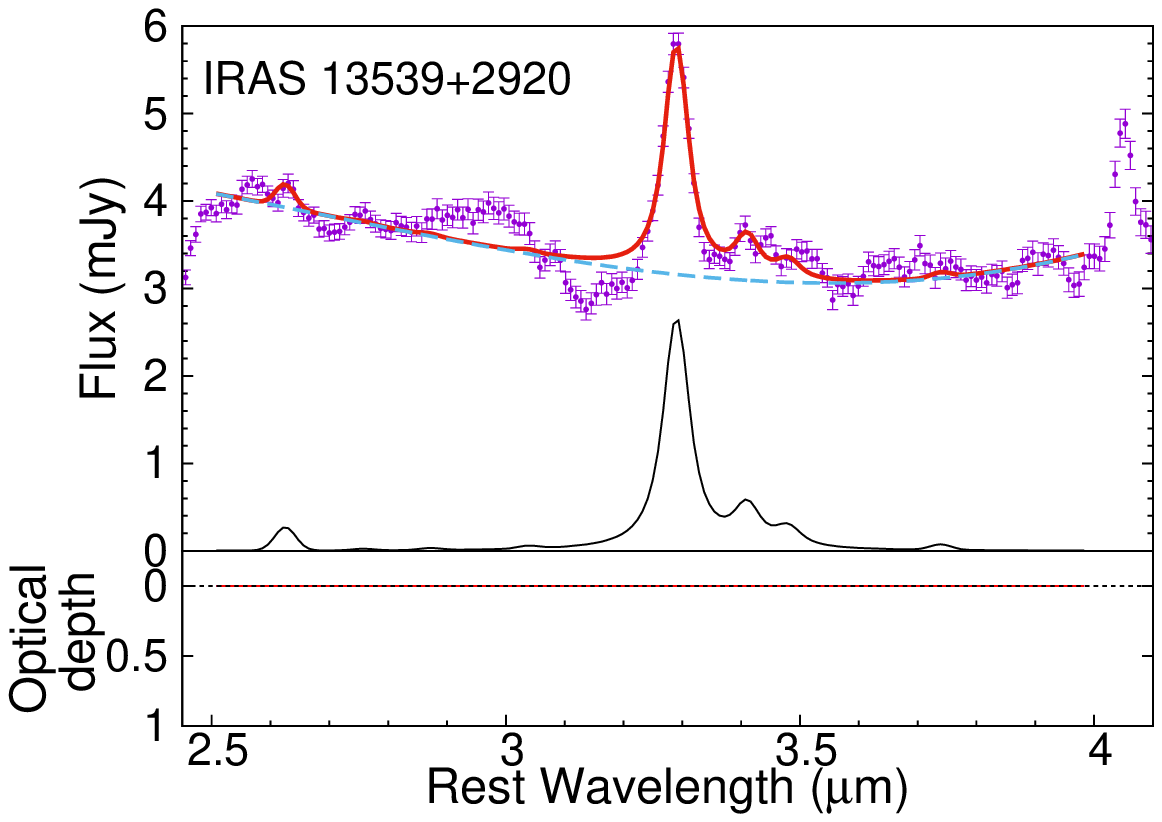}
			\end{center}
		\end{minipage}
		\begin{minipage}{0.28\textwidth}
			\begin{center}
				\includegraphics[width=1.08\textwidth]{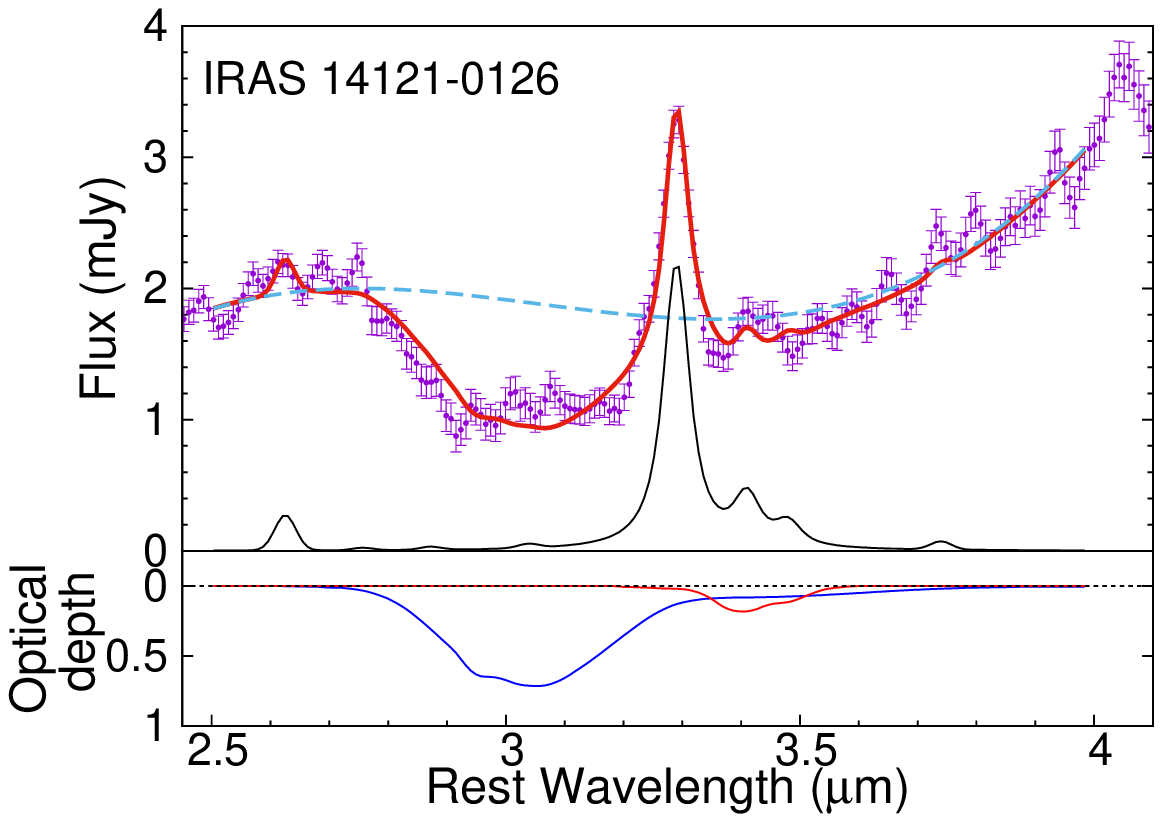}
			\end{center}
		\end{minipage}
		\begin{minipage}{0.28\textwidth}
			\begin{center}
				\includegraphics[width=1.08\textwidth]{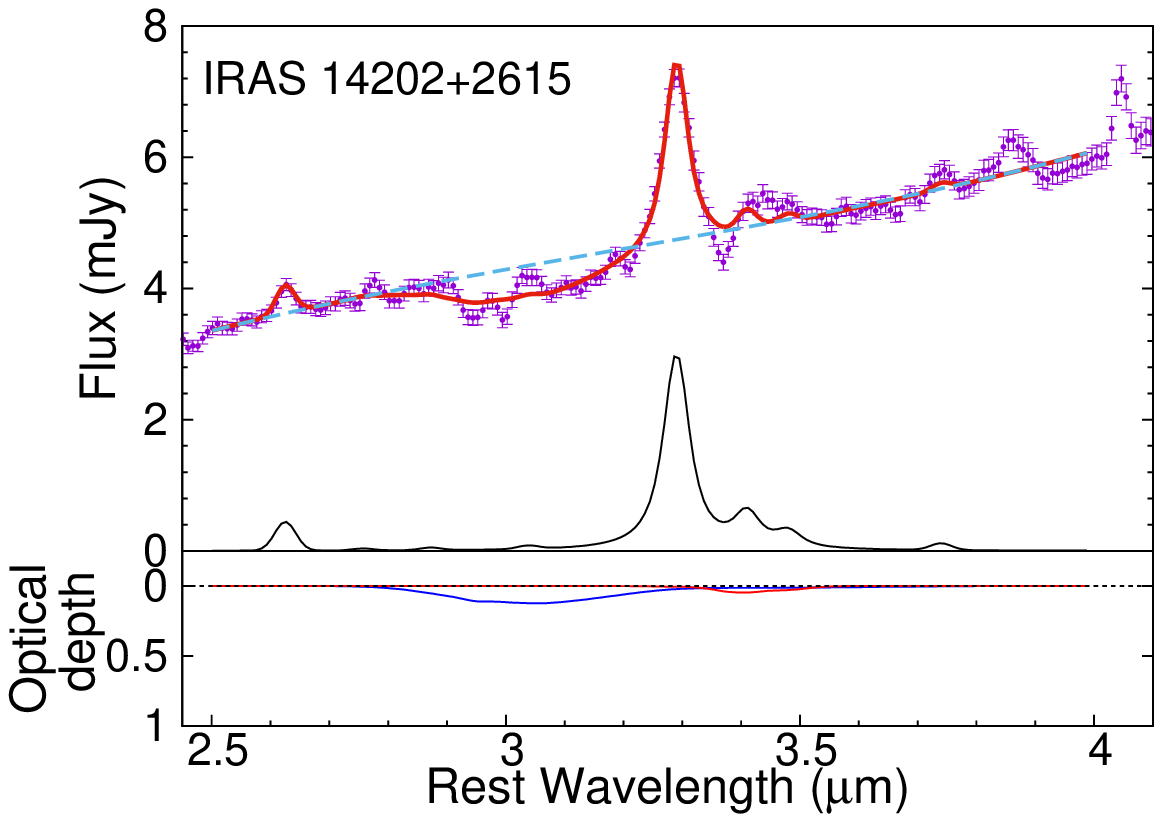}
			\end{center}
		\end{minipage}
	\end{center}
	
	\begin{center}
		\begin{minipage}{0.28\textwidth}
			\begin{center}
				\includegraphics[width=1.08\textwidth]{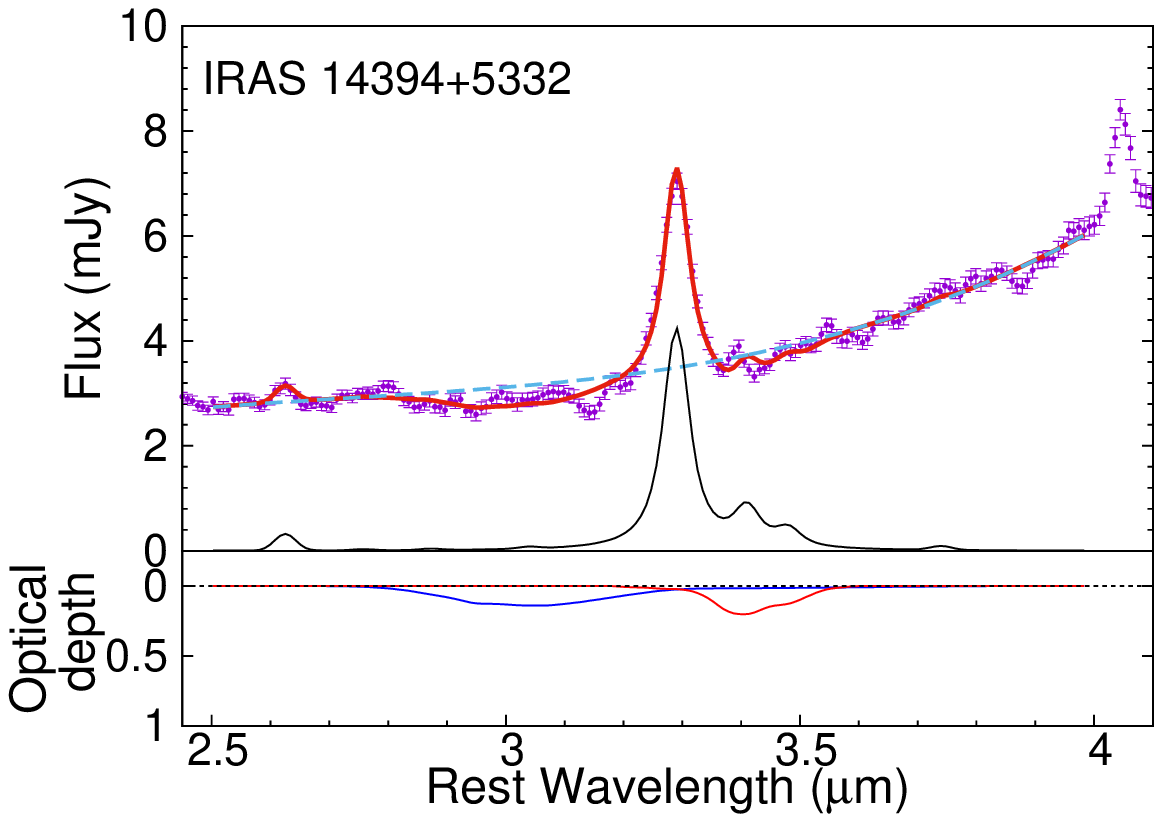}
			\end{center}
		\end{minipage}
		\begin{minipage}{0.28\textwidth}
			\begin{center}
				\includegraphics[width=1.08\textwidth]{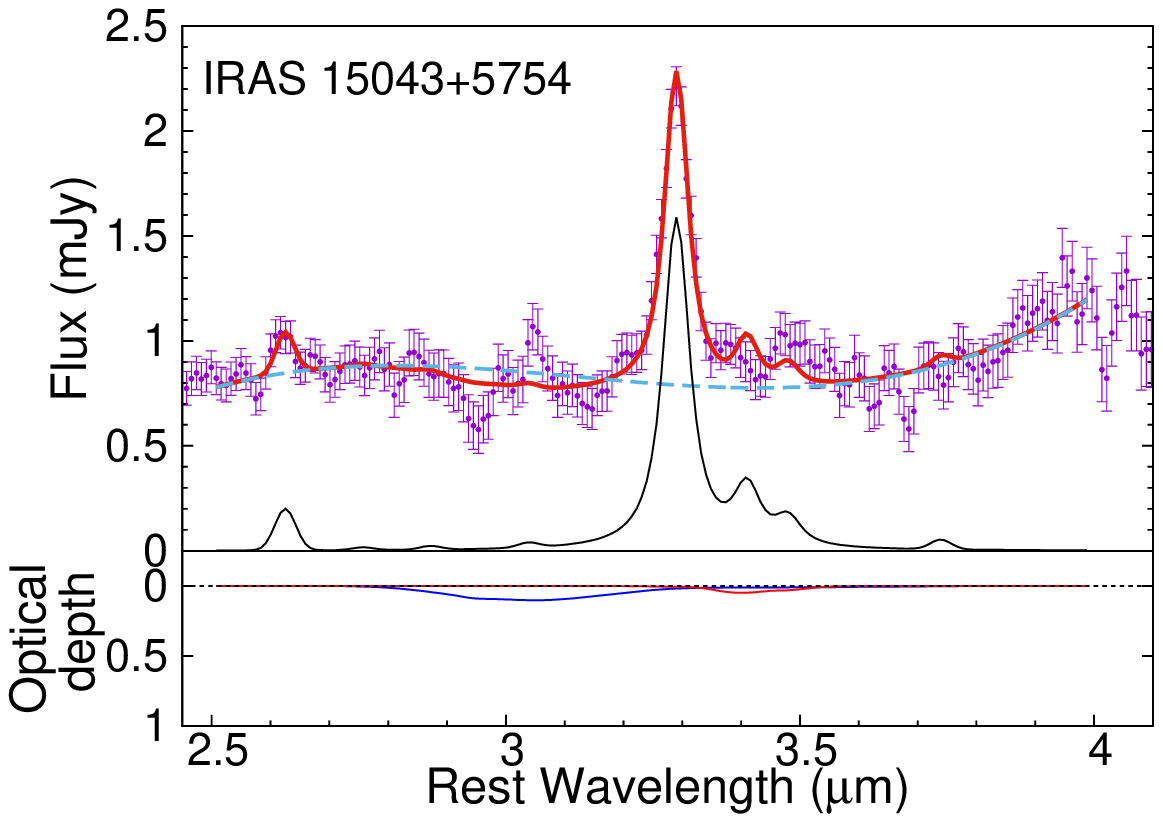}
			\end{center}
		\end{minipage}
		\begin{minipage}{0.28\textwidth}
			\begin{center}
				\includegraphics[width=1.08\textwidth]{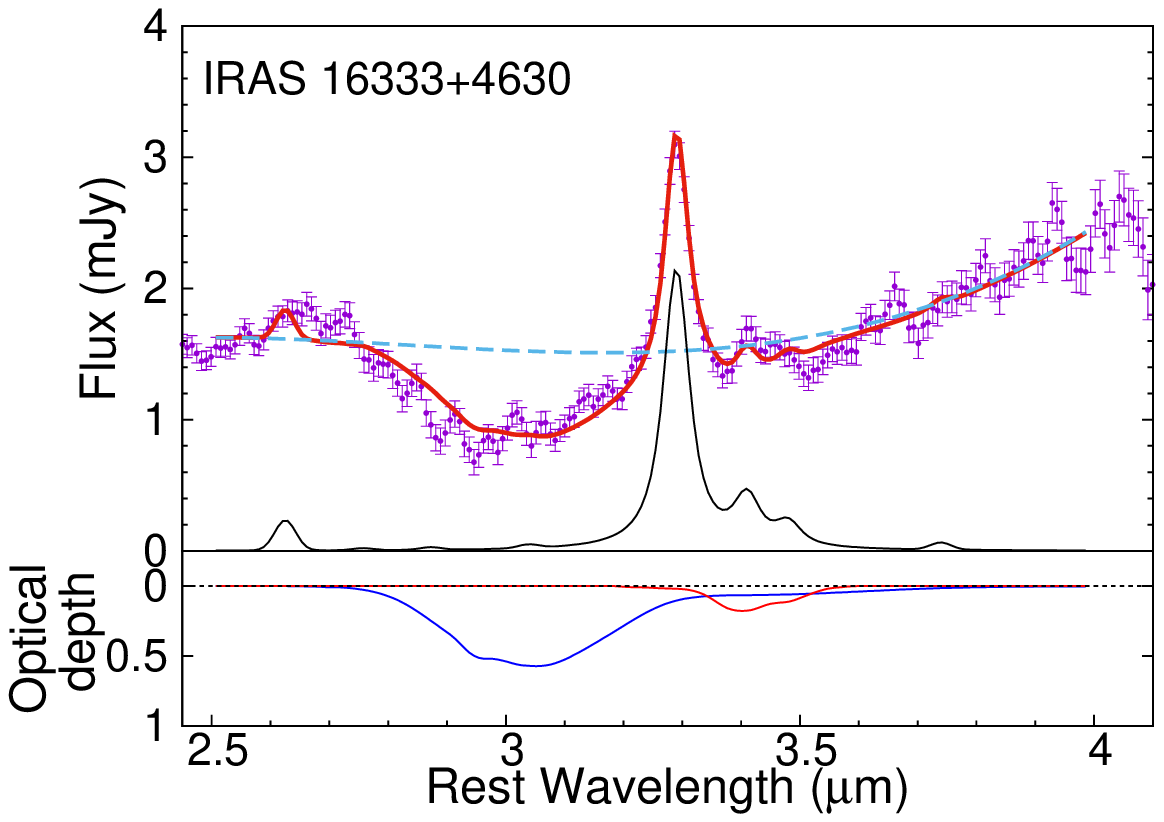}
			\end{center}
		\end{minipage}
	\end{center}
	
	\begin{center}
		\begin{minipage}{0.28\textwidth}
			\begin{center}
				\includegraphics[width=1.08\textwidth]{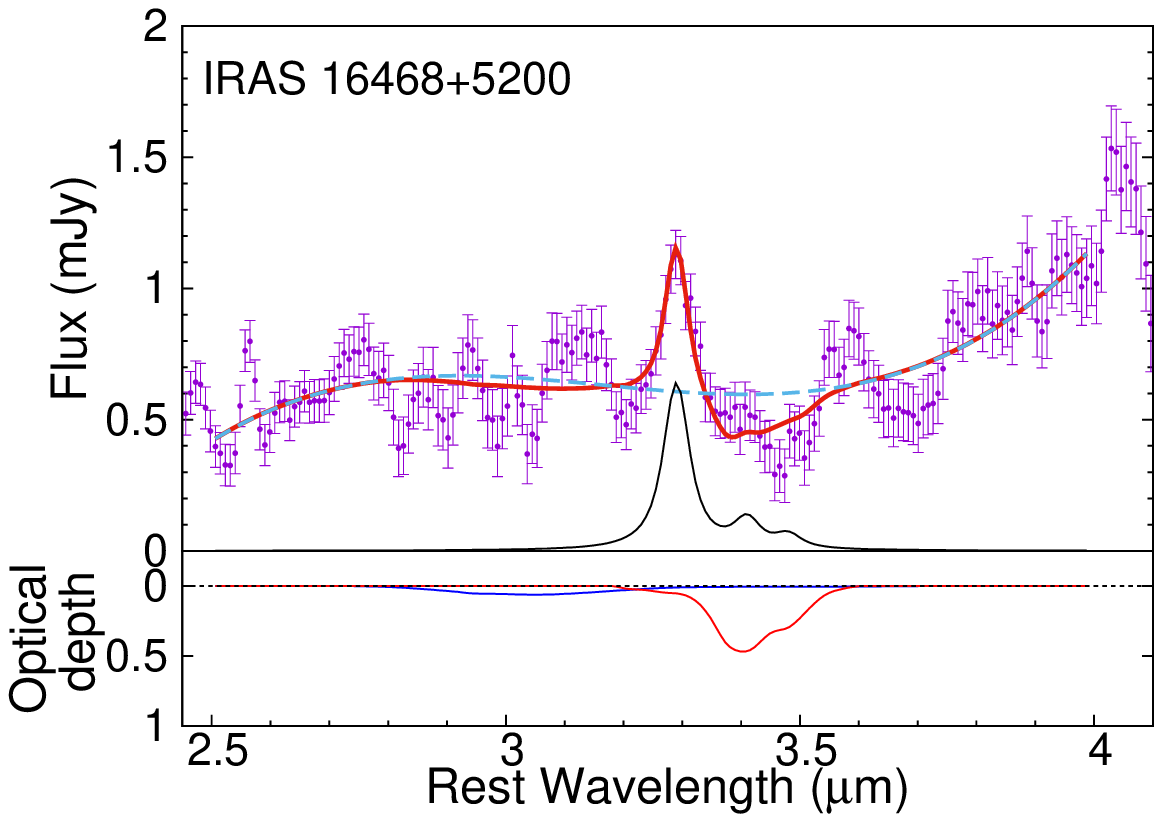}
			\end{center}
		\end{minipage}
		\begin{minipage}{0.28\textwidth}
			\begin{center}
				\includegraphics[width=1.08\textwidth]{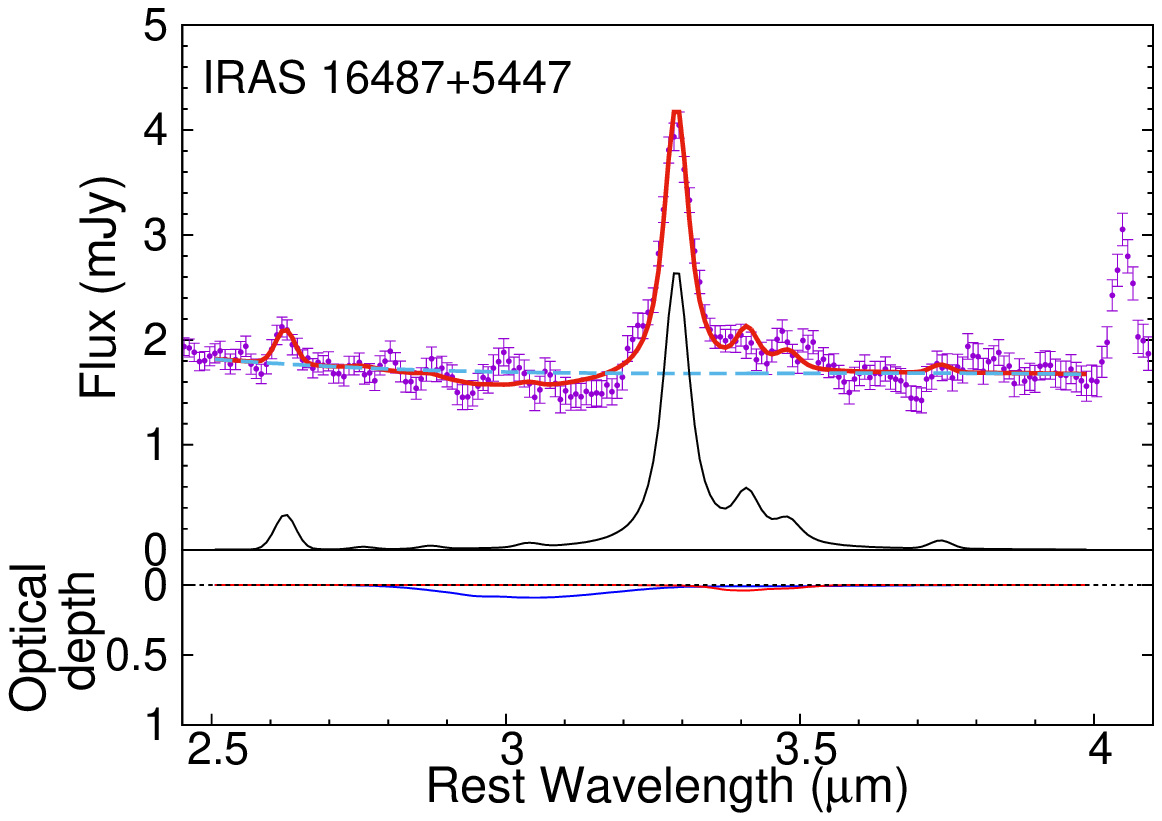}
			\end{center}
		\end{minipage}
		\begin{minipage}{0.28\textwidth}
			\begin{center}
				\includegraphics[width=1.08\textwidth]{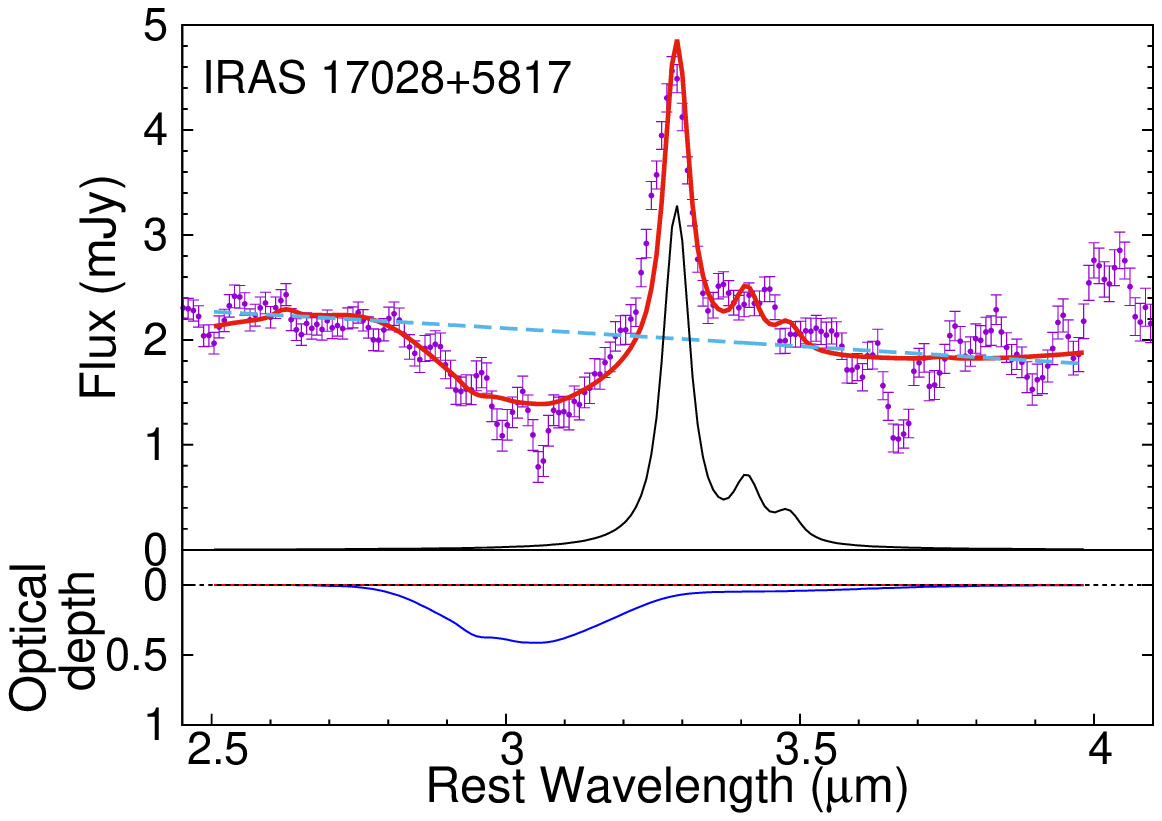}
			\end{center}
		\end{minipage}
	\end{center}
	
	\begin{center}
		\begin{minipage}{0.28\textwidth}
			\begin{center}
				\includegraphics[width=1.08\textwidth]{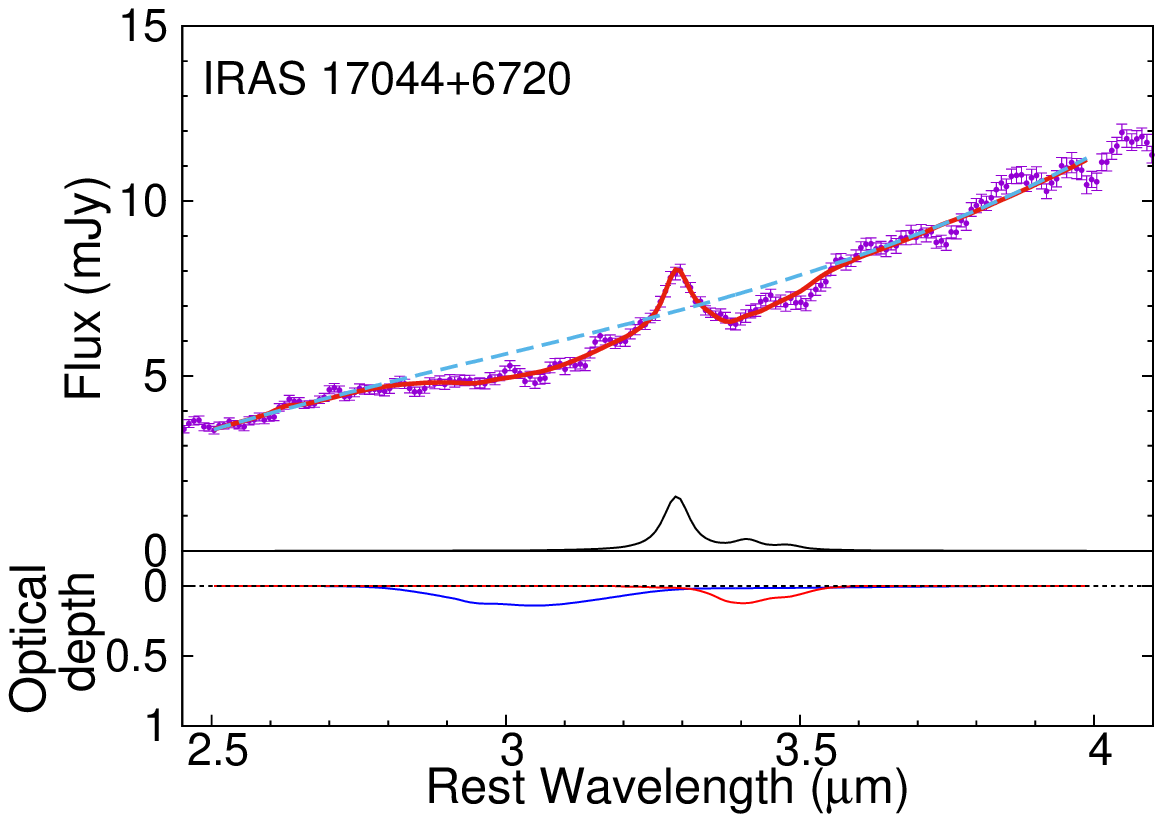}
			\end{center}
		\end{minipage}
		\begin{minipage}{0.28\textwidth}
			\begin{center}
				\includegraphics[width=1.08\textwidth]{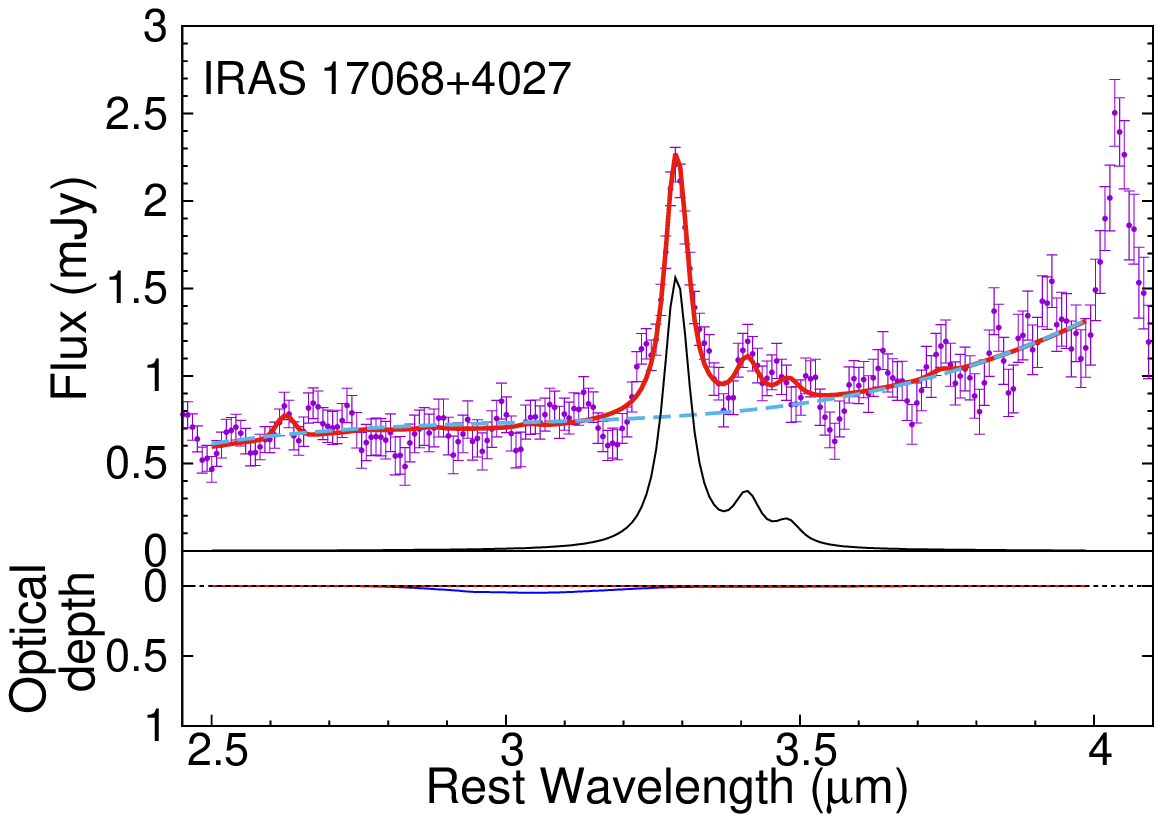}
			\end{center}
		\end{minipage}
		\begin{minipage}{0.28\textwidth}
			\begin{center}
				\includegraphics[width=1.08\textwidth]{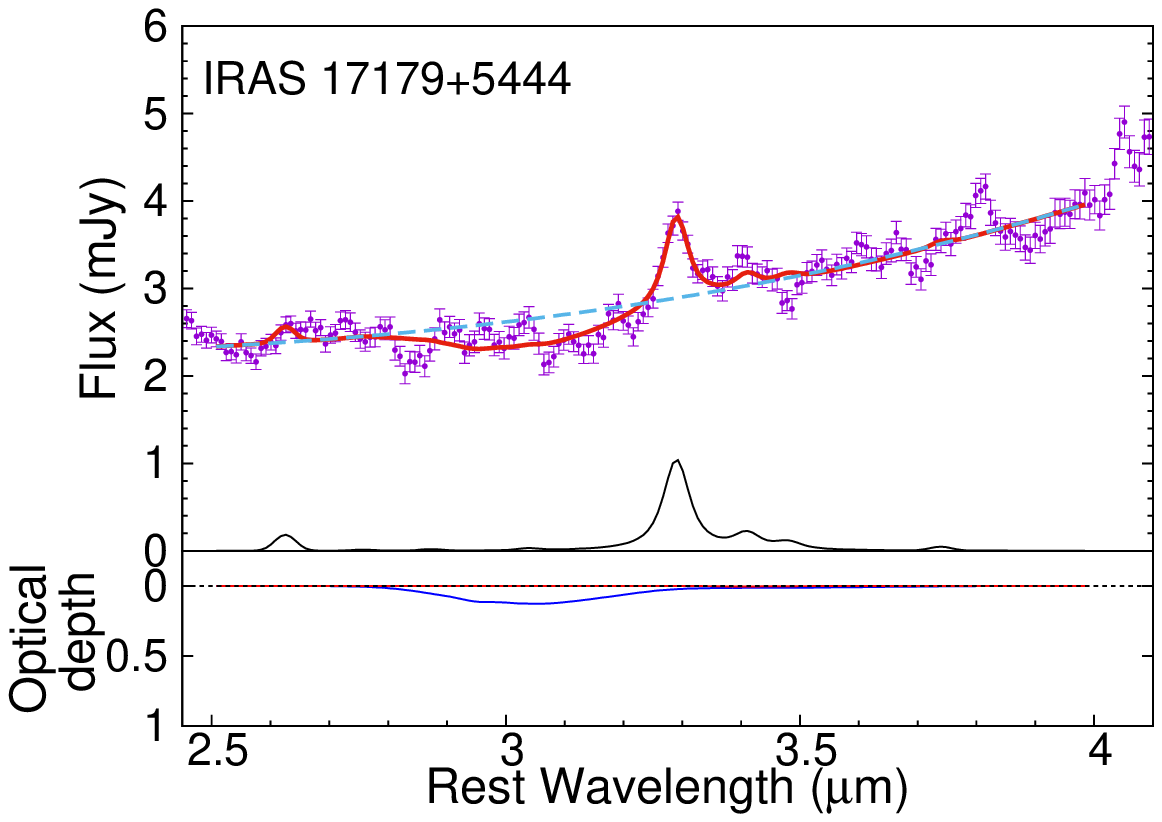}
			\end{center}
		\end{minipage}
	\end{center}
	
	\begin{center}
		\begin{minipage}{0.28\textwidth}
			\begin{center}
				\includegraphics[width=1.08\textwidth]{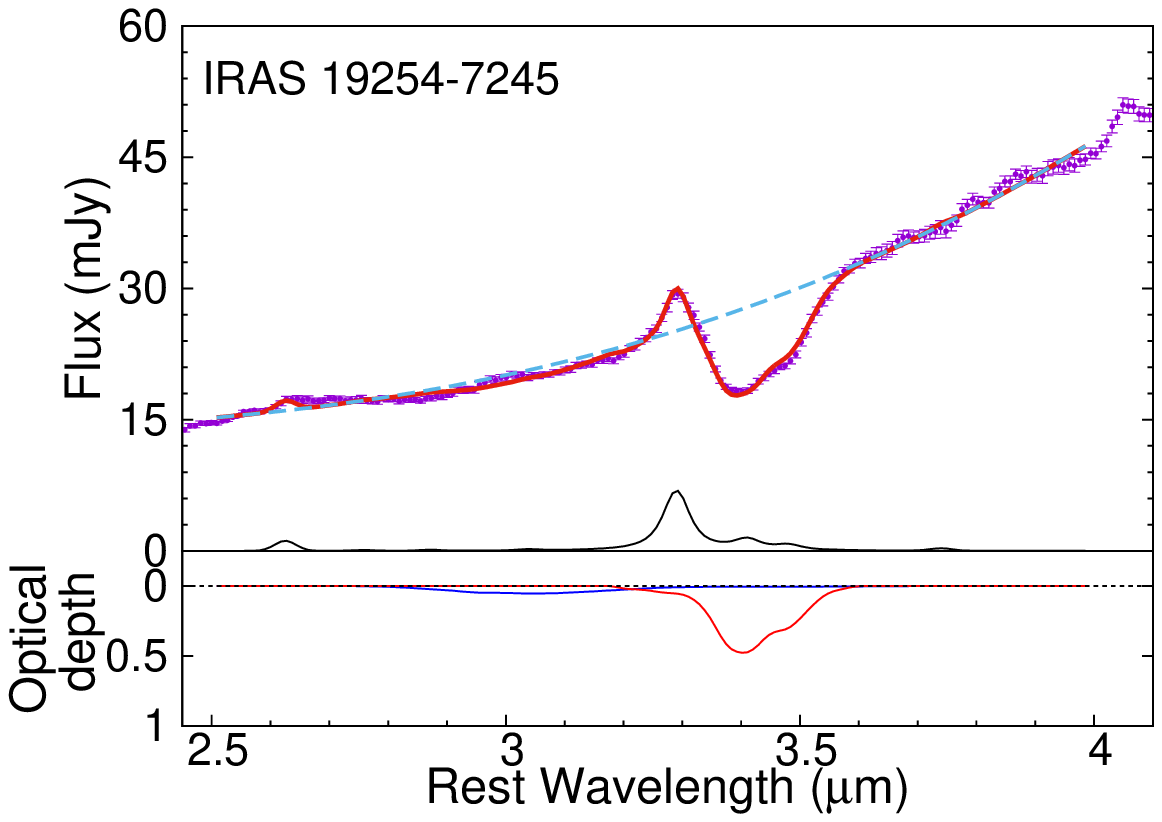}
			\end{center}
		\end{minipage}
		\begin{minipage}{0.28\textwidth}
			\begin{center}
				\includegraphics[width=1.08\textwidth]{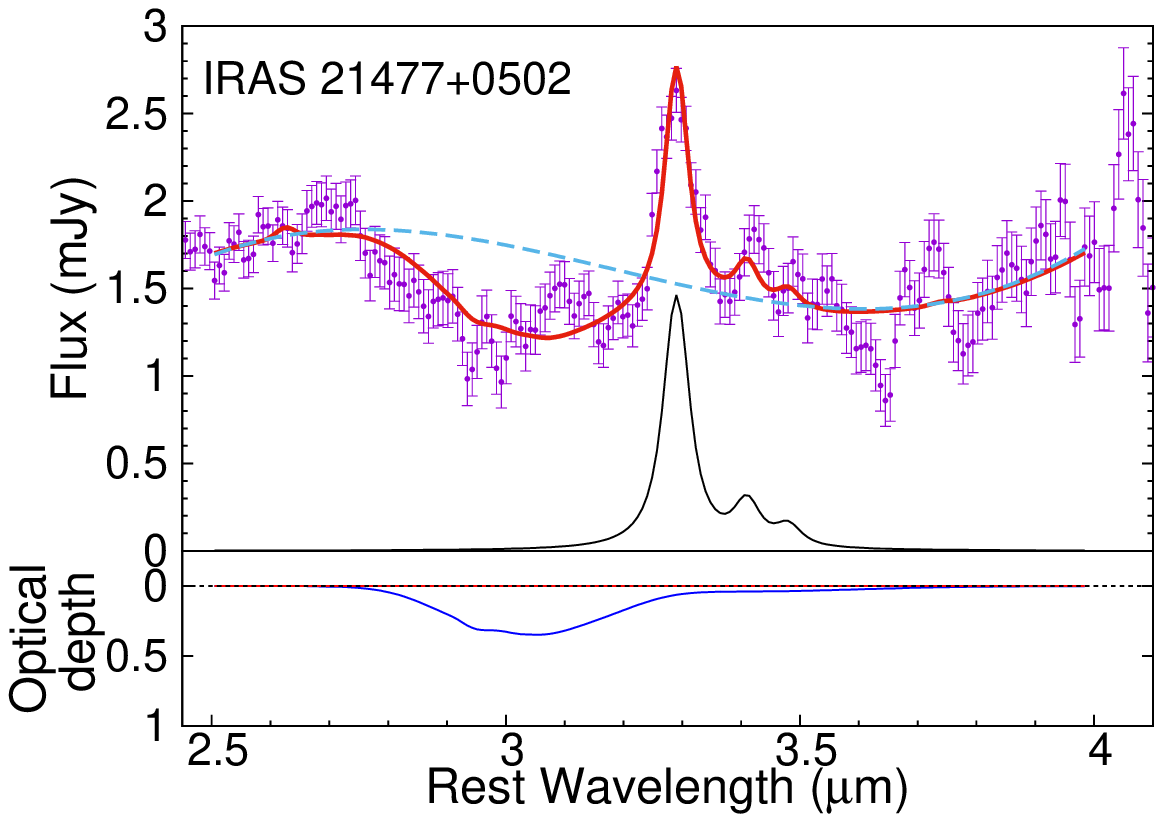}
			\end{center}
		\end{minipage}
		\begin{minipage}{0.28\textwidth}
			\begin{center}
				\includegraphics[width=1.08\textwidth]{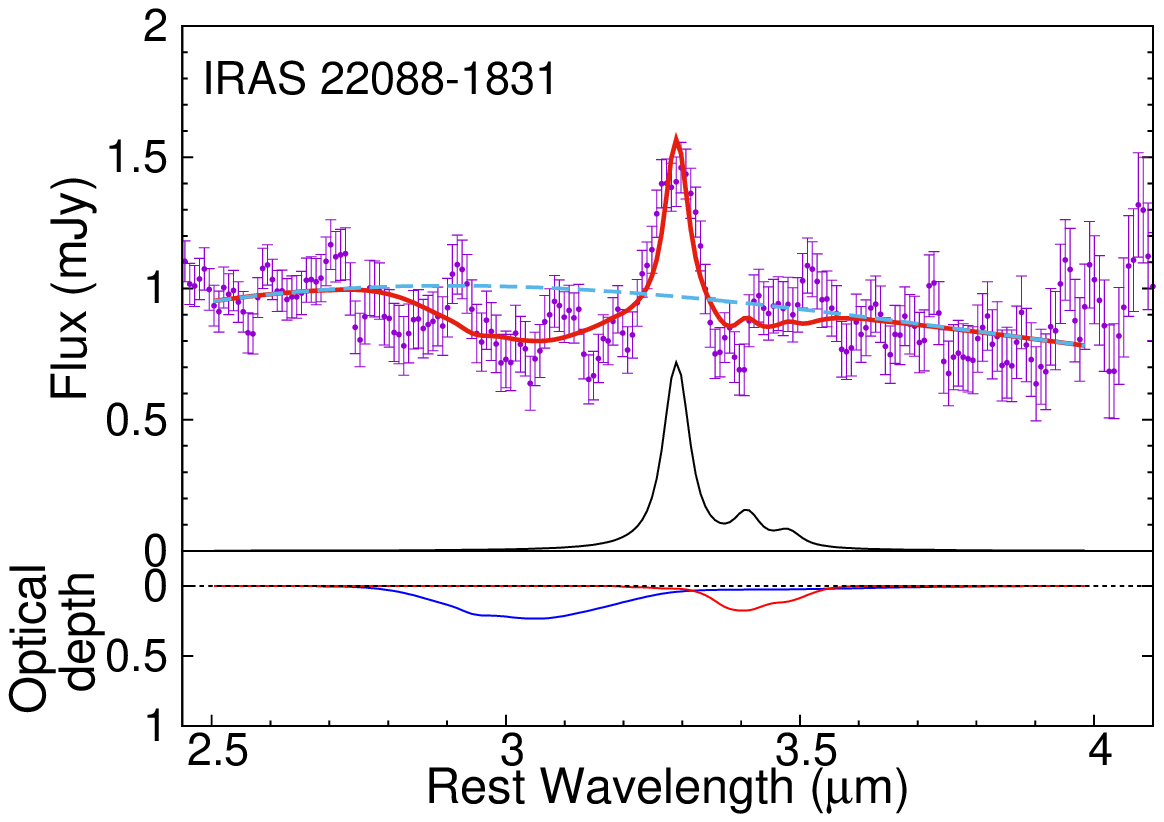}
			\end{center}
		\end{minipage}
		\vspace{2.0mm}
		\caption{(Continued)}
	\end{center}
\end{figure*}

\setcounter{figure}{5}
\begin{figure*}
	
	\begin{center}
		\begin{minipage}{0.28\textwidth}
			\begin{center}
				\includegraphics[width=1.08\textwidth]{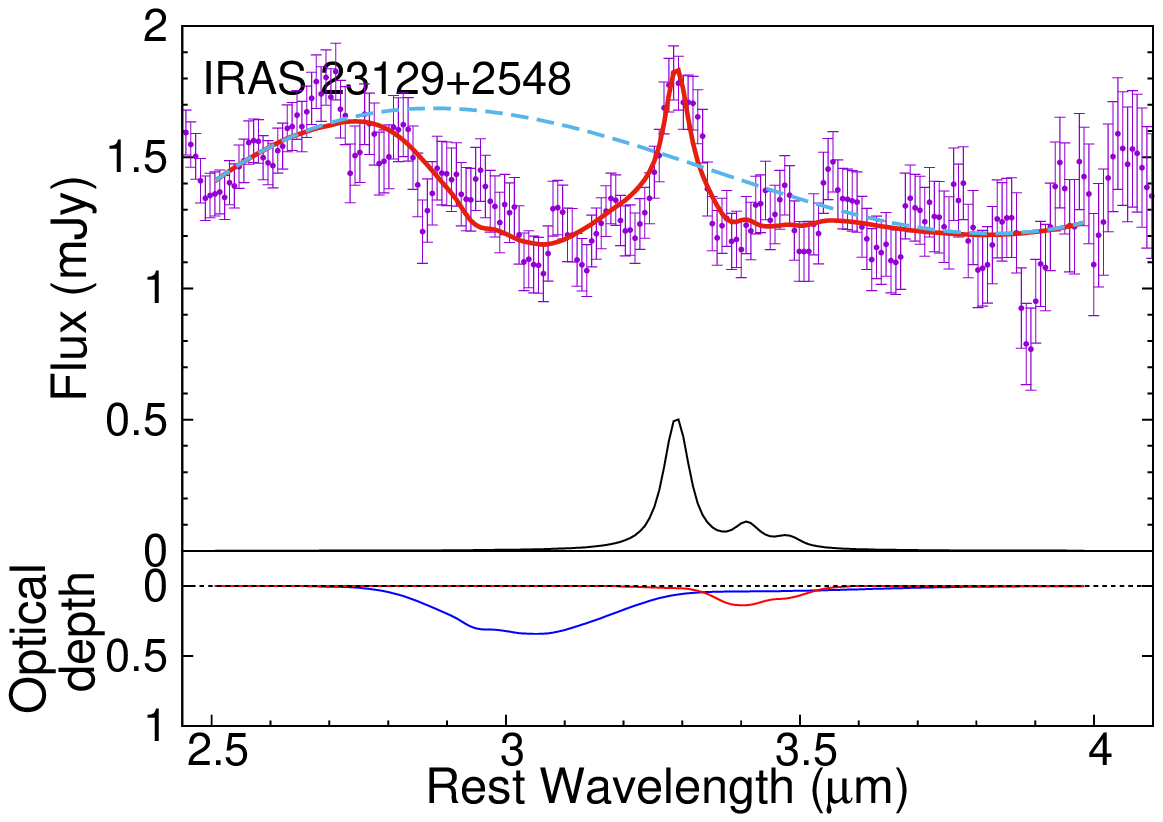}
			\end{center}
		\end{minipage}
		\begin{minipage}{0.28\textwidth}
			\begin{center}
				\includegraphics[width=1.08\textwidth]{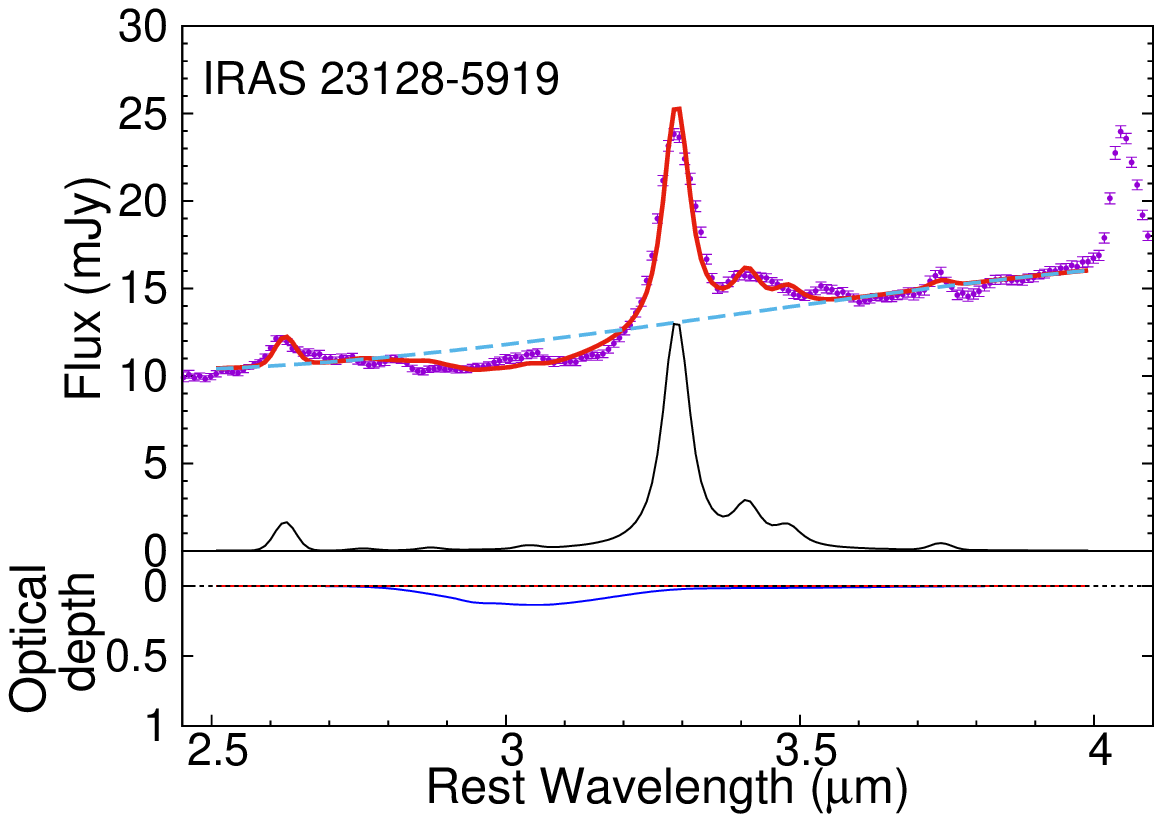}
			\end{center}
		\end{minipage}
		\begin{minipage}{0.28\textwidth}
			\begin{center}
				\includegraphics[width=1.08\textwidth]{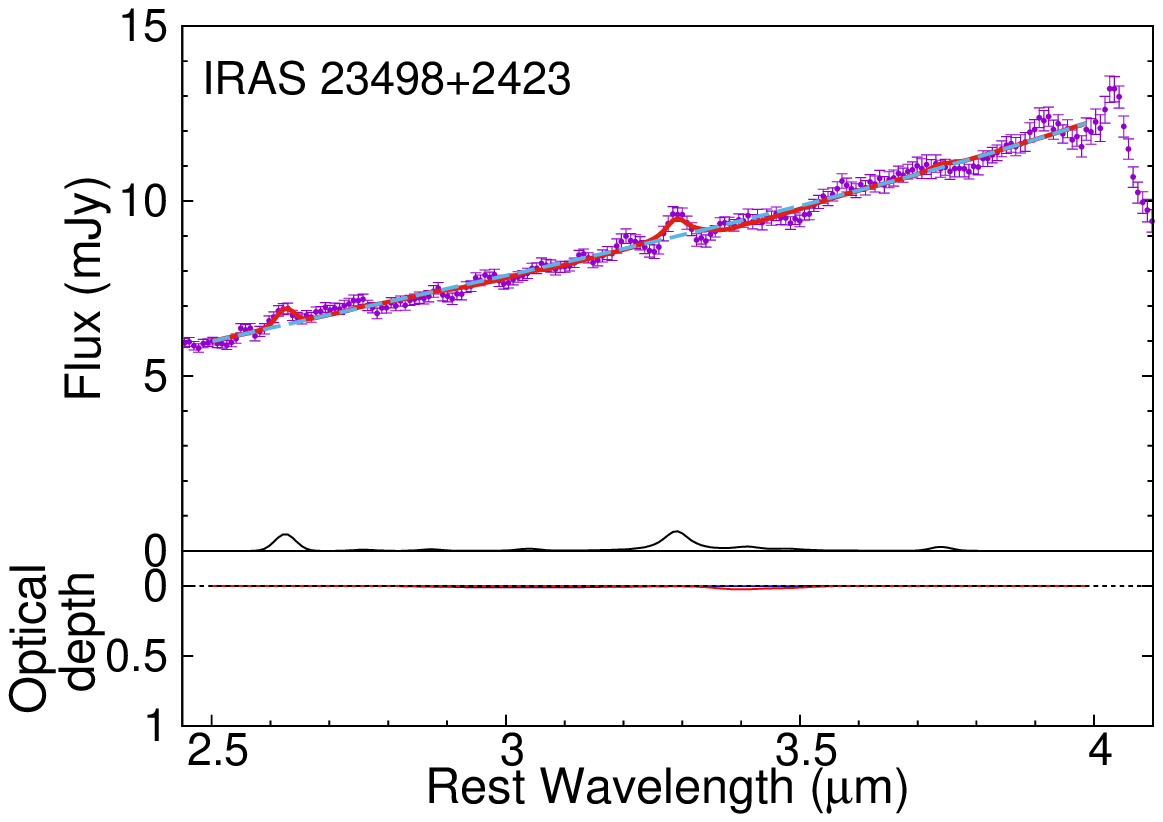}
			\end{center}
		\end{minipage}
		\vspace{2.0mm}
		\caption{(Continued)}
	\end{center}	
	
\end{figure*}

%% file: input/discussion_2_15_draft.tex
%

\subsection{Comparison with the ISM in the Milky Way}
\label{sec:cwmw}

\subsubsection{H$_{2}$O ice}
\label{sec:avh2oavhac}

In star-forming regions in the Milky Way, the interstellar extinction and the optical depth of H$_{2}$O ice are known to correlate with each other.
 \citet{2001ApJ...547..872W} found $\tau_{3.0}=0.072(A_{V}-3.2)$ for the Taurus dark cloud, the most well-studied low-mass star-forming region. H$_{2}$O ice is detected only along the lines of sight with $A_{V}>3.2$ mag. 
To compare this relation with those in ULIRGs, we used the 9.7$\mathrm{\,\mu m}$ silicate optical depth $\tau_{9.7}$ to infer $A_{V}$ in a ULIRG using a relation in the Milky Way $A_{V}/\tau_{9.7}=18.0$ (\citealt{2003dge..conf.....W} and references therein).
Silicate optical depth $\tau_{9.7}$ and the references are summarized in table \ref{tab:result_sil}.

\begin{longtable}{cccccc}
\caption{The estimated optical depths and the silicate optical depth $\tau_{9.7}$.}
	\\ \hline
	\label{tab:result_sil} 
	Object Name & $\tau_{3.0}$ & $\tau_{3.4}$ & $\tau_{9.7}^{*}$ & Uncertainty$^{*}$ & Ref. \\ 
	\hline
	\endfirsthead
	
	\multicolumn{6}{p{140mm}}{(Continued)} \\
	\hline
	Object Name & $\tau_{3.0}$ & $\tau_{3.4}$ & $\tau_{9.7}^{*}$ & Uncertainty$^{*}$ & Ref. \\ 
	\hline
	\endhead
	\multicolumn{6}{p{140mm}}{$^{*}$ The optical depth and the uncertainty are taken from the reference listed in the column 7.
		Reference: (1) Dartois and Mu{\~n}oz-Caro (\citeyear{2007AA...476.1235D}), (2) \cite{2007ApJS..171...72I}, (3) \cite{2009ApJ...694..751I}, (4) \cite{2009ApJ...701..658W}, (5) \cite{2010ApJ...709..801I}, (6) \cite{2011ApJS..193...18W}}
	\endlastfoot
      IRAS 00456$-$2904	&	0.34	$\pm$	0.02	&	0.006	$\pm$	0.014	&	1.2	&	10\%	&	2\\
    IRAS 00482$-$2721	&	0.24	$\pm$	0.09	&	0.11	$\pm$	0.07	&	2.1	&	5\%	&	2\\
    IRAS 01199$-$2307	&	0.69	$\pm$	0.07	&	0.46	$\pm$	0.07	&	2.4	&	5\%	&	5\\
    IRAS 01298$-$0744	&		$<$	0.008	&	0.25	$\pm$	0.04	&	4.0	&	5\%	&	2\\
    IRAS 01355$-$1814	&	0.2	$\pm$	0.1	&	0.05	$\pm$	0.09	&	2.4	&	5\%	&	3\\
    IRAS 01494$-$1845	&	0.48	$\pm$	0.03	&	0.17	$\pm$	0.03	&	1.6	&	10\%	&	3\\
    IRAS 01569$-$2939	&	0.12	$\pm$	0.05	&	0.06	$\pm$	0.04	&	2.8	&	5\%	&	2\\
    IRAS 02480$-$3745	&	0.19	$\pm$	0.08	&	0.23	$\pm$	0.08	&	1.4	&	10\%	&	5\\
    IRAS 03209$-$0806	&	0.25	$\pm$	0.03	&	0.13	$\pm$	0.03	&	1.0	&	10\%	&	5\\
    IRAS 03521$+$0028	&	0.01	$\pm$	0.04	&	0.07	$\pm$	0.04	&	1.3	&	10\%	&	3\\
    IRAS 04074$-$2801	&	0.09	$\pm$	0.04	&	0.25	$\pm$	0.04	&	3.0	&	5\%	&	5\\
    IRAS 04313$-$1649	&	0.3	$\pm$	0.2	&	0.3	$\pm$	0.1	&	2.8	&	5\%	&	3\\
    IRAS 05020$-$2941	&	0.32	$\pm$	0.07	&	0.01	$\pm$	0.04	&	2.4	&	5\%	&	5\\
    IRAS 05189$-$2524	&	0.034	$\pm$	0.007	&	0.027	$\pm$	0.006	&	0.315	&	5\%	&	4\\
    IRAS 06035$-$7102	&	0.38	$\pm$	0.01	&	0.40	$\pm$	0.01	&	2.9	&	5\%	&	1\\
    IRAS 08572$+$3915	&	0.185	$\pm$	0.006	&	0.757	$\pm$	0.006	&	3.8	&	5\%	&	2\\
    IRAS 08591$+$5248	&	0.14	$\pm$	0.02	&	0.05	$\pm$	0.02	&	1.0	&	10\%	&	5\\
    IRAS 09463$+$8141	&	0.25	$\pm$	0.04	&	0.07	$\pm$	0.04	&	2.0	&	5\%	&	2\\
    IRAS 09539$+$0857	&	0.43	$\pm$	0.06	&	0.02	$\pm$	0.05	&	3.5	&	5\%	&	2\\
    IRAS 10035$+$2740	&	0.12	$\pm$	0.02	&	0.22	$\pm$	0.03	&	2.0	&	5\%	&	3\\
    IRAS 10091$+$4704	&	0.02	$\pm$	0.10	&	0.1	$\pm$	0.1	&	2.5	&	5\%	&	3\\
    IRAS 10494$+$4424	&	0.45	$\pm$	0.02	&	0.02	$\pm$	0.02	&	1.7	&	10\%	&	2\\
    IRAS 10594$+$3818	&	0.29	$\pm$	0.04	&	0.002	$\pm$	0.036	&	1.0	&	10\%	&	5\\
    IRAS 11028$+$3130	&	0.26	$\pm$	0.09	&		$<$	0.2	&	2.5	&	5\%	&	3\\
    IRAS 11180$+$1623	&	0.35	$\pm$	0.09	&	0.56	$\pm$	0.10	&	2.0	&	5\%	&	3\\
    IRAS 11387$+$4116	&	0.35	$\pm$	0.03	&	0.12	$\pm$	0.03	&	1.1	&	10\%	&	2\\
    IRAS 12447$+$3721	&	0.21	$\pm$	0.07	&	0.16	$\pm$	0.07	&	1.7	&	10\%	&	5\\
            Mrk 231	&		$<$	0.01	&	0.011	$\pm$	0.006	&	0.64	&	5\%	&	4\\
            Mrk 273	&	0.108	$\pm$	0.007	&	0.072	$\pm$	0.007	&	1.746	&	5\%	&	4\\
    IRAS 13469$+$5833	&	0.24	$\pm$	0.04	&	0.07	$\pm$	0.04	&	1.7	&	10\%	&	3\\
    IRAS 13539$+$2920	&		---		&		$<$	0.03	&	1.6	&	10\%	&	2\\
    IRAS 14121$-$0126	&	0.67	$\pm$	0.03	&	0.18	$\pm$	0.03	&	1.3	&	10\%	&	5\\
    IRAS 14202$+$2615	&	0.115 $\pm$	0.010	&	0.048 $\pm$	0.010	&	0.7	&	10\%	&	5\\
    IRAS 14394$+$5332 &	0.13	$\pm$	0.01	&	0.20	$\pm$	0.01	&	---	&	---	&	--- \\
    IRAS 15043$+$5754	&	0.10	$\pm$	0.04	&	0.05	$\pm$	0.04	&	1.4	&	10\%	&	5\\
    IRAS 16333$+$4630	&	0.53	$\pm$	0.03	&	0.18	$\pm$	0.02	&	1.3	&	10\%	&	3\\
    IRAS 16468$+$5200	&	0.06	$\pm$	0.06	&	0.47	$\pm$	0.08	&	2.5	&	5\%	&	2\\
    IRAS 16487$+$5447	&	0.08	$\pm$	0.02	&	0.04	$\pm$	0.02	&	1.8	&	10\%	&	2\\
    IRAS 17028$+$5817	&	0.38	$\pm$	0.02	&		$<$	0.04	&	1.5	&	10\%	&	2\\
    IRAS 17044$+$6720	&	0.131	$\pm$	0.010	&	0.13	$\pm$	0.01	&	1.8	&	10\%	&	2\\
    IRAS 17068$+$4027	&	0.05	$\pm$	0.04	&		$<$	0.1	&	1.8	&	10\%	&	3\\
    IRAS 17179$+$5444	&	0.12	$\pm$	0.02	&	0.004	$\pm$	0.015	&	---	&	---	&	--- \\
    IRAS 19254$-$7245	&	0.050	$\pm$	0.007	&	0.482	$\pm$	0.006	&	1.345	&	5\%	&	4\\
    IRAS 21477$+$0502	&	0.32	$\pm$	0.03	&		$<$	0.05	&	0.8	&	10\%	&	5\\
    IRAS 22088$-$1831	&	0.22	$\pm$	0.04	&	0.18	$\pm$	0.04	&	2.6	&	5\%	&	5\\
    IRAS 23129$+$2548	&	0.32	$\pm$	0.03	&	0.14	$\pm$	0.03	&	2.6	&	5\%	&	3\\
    IRAS 23128$-$5919	&	0.128	$\pm$	0.006	&		$<$	0.005	&	---	&	---	&	--- \\
    IRAS 23498$+$2423	&	0.010	$\pm$	0.008	&	0.025	$\pm$	0.008	&	0.6	&	10\%	&	6\\
		\hline
\end{longtable}

\begin{figure}
 \begin{center}
 \hspace{-50mm}
 \raisebox{-100mm}{\includegraphics[width=0.9\linewidth]{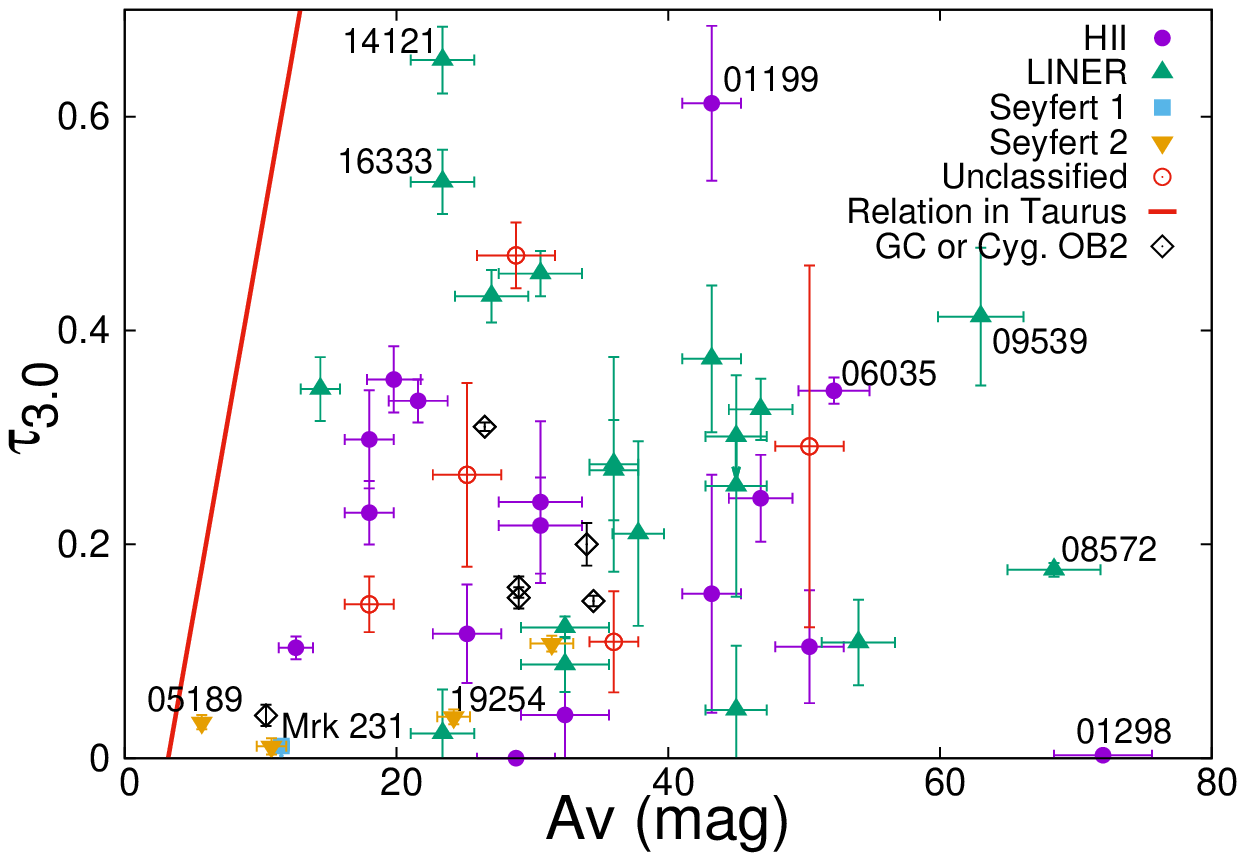}}
 \end{center}
 \caption{The plot of visual extinction $A_{V}$ and the H$_{2}$O ice optical depth $\tau_{3.0}$ . ULIRGs are classified on the basis of optical property (See table \ref{tab:result_sil}; filled purple circles, filled green triangles, filled blue squares, filled yellow inverted triangles, and open red circles respectively denote H\emissiontype{II}, LINER, Seyfert 1, Seyfert 2, and unclassified galaxies). The solid line represents the relation known in the Taurus dark cloud \citep{2001ApJ...547..872W}. The arrows indicate 3$\sigma$ upper limits. Object names are labeled for some representative sources (abbreviated for IRAS objects). Values observed along lines of sight toward the GC or Cygnus OB2 are also shown (open diamonds; \citealt{1999ApJ...514..202F}; \citealt{2002ApJ...570..198C}; \citealt{2004ApJS..151...35G}).}
 \label{fig:av_h2o}
 \end{figure}

Figure \ref{fig:av_h2o} presents a plot of visual extinction $A_{V}$, estimated in the way described above, versus $\tau_{3.0}$. Some values obtained toward the GC sources or Cygnus OB2 are also plotted in figure \ref{fig:av_h2o}. Note that sources in the GC in fact exhibit nonzero $\tau_{3.0}$, because their extinction is partly due to dark clouds local to the sources.
Figure \ref{fig:av_h2o} shows that the observed values clearly deviated from the relation in low-mass star-forming regions and that the observed values are closer to values obtained toward the GC or Cygnus OB2. 
 A ratio of the gradient between the intercept point of the relation in Taurus and an observed value in a ULIRG to the gradient of the relation in Taurus can be connected to the ratio of the amount of low-mass star-forming clouds like Taurus to the total amount of interstellar clouds in a ULIRG. According to this interpretation, low-mass star-forming clouds contribute to no more than 50\% of the interstellar clouds in all the ULIRGs in our sample, and high-mass star-forming clouds are predominant among the interstellar clouds.
 Most dust in local ULIRGs resides in the nuclear (within $\sim$ 1 kpc) regions (e.g., \citealt{2000AJ....119..509S}).
 $\tau_{3.0}$ and $\tau_{9.7}$, both of which are dust features, therefore likely reflect the local condition in the nuclear regions.
 The observed low $\tau_{3.0}/\tau_{9.7}$ indicates that high-mass star-forming activity is predominant in nuclear starbursts in local ULIRGs.
 This is also consistent with the high star-formation density in local ULIRGs \citep{2016AA...591A.136L}, because the high-density environment leads to a high Jeans mass and an intense radiation field that destroys the H$_{2}$O ice mantle.

\begin{figure}
 \begin{center}
 \hspace{-50mm}
 \raisebox{-100mm}{\includegraphics[width=0.9\linewidth]{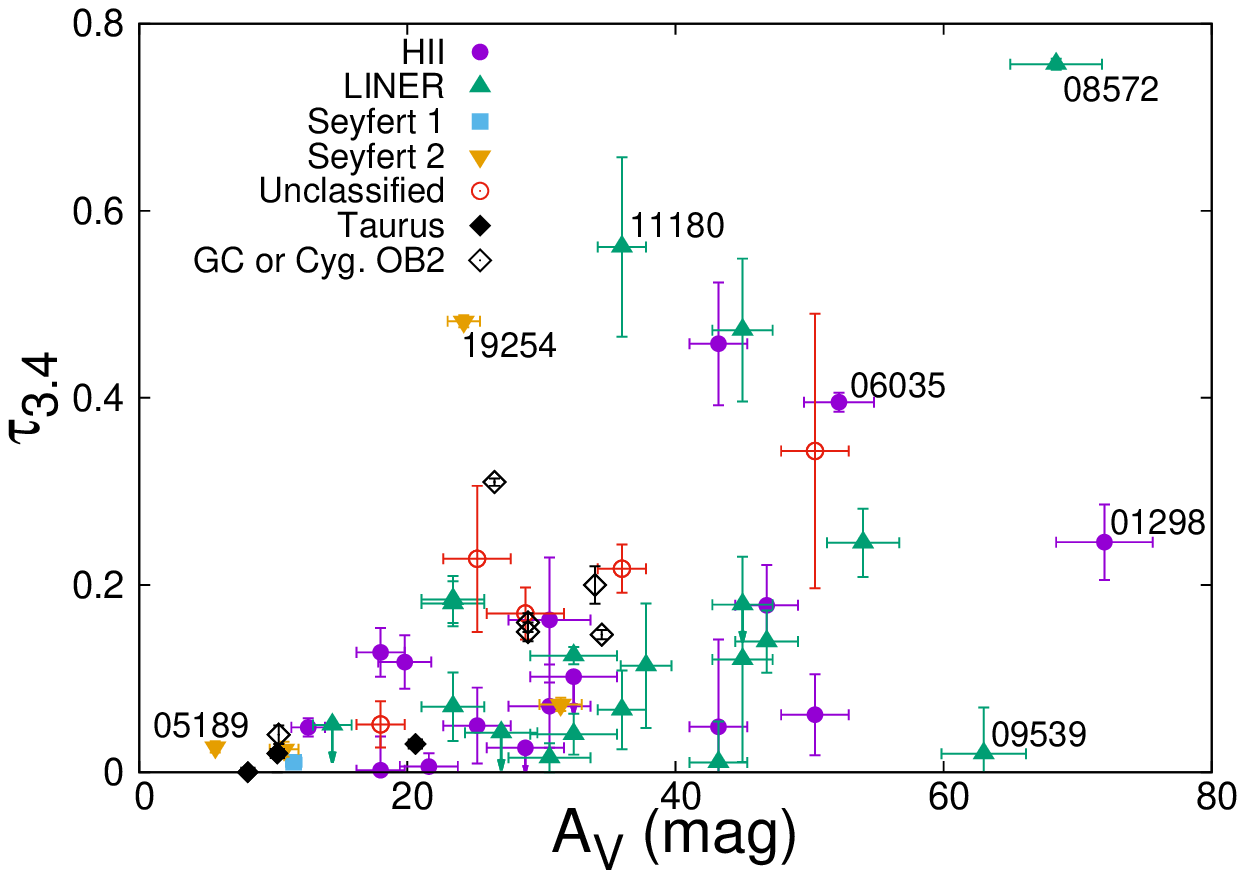}}
 \end{center}
 \caption{The plot of visual extinction $A_{V}$ and the aliphatic carbon optical depth $\tau_{3.4}$. Values observed toward field stars behind the Taurus dark cloud (\citealt{1988MNRAS.233..321W}; \citealt{1997ASPC..122..179P}), stars in the GC and Cygnus OB2 (\citealt{1980Natur.287..518W}; \citealt{1999ApJ...514..202F}; \citealt{2002ApJ...570..198C}; \citealt{2004ApJS..151...35G}) are also shown. Filled black diamonds denote the Taurus dark cloud and the other symbols denote the same as figure \ref{fig:av_h2o}.}
 \label{fig:av_hac}
 \end{figure}

\subsubsection{Aliphatic carbon}
\label{sec:crha}
 Extinction versus the optical depth of aliphatic carbon, $\tau_{3.4}$, are plotted in figure \ref{fig:av_hac}.
 Our ULIRG sample generally exhibits the aliphatic carbon absorption, which also supports the conclusion that a significant number of clouds, other than low-mass star-forming clouds, contribute to population of the interstellar clouds in a ULIRG (section \ref{sec:avh2oavhac}), because the aliphatic carbon absorption is suppressed in Galactic low-mass star-forming regions.
 It is worth noting that some ULIRGs exhibit higher ratios of the optical depth of aliphatic carbon to extinction than do Galactic sources, unlike the H$_{2}$O ice optical depth, for which the entirety of our ULIRG sample exhibits lower ratios to extinction than Taurus (figure \ref{fig:av_h2o}).
 
\begin{figure}
 \begin{center}
 \hspace{-50mm}
 \includegraphics[width=0.9\linewidth]{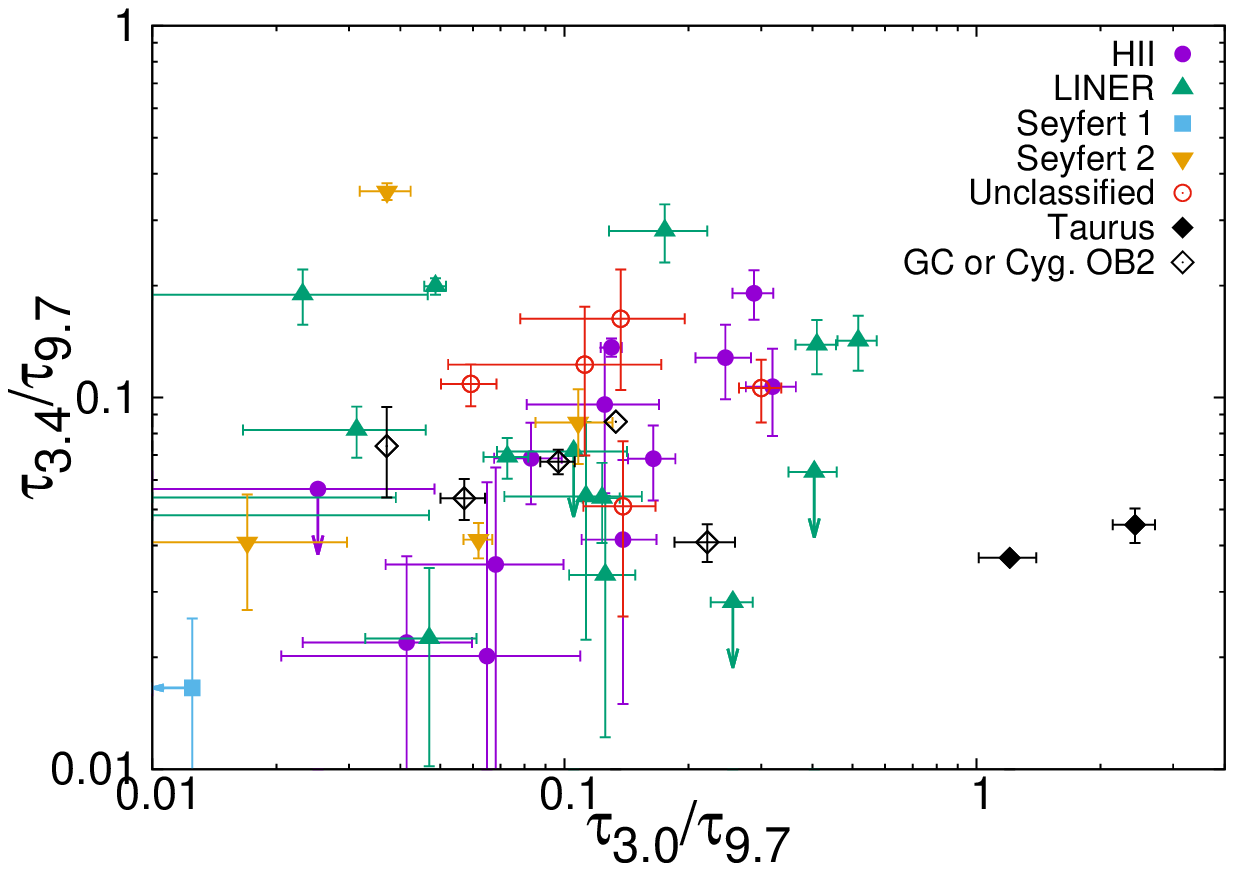}
 \end{center}
 \caption{Distribution of $\tau_{3.0}/\tau_{9.7}$ and $\tau_{3.4}/\tau_{9.7}$. The symbols represent the same as figure \ref{fig:av_hac}.}
 \label{fig:h2ohac}
 \end{figure}

 We present a plot of $\tau_{3.0}/\tau_{9.7}$ versus $\tau_{3.4}/\tau_{9.7}$ in figure \ref{fig:h2ohac} to underline the difference between the distributions of the $\tau_{3.0}/\tau_{9.7}$ ratio and the $\tau_{3.4}/\tau_{9.7}$ ratio.
 Note that in order to avoid uncertainty about the conversion, we hereafter discuss using $\tau_{9.7}$ instead of $A_{V}$ (in ULIRGs we derived from a Galactic relation $A_{V}/\tau_{9.7} = 18.0$; \citealt{2003dge..conf.....W}) as an index of the total amount of dust.
 ULIRGs generally distribute away from the locus of Taurus and have values similar to those of sources in the GC or Cygnus OB2.
 The dust features observed in spectra of GC sources and Cygnus OB2 are widely thought to trace the diffuse ISM (e.g., \citealt{2000ApJ...537..749C}). This may lead to a straightforward interpretation: a resemblance in terms of $\tau_{3.0}/\tau_{9.7}$ and $\tau_{3.4}/\tau_{9.7}$ between GC sources or Cygnus OB2 and ULIRGs indicates that the dust in ULIRGs is dominated by the diffuse ISM, where stars do not form. However, this interpretation clearly contradicts the well-accepted conclusion that ULIRGs are generally dominated by starburst activity (\citealt{2008MNRAS.385L.130N}; \citealt{2010MNRAS.405.2505N}).
 This paradox could be instead interpreted as follows: star formation is ongoing in local ULIRGs but the UV environment there is as intense as the diffuse ISM in the Milky Way.
 As we discussed in section \ref{sec:avh2oavhac}, ULIRGs in our sample consist largely of high-mass star-forming regions.
 Therefore, the resemblance can be interpreted that high-mass star-forming regions in those ULIRGs have a UV environment as intense as in the diffuse ISM in the Milky Way.
 However, it is also clear in figure \ref{fig:h2ohac} that the observed distribution cannot be reproduced by a simple linear combination of two UV environments in the Milky Way, that is, the well-attenuated UV environment observed in Taurus-like low-mass star-forming regions and the intense UV environment observed in the diffuse ISM, which is traced by observations toward the GC or Cygnus OB2.
 The observed distribution exhibits a large scatter especially in the $\tau_{3.4}/\tau_{9.7}$ ratio.
 
  \begin{table}
\caption{Weighted averages of $\tau_{3.0}/\tau_{9.7}$ and $\tau_{3.4}/\tau_{9.7}$.}
\label{tab:weighted_average}
   \begin{center}
	\begin{tabular}{ccc} \hline
	Type 		& $\tau_{3.0}/\tau_{9.7}$	& $\tau_{3.4}/\tau_{9.7}$\\ \hline
	Seyfert 		& 0.045 $\pm$ 0.004				& 0.050  $\pm$ 0.004	\\
	non-Seyfert$^{*}$ 	& 0.110  $\pm$ 0.004			& 0.061  $\pm$ 0.003	\\ \hline
		\multicolumn{3}{p{62mm}}{$^{*}$ IRAS 08572$+$3915 is excluded in calculation because of its high weight (see text).}
	\end{tabular}
   \end{center}
\end{table}

\subsection{Interpretations of $\tau_{3.4}/\tau_{9.7}$ diversity}
\label{sec:iasd}

 In this subsection, we discuss what causes the diversity of the $\tau_{3.4}/\tau_{9.7}$ ratio.
 The diversity of this ratio has been reported in some studies (e.g., \citealt{2000MNRAS.319..331I}; \citealt{2009ApJ...703..270S}).
 We here investigate two effects that could affect the $\tau_{3.4}/\tau_{9.7}$ ratio in our sample ULIRGs: the geometric temperature gradient \citep{2000MNRAS.319..331I} and the diversity of the UV environment in star-forming regions. The former could enhance the $\tau_{3.4}/\tau_{9.7}$ ratio and the latter, in an intense UV environment, could reduce the ratio.
 
 \subsubsection{Obscuring geometry}
 \label{sec:obscg}
 In this subsection, we briefly discuss obscuring geometry of the sources before discussing the $\tau_{3.4}/\tau_{9.7}$ diversity because the geometric temperature gradient model assumes that carriers of an absorption feature veil continuum sources (screen geometry).
 
 In screen geometry, optical depth $\tau^{\mathrm{sc}}$ is simply expressed as $\tau^{\mathrm{sc}} = -\ln{\left( F/F_{0}\right)}$, where $F$ is observed flux and $F_{0}$ is its continuum component.
 However, this conversion is inappropriate when sources and absorbers are mixed. Well-mixed geometry is an extreme case in which sources and absorbers are completely mixed along a line of sight.
 In well-mixed geometry, the optical depth of a feature for given $F/F_{0}$ is calculated by
 \begin{equation}
	 F/F_{0}  = \left(1+r\right)^{-1} \frac{    1-\exp[-(1+r^{-1})\tau^{\mathrm{mix}}]    }{    1-\exp[-r^{-1} \tau^{\mathrm{mix}}]   }, \label{eq:feat_mix}\\
\end{equation}
 
 \noindent
 where $r$ is the extinction coefficient ratio $r=\alpha_{\mathrm{f}}/\alpha_{\mathrm{d}}$. The coefficients $\alpha_{\mathrm{d}}$ and $\alpha_{\mathrm{f}}$ are extinction coefficients for dust extinction and the feature, respectively, in units of (length)$^{-1}$.
 An important aspect of well-mixed geometry is that $F/F_{0}$ cannot be smaller than $\left(1+r\right)^{-1}$ even if $\tau\rightarrow\infty$. In other words, if an observed flux ratio $F/F_{0}$ is lower than $\left(1+r\right)^{-1}$, the system cannot be well-mixed.
 
 The extinction coefficient ratio $r$ for the 3.4 $\mathrm{\,\mu m}$ aliphatic carbon absorption feature is written as
\begin{equation}
 \label{eq:r3.4}
	r_{3.4}=1.086 \frac{\tau_{3.4}}{\tau_{9.7}}\frac{\tau_{9.7}}{A_{V}}\frac{A_{V}}{A_{3.4}}.
\end{equation}
Equation \ref{eq:r3.4} suggests that $r_{3.4}$ is at most 0.2 when the largest $\tau_{3.4}/\tau_{9.7}$ ratio observed toward GCIRS3 ($\tau_{3.4}/\tau_{9.7}$=0.086), $\tau_{9.7}/A_{V}=1/9$ toward the GC \citep{1985MNRAS.215..425R}, and an extinction curve proposed by Weingartner and Draine (\citeyear{2001ApJ...548..296W}) for $R_{V}=3.1$, are assumed.
 Consequently, if $\tau_{3.4}^{\mathrm{sc}}>\ln{(1+0.2)}>0.2$, then it can be conservatively concluded that that system is not well mixed.
 25\% of our sample (12/48 objects) have $\tau_{3.4}$ larger than 0.2.
 
 The extinction coefficient ratio $r$ for the 9.7$\mathrm{\,\mu m}$ silicate absorption feature, $r_{9.7}$, can also be expressed as
\begin{equation}
	r_{9.7}=1.086 \frac{\tau_{9.7}}{A_{V}}    \frac{A_{V}}{A_{9.7}^{\mathrm{dust}}},
\end{equation}
where $A_{9.7}^{\mathrm{dust}}$ represents \textit{featureless} extinction at $9.7\mathrm{\,\mu m}$, which means extinction not due to Si--O bonds in silicate grains.
 Interpolating the sides of the 9.7$\mathrm{\,\mu m}$ extinction feature in the extinction curve for $R_{V}=3.1$, we calculated $A_{V}/A_{9.7}^{\mathrm{dust}} = 0.0166^{-1}$.
 We also assumed $\tau_{9.7}/A_{V}=1/18.0$ \citep{2003dge..conf.....W}. Consequently, $r_{9.7}=3.63$, and this leads to the conclusion that $F/F_{0}$ at 9.7$\mathrm{\,\mu m}$ cannot be smaller than 0.216 in well-mixed geometry.
 If $F/F_{0}> 0.216$ is interpreted in the screen geometry, then the lower limit of the flux ratio corresponds to the upper limit of $\tau_{9.7}^{\mathrm{sc}}<1.53$.
 62\% of our ULIRG sample where $\tau_{9.7}$ is available (28/45 objects) have  $\tau_{9.7}>1.53$, and these ULIRGs are therefore not in well-mixed geometry. 
 
 \subsubsection{Geometric temperature gradient}
 \label{sec:tgtg}
 
 \citet{2000MNRAS.319..331I} observed the 3.4 $\mathrm{\,\mu m}$ absorption feature toward obscured AGN (based on optical or X-ray classification and strong silicate absorption at $\sim$ 10$\mathrm{\,\mu m}$) and found that some exhibited larger $\tau_{3.4}/\tau_{9.7}$ ratios than Galactic sources.
 \citet{2008PASJ...60S.489I} argued that large ($>$ 0.2) $\tau_{3.4}$ was a sign of buried AGN, because the sources presented in Imanishi and Maloney (\citeyear{2003ApJ...588..165I}) that exhibit $\tau_{3.4}>0.2$ are known to have buried AGN. 
 If a luminous central source such as AGN is buried in dust, hot dust responsible for the 3.4$\mathrm{\,\mu m}$ continuum emission is expected to reside closer to AGN than dust emitting the $9.7\mathrm{\,\mu m}$  continuum. This leads to the enhanced $\tau_{3.4}/\tau_{9.7}$ ratio \citep{2000MNRAS.319..331I}.
 \citet{2000MNRAS.319..331I} therefore proposed that the enhanced $\tau_{3.4}/\tau_{9.7}$ ratio indicates a strong, geometric dust temperature gradient.
 \citet{2007ApJS..171...72I} extended this ``geometric temperature gradient'' model to argue that enhanced $\tau_{3.4}/\tau_{9.7}$ in a ULIRG indicates the presence of buried AGN.
 
\begin{figure}
 \begin{center}
 \hspace{-50mm}
 \includegraphics[width=0.9\linewidth]{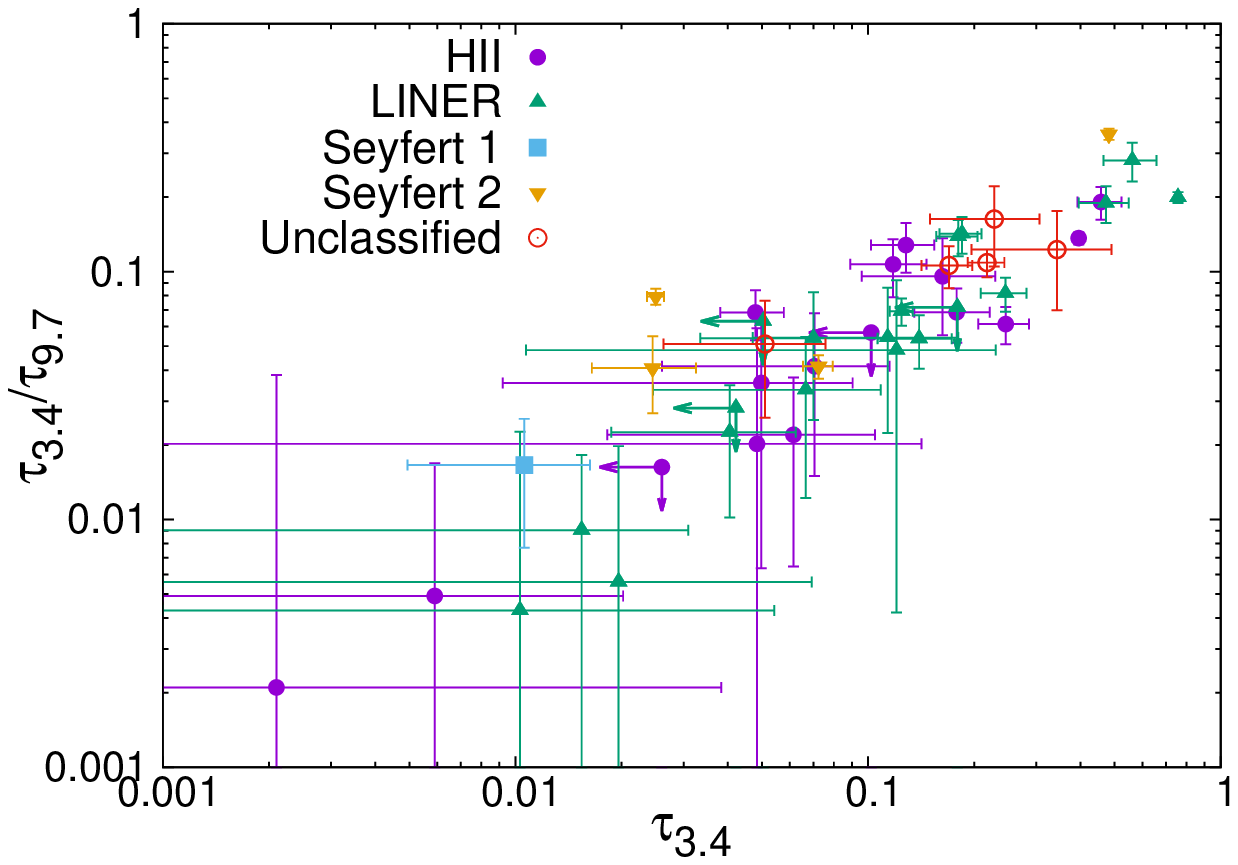}
 \end{center}
 \caption{Distribution of $\tau_{3.4}$ and $\tau_{3.4}/\tau_{9.7}$. The symbols represent the same as figure \ref{fig:av_hac}.}
 \label{fig:HAC-HACSil}
 \end{figure}

 We check if the enhanced $\tau_{3.4}/\tau_{9.7}$ ratio arises only when buried AGN exist in our sample.
 First, we show the relation between the optical depth of aliphatic carbon and its ratio to the silicate dust in figure \ref{fig:HAC-HACSil}.
 Clearly, large $\tau_{3.4}$ ($>$ 0.2), which indicates screen geometry (see section \ref{sec:obscg}), indeed always accompanies a large $\tau_{3.4}/\tau_{9.7}$ ratio.
 This is consistent with the geometric temperature gradient model that needs screen geometry.
 We also note on the silicate absorption feature.
 $\tau_{9.7}$ also cannot be larger than a certain value, 1.5,  in well-mixed geometry of sources and the Galactic ISM (see section \ref{sec:obscg}).
 Out of 12 ULIRGs which exhibit large $\tau_{3.4}$ ($>$ 0.2), 9 ULIRGs also exhibit large $\tau_{9.7}$ ($>$ 1.5), which is consistent with screen geometry. The remaining 3 ULIRGs are IRAS 02480$-$3745 ($\tau_{9.7} =$ 1.4), IRAS 19254$-$7245 ($\tau_{9.7} =$ 1.345), and IRAS 14394$+$5332 (where $\tau_{9.7}$ is not available). $\tau_{9.7}$ in IRAS 02480$-$3745 and IRAS 19254$-$7245 are slightly smaller than 1.5. 
 In summary, the sources which exhibit the enhanced $\tau_{3.4}/\tau_{9.7}$ ratio generally need screen geometry and it is consistent with the geometric temperature gradient model.
 
 Next, we investigate the relations between $\tau_{3.4}$ and (buried) AGN signs.
 The signs we consider here are (1) optical classification (table \ref{tab:result_sil}), (2) NIR color, and (3) the 3.3 $\mathrm{\,\mu m}$ PAH emission.
 
 For (1) optical classification, not all the Seyfert galaxies have enhanced $\tau_{3.4}/\tau_{9.7}$ (e.g., Mrk 273, a Seyfert 2 ULIRG). Moreover, not only Seyfert-type ULIRGs but also non-Seyfert galaxies exhibit large $\tau_{3.4}/\tau_{9.7}$, and the two types of ULIRGs seem to distribute similarly, as for the $\tau_{3.4}/\tau_{9.7}$ ratio (figure \ref{fig:h2ohac}).
  Quantitatively, Seyfert galaxies in our sample have weighted averages of $\left<\tau_{3.0}/\tau_{9.7}\right>=0.045 \pm 0.004$ and $\left<\tau_{3.4}/\tau_{9.7}\right>=0.050 \pm 0.004$ for those with positive $\tau_{3.0}/\tau_{9.7}$ or $\tau_{3.4}/\tau_{9.7}$, whereas non-Seyfert galaxies in our sample have $\left<\tau_{3.0}/\tau_{9.7}\right>=0.072 \pm 0.002$ and $\left<\tau_{3.4}/\tau_{9.7}\right>=0.071 \pm 0.003$.
 Note that for non-Seyfert galaxies, IRAS 08572$+$3915 has by far the highest weight in determination of the weighted averages (10 times as large as the second highest weight). If the weighted average is calculated without IRAS 08572$+$3915, non-Seyfert galaxies have $\left<\tau_{3.0}/\tau_{9.7}\right>=0.110 \pm 0.004$ and $\left<\tau_{3.4}/\tau_{9.7}\right>=0.061 \pm 0.003$. 
 The $\tau_{3.4}/\tau_{9.7}$ ratio exhibits significantly less dependence on the presence of optical Seyferts indications, compared to the $\tau_{3.0}/\tau_{9.7}$ ratio.
  
\begin{figure}
 \begin{center}
 \hspace{-50mm}
 \includegraphics[width=0.9\linewidth]{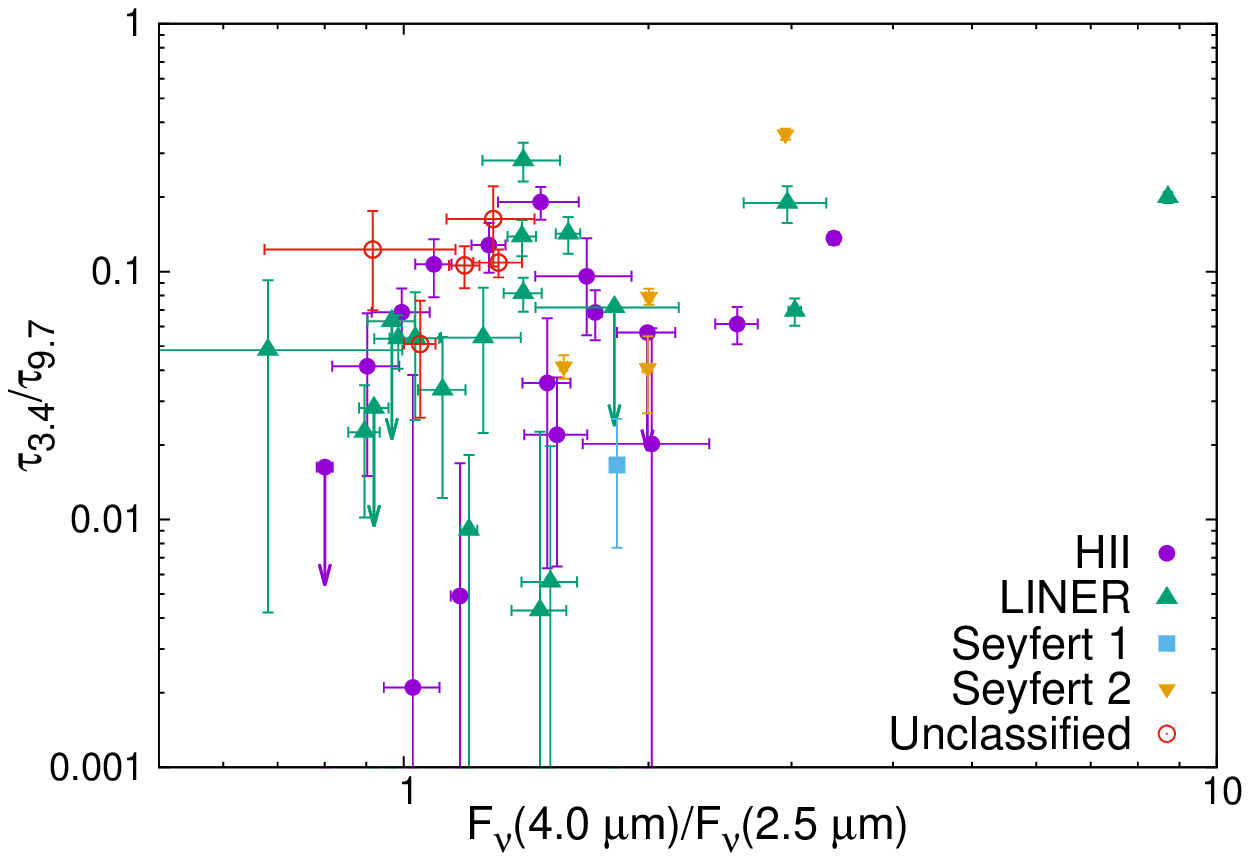}
 \end{center}
 \caption{Observed NIR color and the ratio of the aliphatic carbon absorption to silicate. The symbols represent the same as figure \ref{fig:av_hac}.}
 \label{fig:NIRcol-HACSil}
 \end{figure}

 If AGN are heavily buried in dust, optical classification could miss an AGN sign.
 (2) NIR color is hence sometimes used to discern optically elusive AGN (\citealt{2006MNRAS.365..303R}; \citealt{2008ApJ...675...96S}; \citealt{2010MNRAS.401..197R}), because optically elusive AGN imply that they are enshrouded by a large amount of dust. AGN radiation heats up the enshrouding dust as high as $\sim 10^{2}$ K (e.g., \citealt{2007ApJ...656..148A}). This makes the infrared spectrum peak in the MIR region and reddens NIR color.
 \citet{2010MNRAS.401..197R} classified a ULIRG with spectral slope $\Gamma \lesssim -0.5\,(F_{\lambda}\propto\lambda ^{\Gamma})$, which corresponds to $F_{\nu}(4.0\mathrm{\,\mu m})/F_{\nu}(2.5\mathrm{\,\mu m})\lesssim 2.0$, as pure starburst or unobscured AGN, on the basis of known AGN or starburst.
 Figure \ref{fig:NIRcol-HACSil} shows the observed flux ratio $F_{\nu}(4.0\mathrm{\,\mu m})/F_{\nu}(2.5\mathrm{\,\mu m})$ (in the rest-frame) versus $\tau_{3.4}/\tau_{9.7}$.
 As shown in figure  \ref{fig:NIRcol-HACSil}, ULIRGs in which NIR color indicates the  presence of buried AGN ($F_{\nu}(4.0\mathrm{\,\mu m})/F_{\nu}(2.5\mathrm{\,\mu m}) > 2.0$) indeed exhibit a large $\tau_{3.4}/\tau_{9.7}$ ratio; however, those with starburst-like NIR color ($F_{\nu}(4.0\mathrm{\,\mu m})/F_{\nu}(2.5\mathrm{\,\mu m}) < 2.0$) also have a variety of $\tau_{3.4}/\tau_{9.7}$ ratios, spanning from 0 to $\sim$0.28. The enhanced $\tau_{3.4}/\tau_{9.7}$ ratio is observed regardless of NIR color.
 
\begin{figure}
 \begin{center}
 \hspace{-50mm}
 \includegraphics[width=0.9\linewidth]{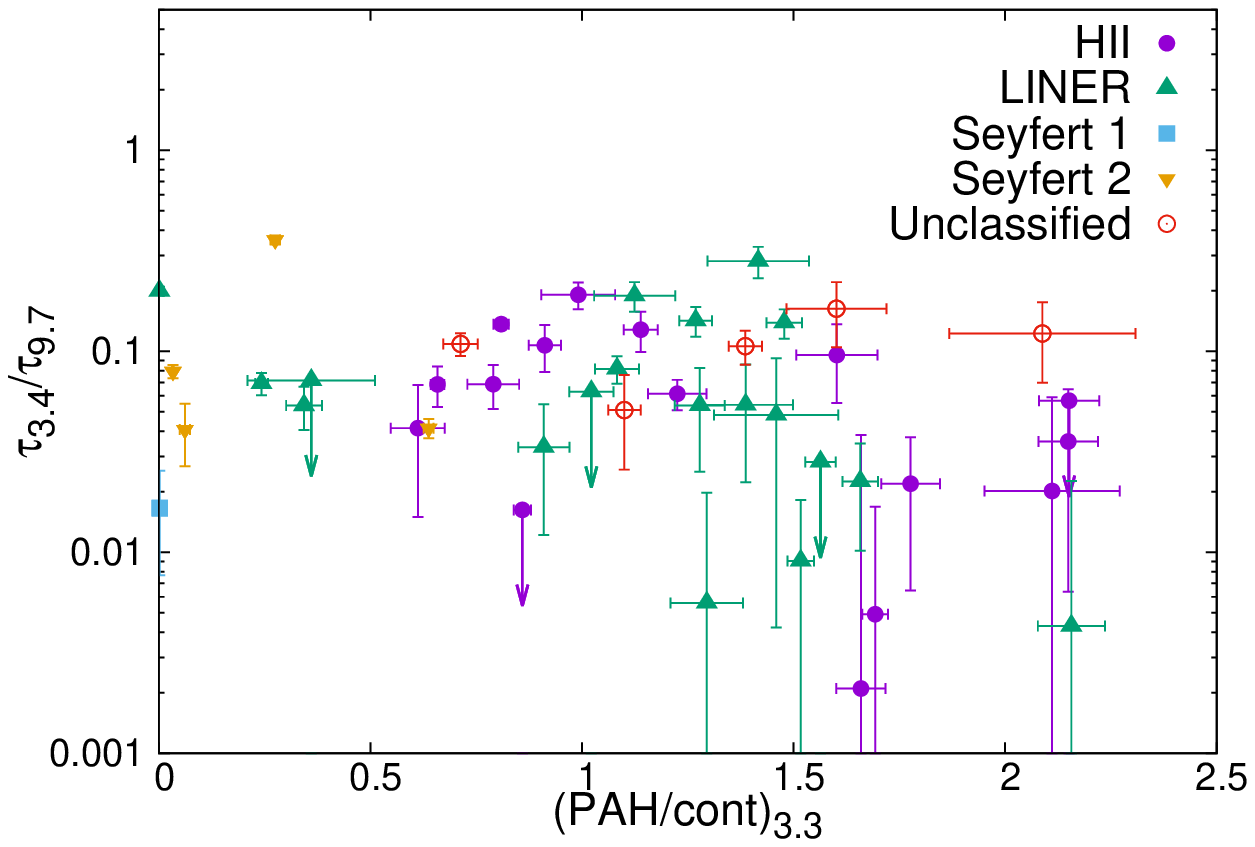}
 \end{center}
 \caption{PAH emission strength to underlying continuum at $3.3\mathrm{\,\mu m}$ and the ratio of the aliphatic carbon absorption to silicate. The symbols represent the same as figure \ref{fig:av_hac}.}
 \label{fig:PAH-HACSil}
 \end{figure}

 Hard UV radiation from AGN has been proposed as a destruction source of PAH molecules (\citealt{1986AA...166....4M}; \citealt{2000ApJ...545..701I}).
 (3) PAH emission is therefore also a useful tool for diagnosing AGN contribution.
 Imanishi and Dudley (\citeyear{2000ApJ...545..701I}) suggested that an equivalent width of the 3.3 $\mathrm{\,\mu m}$ PAH feature $EW_{3.3}$ lower than 30 nm is an indication of AGN. $EW_{3.3}$ = 30 nm corresponds to (PAH/cont)$_{3.3}$ = 0.59, where (PAH/cont)$_{3.3}$ is the PAH/continuum flux ratio at 3.3$\mathrm{\,\mu m}$ (i.e., $f_{\nu}^{\mathrm{PAH}}/f_{\nu}^{\mathrm{cont}}$ at 3.3$\mathrm{\,\mu m}$).
 Figure \ref{fig:PAH-HACSil} shows a plot of  (PAH/cont)$_{3.3}$ versus $\tau_{3.4}/\tau_{9.7}$.
 ULIRGs in which (PAH/cont)$_{3.3}$ suggests the presence of AGN tend to have larger $\tau_{3.4}/\tau_{9.7}$. However, figure \ref{fig:PAH-HACSil} clearly indicates that ULIRGs with significant PAH emissions still exhibit a wide distribution of $\tau_{3.4}/\tau_{9.7}$, as high as $\sim$0.28.
 
 To summarize, for ULIRGs with enhanced $\tau_{3.4}/\tau_{9.7}$, observed $\tau_{3.4}$ and $\tau_{3.4}/\tau_{9.7}$ are consistent with the geometric temperature gradient model, but on the basis of three AGN indications (optical classification, red NIR color, and weak PAH emission at 3.3 $\mathrm{\,\mu m}$), we conclude that the central source that causes geometric temperature gradient is not always AGN.
 
 \subsubsection{Diversity of the UV environment}
 \label{sec:tdotue}
 
 The second scenario that we consider as the cause of the $\tau_{3.4}/\tau_{9.7}$ diversity is the diversity of the UV environment in star-forming regions. \citet{2001AA...367..355M} experimentally showed that aliphatic carbon that produces the 3.4$\mathrm{\,\mu m}$ absorption feature is destroyed by UV radiation. In contrast, silicate dust is not so easily destroyed by UV radiation, so $\tau_{9.7}$ does not change even in a relatively intense UV radiation field.
 The different natures of aliphatic carbon and silicate could provide an alternative explanation for the diversity of the $\tau_{3.4}/\tau_{9.7}$ ratio in the context of the UV environment. Assuming that production of the 3.4 $\mathrm{\,\mu m}$ absorption carrier balances destruction by UV radiation,  we can interpret the diversity of $\tau_{3.4}/\tau_{9.7}$ as an indication of a different UV environment in high-mass star-forming regions in ULIRGs; lower $\tau_{3.4}/\tau_{9.7}$ ratios than those in GC sources or Cygnus OB2 indicate that a UV field stronger than that in Galactic sources irradiates dust in ULIRGs and reduces $\tau_{3.4}$ by destroying aliphatic carbon.
 
\begin{figure}
 \begin{center}
 \hspace{-50mm}
 \includegraphics[width=0.9\linewidth]{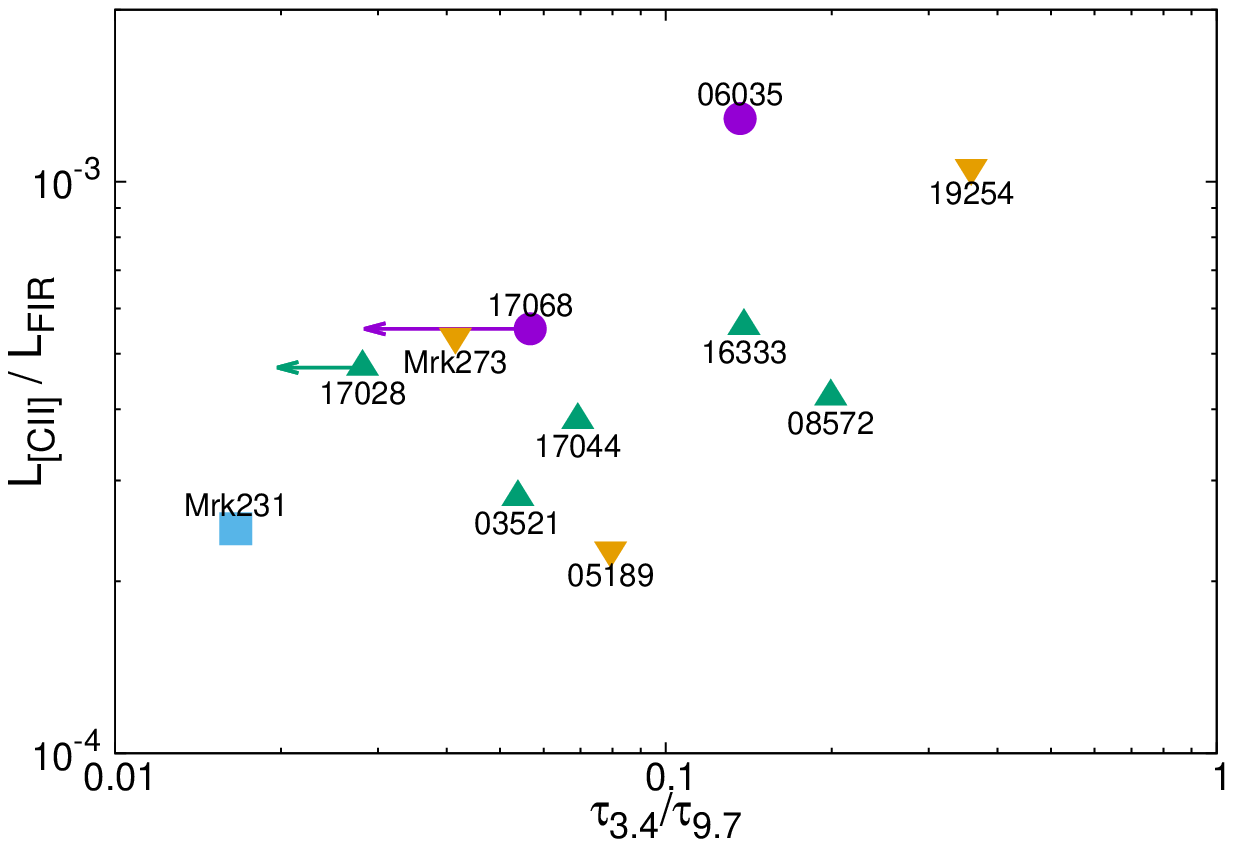}
 \end{center}
 \caption{The plot of the luminosity ratio of the [C \emissiontype{II}] 158 $\mathrm{\,\mu m}$ line  to the far-infrared (\citealt{2003ApJ...594..758L}; \citealt{2012ApJ...755..171S}; \citealt{2013ApJ...774...68D}; \citealt{2013ApJ...776...38F}; \citealt{2014ApJ...790...15S}) versus the ratio of the aliphatic carbon absorption to silicate, with object names (abbreviated for IRAS objects). Far-infrared flux densities $F_{\mathrm{FIR}}$ are calculated from $F_{\mathrm{FIR}} = 1.26\times 10^{-14} (2.58F_{60} + F_{100})$ (W/m$^{2}$)  \citep{1996ARAA..34..749S} where $F_{60}$ and $F_{100}$ are the IRAS flux densities in Jy at 60 and 100 $\mathrm{\,\mu m}$, respectively. The symbols represent the same as figure \ref{fig:av_hac}.}
 \label{fig:CII-HACSil}
 \end{figure}

 To check the validity of the idea that an intense UV environment reduces the $\tau_{3.4}/\tau_{9.7}$ ratio, we assess the correlation between the ratio of the [C \emissiontype{II}] 158 $\mathrm{\,\mu m}$ line luminosity to the far-infrared luminosity, $L_{[\mathrm{C} \emissiontype{II}]}/L_{\mathrm{FIR}}$, and the $\tau_{3.4}/\tau_{9.7}$ ratio.
 The $[\mathrm{C} \emissiontype{II}]$ 158 $\mathrm{\,\mu m}$ line is a major coolant in neutral interstellar gas, whereas FIR emission generally represents gas heating. If the collisional excitation by gas is assumed for the $[\mathrm{C} \emissiontype{II}]$ line, the $L_{[\mathrm{C} \emissiontype{II}]}/L_{\mathrm{FIR}}$ ratio is roughly characterized by  $G/n_{\mathrm{H}}$, where $G$ is the UV energy density.
 It is known that local ULIRGs have significantly lower $L_{[\mathrm{C} \emissiontype{II}]}/L_{\mathrm{FIR}}$ than normal star-forming galaxies (commonly referred to as [C \emissiontype{II}] deficit; e.g., \citealt{1998ApJ...504L..11L}). The mechanisms proposed to explain the [C \emissiontype{II}] deficit, such as ``high $G_{0}/n$ in photo-dissociation regions (PDR)'' (\citealt{2001ApJ...561..766M}; \citealt{2013ApJ...776...38F}; \citealt{2013ApJ...774...68D}), indicate that dust is exposed to a more intense UV environment in the ULIRGs with the [C \emissiontype{II}] deficit.
 Figure \ref{fig:CII-HACSil} shows the correlation between $\tau_{3.4}/\tau_{9.7}$ and $L_{[\mathrm{C} \emissiontype{II}]}/L_{\mathrm{FIR}}$. The $\tau_{3.4}/\tau_{9.7}$ ratio decreases together with the $L_{[\mathrm{C} \emissiontype{II}]}/L_{\mathrm{FIR}}$ ratio (the correlation coefficient $r$ is 0.66). This correlation therefore implies that $\tau_{3.4}/\tau_{9.7}$ diversity reflects the UV environment in star-forming regions: $\tau_{3.4}/\tau_{9.7}$ ratios lower than those observed toward the Galactic sources indicate that destruction by UV radiation prevails. 
 
 \citet{2001AA...367..347M} discussed the balance of the destruction and formation of aliphatic carbon. According to their calculation, if the ratio of the UV energy density to the hydrogen number density $G/n_{\mathrm{H}}>3\,(G_{0}\,\mathrm{cm}^{3})$ ($G_{0}$ corresponds to the UV energy density in Galactic diffuse clouds), the destruction rate of aliphatic carbon surpasses the formation rate of aliphatic carbon in the Milky Way.
 It can therefore be expected that low $\tau_{3.4}/\tau_{9.7}$ ratios observed toward some ULIRGs indicate $G/n_{\mathrm{H}}>3$ and $G\gtrsim300G_{0}$ (if $n_{\mathrm{H}}\sim10^{2}$ cm$^{-3}$;  \citealt{2013ApJ...776...38F}) in their star-forming regions. \citet{2017ApJ...846...32D} also presented the $G/n_{\mathrm{H}}$ ratio in local ULIRGs. By exploiting the intensity ratio of the [C \emissiontype{II}] line to FIR predicted by PDR models, they concluded from the observed [C \emissiontype{II}]/FIR ratio that PDRs in local ULIRGs have $G/n_{\mathrm{H}}>2$. This criterion is in good agreement with our criterion, $G/n_{\mathrm{H}}>3$, indicated by our observed $\tau_{3.4}/\tau_{9.7}$ ratio.
  
  In conclusion, we propose that two separate effects account for the $\tau_{3.4}/\tau_{9.7}$ diversity. One is the geometric temperature gradient scenario, which can enhance the $\tau_{3.4}/\tau_{9.7}$ ratio. ULIRGs with a AGN sign (optical classification, red NIR color, or weak PAH emission at 3.3$\mathrm{\,\mu m}$) tend to exhibit a large $\tau_{3.4}/\tau_{9.7}$ ratio; however, we also found that this effect could be significant in ULIRGs without the AGN signs.
  The other effect is the diversity of the UV environment in star-forming regions, in which scenario the $\tau_{3.4}/\tau_{9.7}$ ratio can be reduced if $G/n_{\mathrm{H}}>3$. The criterion for the reduction of the aliphatic carbon optical depth is comparable to that obtained in previous research to explain the $L_{[\mathrm{C} \emissiontype{II}]}/L_{\mathrm{FIR}}$ ratio in ULIRGs \citep{2017ApJ...846...32D}.
  
%% file: input/conclusion_10.tex
 We have presented observation results for 48 local ULIRGs using AKARI IRC NIR grism spectroscopy. 
 We acquired the NIR spectra (2.5--4.0 $\mathrm{\,\mu m}$ in the rest frame) and estimated the optical depth at 3.0 $\mathrm{\,\mu m}$ by H$_{2}$O ice and at 3.4 $\mathrm{\,\mu m}$ by aliphatic carbon.
  
 The optical depth ratios of H$_{2}$O ice to silicate, $\tau_{3.0}/\tau_{9.7}$, in the ULIRGs are always lower than those in the Taurus dark cloud, a UV-attenuated low-mass star-forming region in the Milky Way, and the ULIRGs have values rather similar to those toward the GC or Cygnus OB2, where dust is exposed to more intense UV environment than that in the Taurus dark cloud.
 This suggests that not only low-mass star-forming regions but also a significant amount of high-mass star-forming regions compose ULIRGs. 
 
 On the other hand, the ULIRGs exhibit a variety of optical depth ratios of aliphatic carbon to silicate, some exhibiting values significantly higher or lower than the Galactic values.
 As \citet{2000MNRAS.319..331I} proposed, the geometric temperature gradient model can qualitatively explain large $\tau_{3.4}/\tau_{9.7}$ ratios: the model proposes that if dust enshrouds a luminous central source (often taken to be AGN), hot dust dominating the 3.4 $\mathrm{\,\mu m}$ continuum resides deeper than does hot dust dominating the 9.7 $\mathrm{\,\mu m}$ continuum.
 On the basis of optical classification, NIR color ($F_{\nu}(4.0\mathrm{\,\mu m})/F_{\nu}(2.5\mathrm{\,\mu m})$), or the PAH emission to continuum ratio (PAH/cont)$_{3.3}$, we examined the connection between the large $\tau_{3.4}/\tau_{9.7}$ ratios and the presence of AGN, as some previous studies have suggested. Our results show that not only AGN but also starburst-dominated ULIRGs exhibit significantly large $\tau_{3.4}/\tau_{9.7}$ ratios; therefore, the geometric temperature gradient also occurs in ULIRGs without the AGN signs.
 Regarding the low $\tau_{3.4}/\tau_{9.7}$ ratio, the correlation between $\tau_{3.4}/\tau_{9.7}$ and $L_{[\mathrm{C} \emissiontype{II}]}/L_{\mathrm{FIR}}$ ([C \emissiontype{II}] deficit) suggests that the high ratio of UV environment to hydrogen density $G/n_{\mathrm{H}}$ ($>$3) in star-forming regions in the ULIRGs causes the destruction of aliphatic carbon, which is the carrier of the 3.4 $\mathrm{\,\mu m}$ absorption, and lowers the $\tau_{3.4}/\tau_{9.7}$ ratio. We propose that the combination of the geometric temperature gradient effect and the high $G/n_{\mathrm{H}}$ ratio in the star-forming regions in ULIRGs is the cause of the diversity of the $\tau_{3.4}/\tau_{9.7}$ ratio.

 We would like to thank the anonymous referee for the thoughtful review and the comments that significantly improved the article.
 This research is based on observations with AKARI, a JAXA project with the participation of ESA.
 This research was supported by JSPS KAKENHI Grant Number JP26247030 (T.N.).